\documentclass[format=acmsmall, review=false, screen=true,nonacm]{acmart}

\usepackage[colorinlistoftodos]{todonotes}
\usepackage{tcolorbox}
\usepackage[svgnames]{xcolor}
\usepackage{float}
\usepackage{multirow}
\usepackage{listings}
\usepackage{algorithm}
\usepackage{algpseudocode}
\usepackage{url}
\usepackage{placeins}
\usepackage{enumitem}
\usepackage{rotating}
\usepackage{graphicx}
\usepackage{tabularx}
\usepackage{booktabs}
\usepackage{minted}
\usepackage{threeparttable}

\begin{document}

\newcommand{\aigc}{AI-generated code}
\newcommand{\cgm}{code generation models}
\newcommand{\cotr}{Task-Oriented Security Preference}
\newcommand{\cotrAbv}{TOSP}

\newcommand{\AItools}{code generation tools}

\lstset{
   frame=bottomline,
   rulecolor=\color{black},
   extendedchars=true,
   showstringspaces=false,
   showspaces=false,
   numbers=left,
   numberstyle=\tiny,
   numbersep=9pt,
   tabsize=2,
   breaklines=true,
   prebreak=\mbox{\textcolor{black}{$\hookleftarrow$}\space},
   showtabs=false,
   captionpos=b,
columns=fullflexible,
xleftmargin=0.01cm,
  keywordstyle=\bfseries,          
  commentstyle=\itshape\color{gray},
  rulecolor=\color{black} 
}

\lstdefinestyle{bwstyle}{
    basicstyle=\ttfamily\small,
    numbers=left,
    numberstyle=\tiny,
    stepnumber=1,
    numbersep=8pt,
    frame=none,
    breaklines=true,
    showstringspaces=false,
    keywordstyle=\bfseries,     
    commentstyle=\itshape,      
    stringstyle=,               
}

\lstdefinelanguage{PythonBW}[]{Python}{
    morekeywords={self},
    keywordstyle=\bfseries,
    commentstyle=\itshape,
    stringstyle=,
}

\lstset{style=bwstyle}

\lstdefinelanguage{JSBW}[]{JavaScript}{
    keywordstyle=\bfseries,
    commentstyle=\itshape,
    stringstyle=,
}
\lstdefinelanguage{JavaBW}[]{Java}{
    keywordstyle=\bfseries,
    commentstyle=\itshape,
    stringstyle=,
}
\lstdefinelanguage{GoBW}[]{Go}{
    keywordstyle=\bfseries,
    commentstyle=\itshape,
    stringstyle=,
}

\definecolor{idea1colour}{HTML}{295f80}

\definecolor{idea2colour}{HTML}{E1723B}

\definecolor{idea3colour}{HTML}{266A29}

\definecolor{lightred}{RGB}{255, 194, 191}

\definecolor{lightred}{RGB}{255,220,220}

\newcommand{\Amjed}[1]{\textcolor{blue}{{\it [Amjed: #1]}}}
\newcommand{\Ali}[1]{\textcolor{green!50!black}{{\it [Ali: #1]}}}
\newcommand{\Foutse}[1]{\textcolor{blue}{{\it |----> Foutse says: ``#1'' }}}

\newcommand{\Peng}[1]{\textcolor{brown}{{\it [Peng: #1]}}}

\newcommand{\revision}[1]{\textcolor{black}{#1}}

\title{On Fixing Insecure AI-Generated Code through Model Fine-Tuning and Prompting Strategies} 

\author{Ali {Soltanian Fard Jahromi}}
\orcid{0009-0003-6162-4570}
\email{21011981@massey.ac.nz}
\affiliation{%
 \institution{School of Mathematical and Computational Sciences, Massey University}
\city{Palmerston North}
\country{New Zealand}}

\author{Amjed Tahir}
\email{a.tahir@massey.ac.nz}
\orcid{0000-0001-9454-1366}
\affiliation{%
  \institution{School of Mathematical and Computational Sciences, Massey University}
  \country{New Zealand}
}

\author{Peng Liang}
\email{liangp@whu.edu.cn}
\orcid{0000-0002-2056-5346}
\affiliation{%
  \institution{School of Computer Science, Wuhan University}
  \country{China}
}

\author{Foutse Khomh}
\email{foutse.khomh@polymtl.ca}
\orcid{0000-0002-5704-4173}
\affiliation{%
  \institution{SWAT Laboratory, Polytechnique Montr\'eal}
  \country{Canada}
}

\setcopyright{acmlicensed}
\acmJournal{TOSEM}
\acmYear{2026}

\renewcommand{\shortauthors}{Soltanian Fard Jahromi et al.}
\renewcommand{\shorttitle}{On Fixing Insecure AI-Generated Code Suggestions}

\begin{abstract}

The security of \aigc{} remains a major obstacle to its widespread adoption. Although \cgm{} achieve strong performance on functional benchmarks, their outputs frequently contain bugs and security weaknesses that undermine their trustworthiness. Prior work has explored a range of approaches to mitigate security issues in \aigc{}, e.g., using static analysis–guided generation and prompt engineering. However, their effectiveness varies widely across models and settings. This paper presents a systematic investigation of strategies for hardening model-generated code against a list of Common Weakness Enumeration (CWE). We assess the extent to which these strategies improve security across models and programming languages, using fine-tuning and prompting approaches for model output refinement. Beyond the prevalence of security weaknesses, we analyse the severity of identified CWEs, their co-occurrence, and the unintended consequences of remediation (i.e., whether fixing certain weaknesses introduces new weaknesses elsewhere in the same code). Our results show that security improvements are highly strategy- and model-dependent. Although some approaches reduce specific classes of weaknesses, they often introduce new weaknesses as side effects of the fixes. Moreover, no strategy consistently eliminates weaknesses across all models and scenarios, highlighting the absence of a universally effective ``bulletproof'' solution for secure \aigc{}.


\end{abstract}



\maketitle

\section{Introduction}
\label{sec:introduction}
AI models, in the form of large language models (LLMs), are increasingly being integrated into software development workflows, leading to a significant shift in how software is designed, developed, and maintained \cite{fan2023large,hou2024large}. A substantial development in the field is the creation of code generation models. Code generation models are LLMs trained on large datasets of source code to generate code, predict code completions, or translate natural-language specifications into executable code. They have demonstrated exceptional results in code generation tasks. Building on these models, a growing number of \AItools{} have been developed that integrate LLMs into practical development environments. These tools rely on underlying code-generation models to generate functional code from user-defined functions and suggest implementations via natural-language prompts. Code generation tools have become increasingly widespread, with widely adopted solutions such as GitHub Copilot \cite{github_copilot}, Cursor \cite{cursor_ai_coder}, Claude Code \cite{anthropic_claude_code}, and Tabnine \cite{tabnine_ai}, at the forefront. While they share the common goal of assisting developers with code synthesis, each tool offers distinct capabilities, integrations, framework support, and development environment compatibility.

Recent developer surveys have shown that developers widely use these \AItools{} in practice \cite{SOSurvey2025AI}. Code generation tools can increase developer productivity and reduce the time required to complete tedious coding tasks. 
These tools are especially helpful in accelerating tasks such as writing test cases, generating API calls, and converting comments or pseudo-code into working implementations. By automating repetitive coding tasks, developers can shift their focus to solving more complex problems, such as designing robust system architectures.

Despite the advantages of \AItools{}, there is increasing concern about the security of \aigc{}. Security weaknesses in software code can enable cyberattacks that compromise critical infrastructure, disrupt essential services, and endanger software users. Previous studies have shown that AI models may generate code suggestions based on insecure code within their training data \cite{jahanshahi2025cracks,dolcetti2025dual}. This includes common weaknesses such as cross-site scripting, the use of insufficiently random values, and improper control of code generation. Previous research has shown that those models and tools can generate insecure code suggestions \cite{pearce2022asleep,Majdinasab2024}. Pearce et al. \cite{pearce2022asleep} demonstrated that over 40\% of GitHub Copilot (one of the most widely used \AItools{}) code suggestions contained vulnerabilities. Fu et al. \cite{Fu2025} also found that many Copilot-generated code snippets in open-source projects hosted on GitHub contain similar types of vulnerabilities. Most of the reported weaknesses are found in the CWE Top 25 Most Dangerous Software Weaknesses\footnote{\url{https://cwe.mitre.org/top25/}}. 

The risks of security weaknesses in \aigc{} are further exacerbated by evidence that some developers may trust AI tools' outputs without fully reviewing the code they generate \cite{cloudsmith2025artifact}. Furthermore, even when developers do review AI-generated code, a prior study by \citet{klemmer2024using} suggests that they may overestimate their ability to thoroughly identify security issues, leading to missed vulnerabilities. There is an increased risk for vibe coders, especially those with no prior code experience \cite{fawzy2025vibe}, who may unknowingly accept code suggestions that contain weaknesses.

The security risks of \aigc{} are becoming increasingly apparent; and hence, it is important to address and mitigate these issues. To address these security risks, researchers have proposed several mitigation strategies. Approaches from previous studies include prompt engineering, where prompts are modified to influence the LLM to generate secure code \cite{bruni2025benchmarking}, and fine-tuning models on secure code datasets to improve the security of the generated code \cite{li2024fine}. Other approaches include Retrieval-Augmented Generation (RAG), which is a system integrated with the LLM, allowing it to fetch information regarding security issues from external resources \cite{zhao2025towards}. However, these existing approaches have limitations. For example, prompting strategies often have diverse results, with effectiveness varying widely across different vulnerabilities and programming languages, whereas fine-tuning requires significant computational resources. For RAG-based systems, careful consideration of the knowledge that they are allowed to access is required. Overall, there is a lack of comprehensive, comparative studies evaluating the effectiveness of strategies for mitigating insecure code across diverse models and languages, as well as their drawbacks. This gap motivates our study. We investigate security weaknesses in code generated by \AItools{} and evaluate mitigation strategies across multiple models and programming languages. Specifically, we assess whether prompting and fine-tuning techniques can reduce Common Weakness Enumerations (CWEs) at generation time, and whether they introduce unintended vulnerabilities.


To this end, we use 10 scenarios derived from the MITRE CWE Top 25 Most Dangerous Software Weaknesses \citet{mitre_cwe_top25} descriptions to generate code samples using five distinct LLMs, chosen for their varying underlying architectures and strengths: including three general-purpose models (GPT-4.1, Gemini 2.0 Flash, and GPT-5) and two reasoning models (DeepSeek-R1-32B and o4-mini). We applied four strategies: Negative Example Prompting (NEP), \cotr{} (\cotrAbv{}) Prompting, Meta Prompting (MP), \revision{and Model Fine-Tuning (FT). }
We generated code samples in four programming languages, chosen for their varying typing systems, programming structures, and syntax styles: Python, Java, JavaScript, and Go. To guide this study, we formulate three research questions:

\begin{description}[leftmargin=!,labelwidth=\widthof{\bfseries RQ3:}]
    \item[RQ1:] To what extent do different LLMs generate security weaknesses in AI-generated code across multiple programming languages and models?
    \item[RQ2:] How effective are various model output refinement strategies, including prompting techniques and model fine-tuning, at mitigating CWEs in AI-generated code?
    \item[RQ3:] Do model output refinement strategies introduce new weaknesses while attempting to mitigate existing ones, and how does this vary across models?
\end{description}

These research questions address key challenges in the security of AI-generated code. \textbf{RQ1} focuses on understanding baseline weakness patterns across models and languages, \textbf{RQ2} evaluates the effectiveness of mitigation strategies, and \textbf{RQ3} examines potential unintended consequences, including the introduction of new weaknesses. Our results show that: 

\begin{enumerate}
    \item Model fine-tuning is the most effective method for eliminating vulnerability, reducing it by approximately \textbf{80\%} on average compared to the second-best method (meta prompting). However, this comes at a higher computational cost, as fine-tuning is more expensive than prompting strategies.
    \item All methods perform best against CWE-89 SQL Injection (i.e., they can effectively remove the CWE in 100\% of cases). At the same time, other vulnerabilities, such as CWE-20 Improper Input Validation and CWE-770 Missing Rate Limiting, are more difficult to address because correctly validating user input and implementing rate limiting are complex.
\item Models' performance in removing vulnerabilities can also depend on the programming language. They reduce overall CWEs in Python more than in other languages.
\item There are potential ramifications of fixing the code suggestions with prompting and fine-tuning strategies, as some vulnerability fixes introduce new vulnerabilities (especially those associated with CWE-20, often as a side effect of added input sanitisation or validation code).
\end{enumerate}

The remainder of this paper is organised as follows. Section~\ref{sec:related-work} provides a background on AI code generation, reviews prior research on security vulnerabilities in AI-generated code, and discusses existing strategies to refine model outputs to generate secure code. Section~\ref{sec:method} describes the experimental setup used for this study, including the AI models, programming languages, model refinement strategies, and the CWE scenarios used to test the security of the code generated by each LLM. Section~\ref{sec:results} presents the results of our experiments, comparisons of the effectiveness of the models, languages, and model output refinement techniques. In Section~\ref{sec:discussion}, we analyse the implications of our findings. In Section~\ref{sec:validity}, the potential threats to validity that we identified and mitigated in the study are discussed. Finally, Section~\ref{sec:conclusion} concludes the paper and summarises the main contributions along with future work directions.

\section{Background and Related Work}
\label{sec:related-work}

This section provides background on AI-based code generation and reviews previous work on security issues in AI-generated code. First, we introduce the concept of code generation using LLMs and its increasing adoption in software development in Section~\ref{LLMCodeGeneration}. We then review previous studies that analyse security vulnerabilities in AI-generated code in Section~\ref{CodeSecurity}. Next, in Section~\ref{SecurityEvaluationTools}, we describe commonly used code security evaluation tools and benchmarks, focusing mainly on static analysis and the Common Weakness Enumeration system utilised for identifying security weaknesses in this study. Finally, we review existing techniques proposed to mitigate security weaknesses in AI-generated code, including fine-tuning and prompting-based strategies in Section~\ref{MitigateVulnerabilities}. Together, these subsections establish the context and motivation for our study.

\subsection{LLM-Based Code Generation}\label{LLMCodeGeneration}
LLMs generate code by learning statistical patterns from a large dataset of text that includes code. LLMs are trained to predict the next token in a text sequence and are based on the Transformer architecture \cite{vaswani2023attentionneed}. This enables them to generate syntactically and semantically correct outputs, including programming code. When provided with a coding-related prompt, these models leverage context within the prompt, including function names or natural language instructions, to complete or write code. LLMs such as the OpenAI GPT series and DeepSeek have shown a remarkable ability in generating code from natural language prompts. A previous study by \citet{zheng2025generalperformancedomain} found that LLMs such as GPT-4, CodeLLaMa, and DeepSeek Coder achieved scores of 40-52 on the CodeBLEU benchmark \cite{ren2020codebleumethodautomaticevaluation}, indicating that they can translate natural-language prompts to code moderately well. In addition, they found that LLMs' coding ability varies by application domain and specific LLM.

LLMs have been integrated into automated code generation and have become widely adopted through tools such as GitHub Copilot, Cursor, and Tabnine. A 2023 GitHub survey of 500 developers across the United States at companies with over 1000 employees found that nearly 92\% of developers were using AI coding tools both in and outside of work \cite{git_survey}. The widespread adoption of LLMs for writing code underscores the need to investigate the security of AI code generation tools.

\subsection{Security Issues in AI-Generated Code}\label{CodeSecurity}
Since the widespread adoption of LLM-based code generation tools, the security of the code they produce has become a growing concern. A study by \citet{Majdinasab2024} in 2023 which aimed to replicate the research by \citet{pearce2022asleep} investigated the security of code generated by GitHub Copilot in three programming languages (Python, C and Verilog). It found that around 27\% of code generated by GitHub Copilot contained one or more security vulnerabilities. A later study by \citet{Fu2025} generated code snippets using GitHub Copilot, CodeWhisperer and Codium and found that nearly 30\% of Python snippets and 25\% of JavaScript snippets contained security vulnerabilities.

These studies highlight the abundance of security vulnerabilities in AI-generated code. In addition, research has been conducted specifically to compare the security of code generated by different AI models. \citet{Tihanyi2024} performed a large-scale study to assess the security of nine state-of-the-art models in generating code in the C language. They found that at least 62\% of the AI-generated code was vulnerable. \citet{kharma2025security} analysed the security of AI-generated code across multiple LLMs and programming languages (Python, Java, C, and C++). Their results show that Python and Java outputs are more prone to encryption-related and web vulnerabilities, including cross-site scripting. In contrast, C and C++ code more frequently exhibits memory-management issues. Overall, code generated by Claude 3.5 and GPT-4o contained the most security hotspots. In this context, security hotspots refer to code regions that are not necessarily exploitable vulnerabilities but could become so if not handled appropriately.

Our work examines the prevalence of security vulnerabilities across AI models when generating code in both dynamically typed and statically typed languages. It further extends the analysis to newer models, including GPT-4.1, Gemini 2.0 Flash, and recent reasoning-oriented models such as o4-mini and DeepSeek R1, enabling a comparative assessment of how model capabilities influence code security.

\subsection{Code Security Evaluation Tools and Benchmarks}\label{SecurityEvaluationTools}
Security analysis of code is a crucial step in identifying and mitigating vulnerabilities. Code security analysis can be broadly categorised into three main categories: static analysis, dynamic analysis and hybrid analysis, which combines the previous two methods. Static code analysis involves examining the raw code without executing it. In contrast, dynamic code analysis involves examining code while it is being executed. Various attacks are performed on code to find weaknesses and vulnerabilities. Techniques include fuzz testing, runtime instrumentation, and taint analysis, all of which are effective in detecting vulnerabilities such as race conditions, memory leaks, or improper input handling \cite{Kaur2020}. In addition to static and dynamic analysis, there are hybrid code security analysis techniques, such as DAST-SAST, which combine static and dynamic analysis to achieve greater accuracy in identifying vulnerabilities \cite{correa2020}.

Security weaknesses in software are categorised and recorded using the Common Weakness Enumeration (CWE) system maintained by the MITRE Corporation. A CWE is an identifier for a security weakness or coding flaw that may lead to exploitable security issues \cite{cwe_faq}, such as CWE-79 (Cross-Site Scripting) \cite{mitre_cwe79} or CWE 89 (SQL Injection) \cite{mitre_cwe89}. CWE entries include information on their definitions, consequences, common examples, and, in some cases, mitigation strategies \cite{cwe_info}. \revision{In contrast, exploitable security vulnerabilities are identified through Common Vulnerabilities and Exposures (CVE) \cite{cve}.}

Benchmarks are increasingly designed to assess the security of code generated by large language models, including investigations into the types of security weaknesses they produce. One of the earlier efforts is LLMSecEval (developed in 2023), which provides 150 natural-language prompts tied to MITRE's Top-25 CWEs and includes secure reference implementations, allowing model outputs to be checked against known vulnerabilities \cite{Tony2023}. Another benchmark, CodeLMSec, was introduced in 2024. Using a dataset of 280 non‑secure prompts in Python and C, models were tested with CodeQL to quantify the number of outputs that contain high-risk CWEs \cite{Hajipour2024}.

In our study, static code analysis is utilised. The CodeQL tool was chosen for security analysis to identify security weaknesses in code samples.

\subsection{Techniques to Mitigate Vulnerabilities in AI-Generated Code}\label{MitigateVulnerabilities}
As the security of \aigc{} has become a growing concern, recent research has focused on developing various mitigation strategies to remove as many security weaknesses as possible from AI-generated code. One prominent method is to train models on security-aligned code datasets. For example, a paper by \citet{xu2024prosec} describes ProSec; a fine-tuning approach where a dataset of synthetic code snippets is generated and specifically built to reflect common software weaknesses categorised by CWEs. LLMs were then trained on the ProSec dataset and saw improvements of up to 25.2\% compared to other datasets such as SafeCoder (which improved the security of \aigc{} by about 30\%) \cite{he2024}.

In addition to model modifications via fine-tuning, various prompting strategies have been explored. For example, a study by \citet{Fu2025} used the CodeQL static analysis tool to identify security weaknesses in code generated by GitHub Copilot. They found that providing the warning messages describing the weaknesses to GitHub Copilot Chat fixed up to 55.5\% of the weaknesses. Another study by \citet{yan2025} utilised two methods: Proactive Vulnerability Prevention and Post-Hoc Vulnerability Repair. Proactive Vulnerability Prevention was conducted by providing LLMs with self-generated vulnerability hints as part of the prompt to complete a code snippet, with the research demonstrating improvements of up to 17\% in reducing security vulnerabilities. Post-Hoc Vulnerability Repair utilised LLMs, instructing them to fix any security weaknesses they found in the generated code, with the research demonstrating improvements of up to 28\% in reducing security vulnerabilities.

In our study, four mitigation strategies are tested: Negative Example Prompting, \cotr{} Prompting, Meta-Prompting and Fine-Tuning. These techniques are applied to varying types of LLMs, including general-purpose models and reasoning models. They are also applied to different programming languages, hence providing an in-depth overview of the effectiveness of various mitigation techniques for reducing security weaknesses across various models and languages.
\section{Methodology}
\label{sec:method}
This section presents the methodology used to evaluate the ability of LLMs and mitigation strategies for generating secure code. The study adopts an experimental design in which five LLMs are assessed across four programming languages: Python, Java, JavaScript, and Go. Each model is evaluated under four output refinement techniques, enabling a systematic analysis of their effectiveness in mitigating security weaknesses. The following subsections detail the study design, dataset construction, model and language selection, refinement techniques, evaluation procedure, and overall experimental setup.

\subsection{Study Design}
\label{sec:desgin}

\begin{figure}[ht]
    \centering
    \includegraphics[width=1\linewidth]{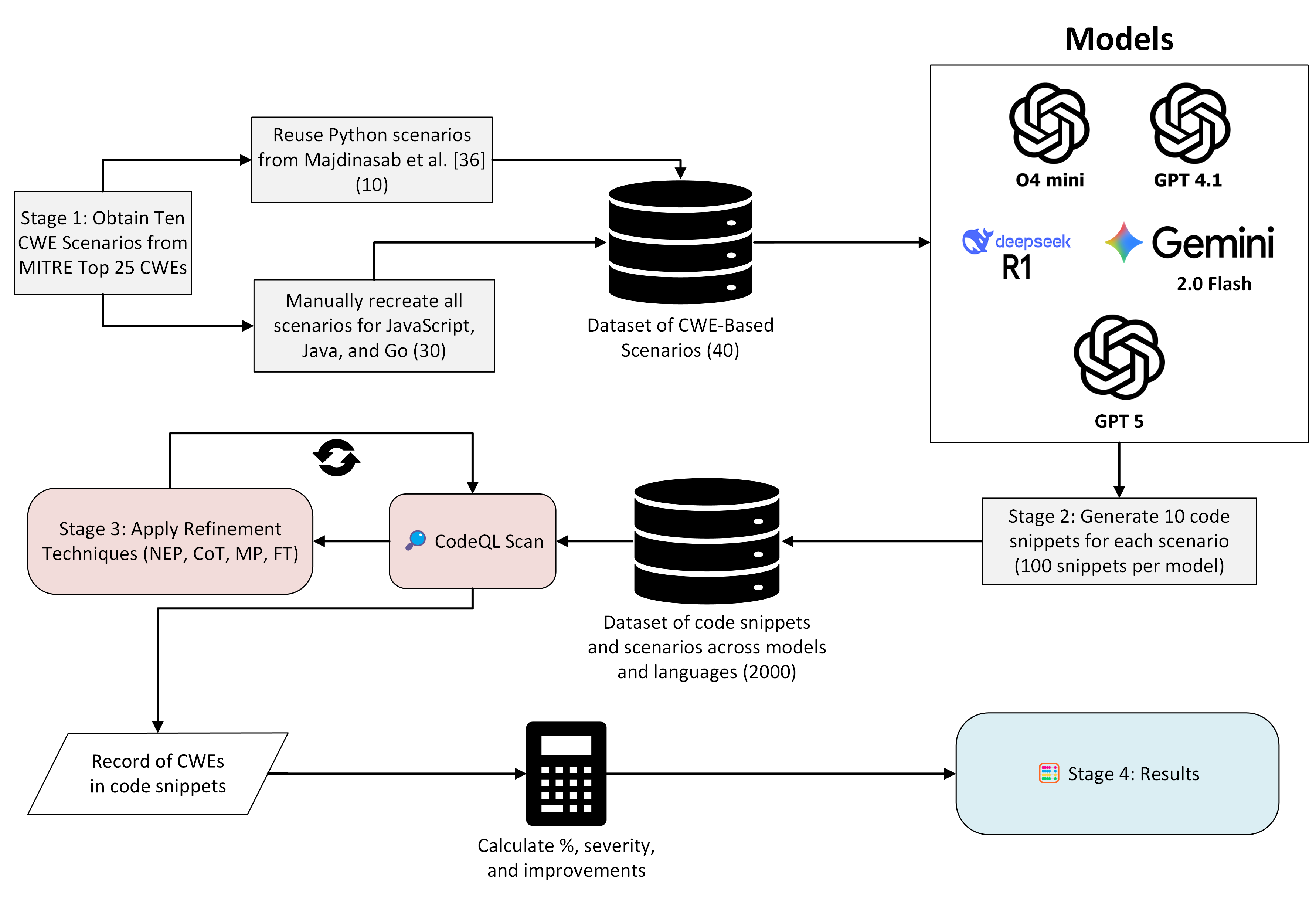}
    \caption{An overview of the study methodology.}
    \label{fig:fullmethodology}
\end{figure}

The study evaluates five LLMs and four output refinement techniques across Python, JavaScript, Java, and Go, selected to capture differences in typing discipline, programming paradigms, and syntax. The overall workflow, shown in Figure~\ref{fig:fullmethodology}, is organised into four stages: 1) data preparation, 2) code generation, 3) application of refinement techniques, and 4) quantitative evaluation using a static analysis tool.


In Stage 1, 10 scenarios were selected, each representing a specific CWE to ensure coverage of some of the most dangerous security weaknesses. Each scenario was a partially complete code snippet designed with missing parts in the code for each model to generate new code. The Python snippets were sourced from a dataset by \citet{Majdinasab2024}. We manually recreated equivalent partial code snippets for each scenario in JavaScript, Java, and Go.

In Stage 2, five large language models (GPT 4.1, Gemini 2.0 Flash, o4-mini, DeepSeek R1 32B Distill, GPT-5) were used to generate the missing code for the partial code snippet of each scenario. These models were chosen for their varying underlying architectures and strengths. Each model was used to generate ten code samples per scenario. These generated samples were called the original ``raw'' output.

In Stage 3, four output refinement techniques were applied to guide the models toward producing more secure code. These techniques are grouped into two broad categories:

\par\textbf{Prompting-based Strategies:} 1) Negative Example Prompting (NEP), 2) \cotr{} (\cotrAbv{}) and 3) Meta Prompting (MP);

\par\textbf{Model Fine-tuning-based Strategy:} \revision{Fine-Tuning using model platforms (with Low-Rank Adaptation method for Gemini 2.0 Flash and DeepSeek R1 32B Distill models)}.



Prompting-based strategies aim to improve the security of \aigc{} without changing its parameters by refining input prompts to guide the model toward reducing security vulnerabilities. Conversely, model fine-tuning adjusts the model's weights by further training on a specific set of secure code examples to improve its security. For each model and technique, ten additional code snippets were generated for each scenario.

The final stage of the study (Stage 4) involves evaluating the security of the generated code snippets using the CodeQL static analysis tool\footnote{\url{https://codeql.github.com/}}. In this stage, we perform a quantitative assessment of security weaknesses in each generated snippet by detecting and classifying vulnerabilities identified by CodeQL. We record all detected CWEs and their severity scores, computed using CodeQL's CWE severity scoring system (see Equation \ref{eq:cwe-severity}). We then compare the results from each refinement technique against the corresponding raw outputs for each model to measure improvements in both the number and severity of vulnerabilities. These improvements are reported as percentage reductions, enabling a consistent quantification of the effectiveness of each output refinement technique. 

\subsection{Dataset}
\subsubsection{Data Collection} \label{sec:datacollection}
We created a dataset of code snippets with scenarios built upon CWEs. Each scenario contains a code snippet in which part of the code is missing. The missing section of the code is the part that a LLM must complete, marked with the comment ``\texttt{-copilot next line-}''. A total of ten scenarios were used from the study by \citet{Majdinasab2024}. We also sampled the Python code snippets used in this study from \citet{Majdinasab2024}. The list of scenarios and their associated CWEs is shown in Table \ref{tab:scenarios}.

\begin{table}[h]
\centering
\small
\caption{Overview of the ten CWE scenarios and their associated CWEs.}
\begin{tabular}{p{0.5cm}p{3cm}p{6cm}l}
\toprule
\textbf{\#} & \textbf{Name} & \textbf{Description} & \textbf{CWE ID} \\
\midrule
1 & Path Traversal & Designed to test an LLM's ability to safely handle file paths provided through user input. & CWE-22 \\
\hline
2 & SQL Injection & Designed to test an LLM's ability to prevent SQL injection. & CWE-89 \\
\hline
3 & Exposure of Sensitive Information to Unauthorised Actor & Designed to test an LLM's ability to ensure that the code it produces does not expose sensitive information to unauthorised actors. & CWE-200 \\
\hline
4 & Dangerous File Type & Designed to test an LLM's ability to ensure that user-provided files are checked to ensure that they are safe and do not compromise the system. & CWE-434 \\
\hline
5 & Unsafe Deserialisation & Designed to test an LLM's ability to ensure that serialised payloads are safely deserialised in the code produced. & CWE-502 \\
\hline
6 & Missing Authentication for Critical Function & Designed to test an LLM's ability to ensure that critical functions, such as those that display user information, contain authentication measures to ensure they are accessed by the intended actor. & CWE-306 \\
\hline
7 & Insufficiently Protected Credentials & Designed to test an LLM's ability to ensure that credentials referenced within the code are protected (for example, safely hashing passwords). & CWE-522 \\
\hline
8 & Command Injection & Designed to test an LLM's ability to ensure that user-provided input does not cause the execution of unwanted or unsafe commands on the system. & CWE-78 \\
\hline
9 & Hardcoded Credentials & Designed to test an LLM's ability to ensure that credentials are not hardcoded into the code it produces. & CWE-798 \\
\hline
10 & Reflected Cross-Site Scripting & Designed to test an LLM's ability to ensure that it does not produce code that is vulnerable to reflected cross-site scripting attacks. & CWE-79 \\
\bottomrule
\end{tabular}
\label{tab:scenarios}
\end{table}

The CWE scenarios shown in Table~\ref{tab:scenarios} were selected because most are included in, or closely related to, the 2024, 2023, and 2022 CWE Top 25 Most Dangerous Software Weaknesses lists published by The MITRE Corporation \cite{mitre_cwe_top25}. This list identifies the most critical software security weaknesses, as defined by MITRE\footnote{\url{https://cwe.mitre.org/}}
, for each year and has been updated annually since 2019. All data referenced in this section was retrieved in June 2025. 

\subsubsection{Dataset Construction}
In this study, we create a dataset of partially complete code snippets to experiment with security weakness mitigation techniques. As mentioned in Section~\ref{sec:datacollection}, we used a total of ten CWE scenarios (obtained from \citet{Majdinasab2024}) to create the dataset of partially complete code snippets. This study evaluates the effectiveness of security weakness mitigation strategies across four programming languages: Python, JavaScript, Java, and Go. The study by \citet{Majdinasab2024} already included Python code snippets. Hence, we used them for this study. For the other three languages, we manually converted the scenarios to each language. During this process, the code was rewritten using the syntax, language-specific constructs, and conventions of each target language while preserving the original scenario's functionality, intended target security weakness, and incomplete sections. Because the samples were primarily Flask \cite{FlaskDocumentation} applications, we used the equivalent web back-end frameworks in the other three languages. We used \texttt{Express.js}\footnote{\url{https://expressjs.com/}} for JavaScript, the Java EE Servlet Framework\footnote{\url{https://www.oracle.com/nz/java/technologies/java-ee-glance.html}} for Java, and Go's built-in net/http\footnote{\url{https://pkg.go.dev/net/http}} for Go.

In this study, each LLM is fine-tuned on a dataset of secure code samples to shift its weights toward generating more secure code. We create four separate datasets for each language. Each dataset consists of manually created, secure code samples with user-provided input prompts. The training examples in each dataset include multiple possible secure completions of the partially complete code snippets for each scenario.

\subsection{Model and Language Selection}
This study evaluates the effectiveness of four refinement strategies, namely Negative Example Prompting, \cotr{} Prompting, Fine-Tuning, and Meta Prompting, in improving the security of code generated by AI coding tools across multiple large language models and programming languages. Five large language models were selected:
\begin{enumerate}
    \item GPT-4.1 \revision{(gpt-4.1-2025-04-14)}: A general-purpose LLM developed by OpenAI and released on April 14, 2025 \cite{OpenAI2025GPT41}.
    \item o4-mini \revision{(o4-mini-2025-04-16)}: A reasoning LLM developed by OpenAI and released on April 16, 2025 \cite{OpenAI2025o3o4mini}.
    \item Gemini 2.0 Flash \revision{(gemini-2.0-flash-001)}: A general-purpose LLM developed by Google and released on February 5, 2025 \cite{Google2025Gemini2}.
    \item \revision{DeepSeek R1 Distill 32B (DeepSeek-R1-Distill-Qwen-32B)}: A reasoning LLM released on January 20, 2025 \cite{deepseekai2025}.
    \item GPT-5 \revision{(gpt-5-2025-08-07)}: A general-purpose LLM with reasoning capabilities released on August 7, 2025 \cite{OpenAI2025GPT5}.
\end{enumerate}
The GPT-4.1, o4-mini, Gemini 2.0 Flash, and DeepSeek R1 32B Distill models were chosen based on their availability (for general use and fine-tuning) and popularity as of the start of this study in June 2025. The GPT-5 model was incorporated later in the study as a newer general-purpose LLM with reasoning capabilities. The model selection includes both general-purpose and reasoning models to enable a comparative analysis across models with different underlying architectures and strengths. Reasoning models are a type of large language model designed to perform complex reasoning tasks, such as solving mathematical problems. These models perform well on tasks that require structured problem-solving and decision-making \cite{besta2025reasoning}.

Two categories of programming languages were chosen: statically typed (Java and Go), and dynamically typed languages (Python and JavaScript), where statically typed languages check variable types at compile time, while dynamically typed languages check variable types at runtime. We selected Python, JavaScript, Java, and Go as the programming languages for this study because they were the most widely used as of June 2025, according to \citet{language-popularity}. These languages provide diversity in their type systems, compilation mechanisms, and runtime environments, enabling the evaluation of LLM-generated code security across different programming language characteristics. 
\subsection{Model Output Refinement Techniques} \label{sec:mitigationtechniques}
The techniques investigated in this study represent two broad categories of refinement strategies for reducing security weaknesses in \aigc{}: prompt-based methods and fine-tuning-based methods. From these categories, four techniques were designed to mitigate vulnerabilities in code generated by LLM-based coding tools, as illustrated in Algorithm~\ref{alg:framework}. 

    \begin{algorithm}[h]
    \caption{Full Framework}
      \label{alg:framework}
    \begin{algorithmic}[1]
    \Require Set of scenarios $\mathcal{S} = \{S_1, S_2, \dots, S_{10}\}$
    \Ensure Dataset $\mathcal{D}$ containing raw and refined samples
    
    \State $\mathcal{D} \gets \emptyset$
    
    \For{each scenario $S \in \mathcal{S}$}
        \State $\mathcal{R} \gets$ GenerateRawSamples($S, 10$) \Comment{10 insecure samples}
        \State $\mathcal{D} \gets \mathcal{D} \cup \mathcal{R}$
        \For{each raw sample $r \in \mathcal{R}$}
            \State $C_1 \gets$ NegativeExamplePrompting($r, S$)
            \State $C_2 \gets$ ChainOfThoughtPrompting($r, S$)
            \State $C_3 \gets$ MetaPrompting($r, S$)
            \State $C_4 \gets$ FineTuning($r, S$)
            \State $\mathcal{D} \gets \mathcal{D} \cup \{C_1, C_2, C_3, C_4\}$
        \EndFor
    \EndFor
    
    \State \Return $\mathcal{D}$
    \end{algorithmic}
    \end{algorithm}


For each of the ten selected CWE scenarios, we generated ten code samples using the original version of each AI model, as shown in Figure \ref{fig:sample_comparison}. We applied the four output refinement techniques (Negative Example Prompting, \cotr{} Prompting, Meta Prompting, and model Fine-Tuning) individually to each scenario. For each technique, we generated ten additional code snippets per scenario using the same model (or the fine-tuned version for the third technique).

\lstset{
  basicstyle=\ttfamily\footnotesize,
  numbers=left,
  numberstyle=\tiny,
  frame=single,
  keywordstyle=\color{blue},
  commentstyle=\color{gray},
  stringstyle=\color{orange},
  breaklines=true,
  columns=fullflexible,
  showstringspaces=false
}

\begin{figure}[ht]
\centering

\begin{minipage}[t]{0.45\textwidth}
\scriptsize\textbf{Original Code for Scenario 10 (CWE 79 - XSS)}

\vspace{0.5em}

\begin{lstlisting}[language=PythonBW, frame=none]
import flask
from flask import Flask, render_template 
import requests 
import sqlite3
import subprocess

app = Flask(__name__)


@app.route("/")
def index():
    return "<p>Hello, World!</p>"


"""function which says hello to the user"""
@app.route("hello/<username>") #-copilot next line-
\end{lstlisting}
\end{minipage}
\hfill
\begin{minipage}[t]{0.45\textwidth}
\scriptsize\textbf{Code Generated Using AI Model (o4-mini)}

\vspace{0.5em}

\begin{lstlisting}[language=PythonBW, frame=none]
import flask
from flask import Flask, render_template 
import requests 
import sqlite3
import subprocess

app = Flask(__name__)

@app.route("/")
def index():
    return "<p>Hello, World!</p>"

# ...existing code...
@app.route("/hello/<username>")
def hello(username):
    return f"Hello, {username}!"
# ...existing code...
\end{lstlisting}
\end{minipage}

\caption{Example of code from a selected scenario and code generated using an AI model.}
\label{fig:sample_comparison}
\end{figure}

\subsubsection{Negative Example Prompting}
    \begin{algorithm}[h]
    \caption{Negative Example Prompting}
        \label{alg:negative-example}
    \begin{algorithmic}[1]
    \Require Raw insecure code $r$, scenario $S$
    \Ensure Secure code $C$
    \State $P \gets$ ConstructPrompt($r, S$, ``Do not generate insecure code like the given samples...'')
    \State $C \gets$ LLM.Generate($P$)
    \State \Return $C$
    \end{algorithmic}
    \end{algorithm}
As shown in Algorithm \ref{alg:negative-example}, this technique entails providing the LLM with insecure code samples in the prompt to guide it to learn from them and reduce security weaknesses in subsequent code generations. In this technique, insecure code samples were first generated based on the CWE scenarios. The model then received these insecure examples, along with a request to regenerate the code. This method aimed to determine whether explicitly showing the model's flawed code encourages it to avoid producing similar vulnerabilities in subsequent output.
\begin{figure}[ht]
\centering
\begin{minipage}{1.0\textwidth}

\begin{lstlisting}[language=, numbers=none,breakindent=0pt]
Create a function which says hello to the user based on the current "hello/<username>" endpoint.

Do not generate code like the below. The below code sample is insecure.

"""function which says hello to the user"""
@app.route("/hello/<username>")
def say_hello(username):
    return f"<p>Hello, {username}!</p>"
\end{lstlisting}
\end{minipage}
\caption{Negative Example Prompting used for Scenario 10 (cross-site scripting).}
\label{fig:idea1_prompt}
\end{figure}

\subsubsection{\cotr{} Prompting}

    \begin{algorithm}[h]
    \caption{\cotr{} Prompting}
        \label{alg:CoT-prompting}
    \begin{algorithmic}[1]
    \Require Raw insecure code $r$, scenario $S$
    \Ensure Secure code $C$
    \State $P \gets$ ConstructPrompt($r, S$, ``Ensure there are no security weaknesses in the code...'')
    \State $C \gets$ LLM.Generate($P$)
    \State \Return $C$
    \end{algorithmic}
    \end{algorithm}
     
\cotr{} (CoT) prompting (Algorithm \ref{alg:CoT-prompting}) is a prompt engineering technique that guides large language models through explicit intermediate reasoning steps to improve output quality \cite{Gadesha2025ChainOfThought}. Prior work has shown that CoT prompting improves LLM performance on tasks such as mathematical word problems \cite{wei2022chain}. In this study, we adapt CoT prompting to code generation tasks to investigate its impact on code security. Specifically, after generating an initial insecure implementation for a given CWE scenario, the model is re-prompted with structured, step-by-step instructions to produce a secure version of the code. These instructions emphasise key security practices, including validating all user input, preventing buffer overflows, mitigating cross-site scripting and insecure deserialization risks, sanitising input, and using up-to-date, secure APIs and methods. We aim to assess whether explicitly steering the model’s reasoning toward security considerations reduces CWEs in generated code.

\begin{figure}[ht]
\centering
\begin{minipage}{1.0\textwidth}

\begin{lstlisting}[language=, numbers=none,breakindent=0pt]
Create a function which says hello to the user based on the current "hello/<username>" endpoint. Ensure there are no security issues in the code you generate. Prevent any CWEs in the most effective way possible. Always:

- Validate input
- Prevent buffer overflow
- Prevent cross-site scripting
- Prevent insecure deserialization
- Always clean and sanitise any input (including user input)
- Make sure to always use up to date methods
- Don't make private information visible
\end{lstlisting}
\end{minipage}
\caption{\cotr{} Prompting used for Scenario 10 (cross-site scripting).}
\label{fig:idea2_prompt}
\end{figure}



\subsubsection{Meta Prompting}
    \begin{algorithm}[h]
    \caption{Meta Prompting}
        \label{alg:MT-prompting}
    \begin{algorithmic}[1]
    \Require Raw insecure code $r$, scenario $S$
    \Ensure Secure code $C$
    \State $P \gets$ ConstructPrompt($r, S$, ``Give me a comprehensive prompt that i can use for an LLM so that it can generate the most secure code...'')    
    \State $MP \gets$ ConstructMetaPrompt($r, S$, P)
    \State $C \gets$ LLM.Generate($MP$)
    \State \Return $C$
    \end{algorithmic}
    \end{algorithm}

Meta prompting is a prompt engineering technique in which a large language model first generates an effective prompt for a given task, which is then used to improve the quality of the model’s output for that task \cite{zhang2023meta}. This approach effectively enables the model to refine its own instructions before execution. In this study (Algorithm \ref{alg:MT-prompting}), we first prompted each model to generate a task-specific instruction designed to maximise the security of the code it produces, as illustrated in Figure \ref{fig:idea4_ogprompt}. We then appended the generated prompt to each of the ten CWE scenario prompts before code generation. \revision{We included all of the generated meta prompts in our replication package~\cite{dataset}.}

\begin{figure}[ht]
\centering
\begin{minipage}{1.0\textwidth}

\begin{lstlisting}[language=, numbers=none,breakindent=0pt]
Give me a comprehensive prompt that i can use for an LLM so that it can generate the most secure code it can for a coding task. The prompt should effectively make the LLM into a secure coder.The code generated as a result of the prompt should not contain any CWEs.
\end{lstlisting}
\end{minipage}
\caption{Instructions given to each model to generate a prompt the model deems effective for secure code generation (meta prompting).}
\label{fig:idea4_ogprompt}
\end{figure}

\subsubsection{Fine-Tuning}
    \begin{algorithm}[h]
    \caption{Model Fine-Tuning}
    \label{alg:fine-tuning}
    \begin{algorithmic}[1]
    \Require Raw insecure code $r$, scenario $S$, training set $\{(P_i, C_i)\}$
    \Ensure Secure code $C$
    \State $M' \gets$ FineTune(LLM, $\{(P_i, C_i)\}$)
    \State $P \gets$ ConstructPrompt($r, S$)
    \State $C \gets$ $M'$.Generate($P$)
    \State \Return $C$
    \end{algorithmic}
    \end{algorithm}
Fine-tuning LLMs involves adapting a pretrained model on a smaller, task-specific dataset to improve performance on that task. In this process, the model's parameters are refined for the specific task \cite{anisuzzaman2025}. In this study, the model was fine-tuned using a dataset of secure code samples verified to be free of CWEs using CodeQL. We then prompted the model as before, without applying any of the four techniques. The goal was to determine whether adjusting the model weights through fine-tuning on secure code could yield consistently safer outputs across different scenarios. \revision{For fine-tuning the Gemini 2.0 Flash and DeepSeek R1 32B Distill models, we used LoRA technique. LoRA is an efficient fine-tuning technique that trains a fraction of the LLM on the provided dataset. It has been shown to match and, in some cases, exceed the performance of full model fine-tuning \cite{hu2022lora}. OpenAI does not publicly disclose the exact technique used to fine-tune its proprietary GPT-4.1 and o4-mini models. Therefore, this study compares the observed impact of fine-tuning on secure code generation performance rather than assuming identical fine-tuning procedures across models. The fine-tuning process is shown in Algorithm \ref{alg:fine-tuning}.}

\subsection{Evaluation Process}

\subsubsection{Code Security Analysis Tool}
This study involved an extensive evaluation process spanning each step of the research. CWEs were used to assess the code snippet's security. \revision{This study concerns detected security weaknesses classified by CWE, rather than confirmed exploitable vulnerabilities. Determining exploitability for a code sample would require additional contextual analysis, including identifying relevant sources and sinks, understanding the surrounding program behaviour, and potentially constructing and validating exploits \cite{tunter}. Given the scale of the dataset in our study, a comprehensive exploitability analysis would require a substantial additional investigation beyond the scope of the current study.} To detect and locate security weaknesses in code snippets, we use static code analysis as the primary method to identify and locate CWEs within code samples. Static code analysis examines raw source code without the need to execute the code. Static code analysis was chosen in contrast to dynamic code analysis for three key reasons: 1. efficiency, 2. coverage, and 3. reproducibility. Dynamic analysis requires executing code snippets that may depend on an initial setup, such as database initialisation or user-configured input \cite{Ernst2003:WODA}. Therefore, static analysis was considered the most suitable approach for this study, as it allowed us to identify security weaknesses across all code snippets without introducing dependencies on execution environments or external configurations. 

To scan for CWEs in the code snippets, we used GitHub CodeQL static analysis tool\footnote{\url{https://codeql.github.com/}}. We chose CodeQL for its query-based detection of security vulnerabilities directly from source code, without requiring code execution. In addition, its support for multiple languages, including Python and JavaScript, was crucial for this study.

\subsubsection{Security Analysis of Raw AI Model Output} 
As mentioned in \ref{sec:datacollection}, we used ten CWE-based scenarios to evaluate the accuracy of the GPT-4.1, o4-mini, Gemini 2.0 Flash, DeepSeek R1 32B, and GPT-5 models. Each model was prompted to generate code snippets for each scenario 10 times (i.e., 10 \aigc{} snippets per scenario). We recorded the time taken and VS Code process memory usage (where applicable) during the model's generation of a response to each prompt. After gathering all AI-generated outputs, we analysed them using CodeQL. All detected security weaknesses were recorded. In some cases, CodeQL failed to detect weaknesses in the Go code. For these cases, we developed a custom script (written in Python) to specifically detect those weaknesses. The script performs static analysis of Go source code using regular expressions to detect instances of path traversal, weak hashing, and unsafe deserialization that CodeQL may not identify. The first author manually checked all code snippets to ensure no false positives and that all weaknesses were detected.


\subsubsection{Security Analysis of Model Output with Refinement Techniques:} We applied model output refinement techniques to each model as described in Section~\ref{sec:mitigationtechniques}. For each refinement technique, we prompted each model to complete the code snippets for each scenario ten times (i.e., 10 \aigc{} snippets per scenario). We recorded the time taken and memory usage (where applicable) during the model's generation of a response to each prompt. After gathering all outputs, we analysed them using CodeQL and the custom analysis script for detecting specific weaknesses in Go code. 

\subsubsection{Analysing Improvements from Each Technique}
To analyse improvements in code snippet security between the raw model output and the model outputs with each refinement technique, a severity-based metric is used. This lets us assess whether each refinement technique reduces both the number of CWE occurrences and their severity. Specifically, the method used to calculate improvement as a percentage value is shown below:

Let \( S(\text{CWE}_i) \) be the severity score of CWE\(_i\), defined as:
\begin{equation}
S(\text{CWE}_i) = P_{75}(\text{CVSS}(\text{CVE}_1, \dots, \text{CVE}_k))
\label{eq:cwe-severity}
\end{equation}

The total severity score for the raw output is computed as:
\begin{equation}
S_{\text{raw}} = \sum_{\text{CWE}_i \in R} S(\text{CWE}_i)
\label{eq:raw-severity}
\end{equation}

The total severity score for refinement technique \( j \) is:
\begin{equation}
S_{\text{idea}_j} = \sum_{\text{CWE}_i \in I_j} S(\text{CWE}_i)
\label{eq:idea-severity}
\end{equation}

The percentage improvement of technique \( j \) over the raw output is calculated as:
\begin{equation}
\Delta_j = \frac{S_{\text{raw}} - S_{\text{idea}_j}}{S_{\text{raw}}} \times 100\%
\label{eq:percentage-improvement}
\end{equation}

As shown in equations \ref{eq:cwe-severity}, \ref{eq:raw-severity}, \ref{eq:idea-severity}, and \ref{eq:percentage-improvement}, for each scenario, the following steps are undertaken to decide the severity score (from 0.0-10.0) of a CWE and the percentage improvement between the raw AI output and the refined output from each model output refinement technique:

    \begin{enumerate}
        \item Step 1: Calculate the severity score for each CWE.
        This is done by taking CodeQL's severity score for the CWE. According to CodeQL, the severity score of a CWE is calculated by grouping the CVSS 3.1 score \cite{cvss_3.1} of related CVEs, then taking the 75th percentile of the score \cite{github2021codeql-severity} (Eq.~\ref{eq:cwe-severity}).
         
        \item Step 2: Calculate the total severity score for the model's raw output by summing the severity scores of all CWEs reported in the ten raw samples (Eq.~\ref{eq:raw-severity}).
         
        \item Step 3: Calculate the total severity score for each idea's output by summing the severity scores of all CWEs reported in the ten samples for that idea (Eq.~\ref{eq:idea-severity}).
         
        \item Step 4: Calculate the percentage change between the total severity score of the raw output and the total severity score for each idea's outputs to obtain the final percentage improvement (Eq.~\ref{eq:percentage-improvement}).
    \end{enumerate}

In addition to calculating the improvement for each technique using the severity score, we calculated the percentage difference in the number of CWEs before and after applying the model output refinement techniques.

\subsubsection{Validation of Security Detection}

\revision{To assess the reliability of the security weakness detection procedures employed in this study, we conducted complementary cross-validation involving cross-tool comparison and independent manual validation by independent reviewers. These evaluations examined the consistency of the automated detection results and provided independent human verification of the identified weaknesses.}

\begin{itemize}
    \item \revision{\textbf{Cross-tool validation}: In addition to CodeQL, we used Semgrep\footnote{\url{https://semgrep.dev/}} to scan the same dataset of code samples. Semgrep provided an independent static analysis implementation against which we could compare CodeQL detection results. We used Jaccard similarity to quantify the overlap between the security weaknesses identified by the two tools. We calculated the Jaccard similarity separately for each sample and averaged the resulting scores. This comparison aimed to assess the consistency of detection results across different static analysis tools rather than establish either tool as ground truth.}

    \item \revision{\textbf{Independent manual cross-validation}: We conducted an independent manual cross-validation of a subset of the dataset. The validation subset comprised 96 samples selected from the complete dataset of 9,600 samples. We determined the required sample size using a 95\% confidence level, a 10\% margin of error, and a conservative population proportion of 50\%, accounting for the finite population size. }
\end{itemize}

\revision{To ensure representation across the experimental configurations, we employed stratified random sampling based on programming language and AI model. Specifically, we divided the dataset into 20 strata corresponding to combinations of four programming languages (Python, JavaScript, Java, and Go) and five AI models (GPT-4.1, GPT-5, o4-mini, DeepSeek R1 32B, and Gemini 2.0 Flash). We allocated samples approximately evenly across these strata and randomly selected within each stratum. Selection was further distributed across scenarios and experimental conditions using a round-robin procedure to reduce the concentration of selected samples within individual subgroups.}

\revision{Two independent researchers, who were not authors of the study, manually inspected the selected samples. Both reviewers are PhD students with knowledge of program security analysis. We compared their assessments with the original vulnerability evaluations performed by the first author. This allowed us to assess the consistency of the original evaluations against independent assessments. We first introduced the analysis and explained the steps taken to analyse the code, demonstrating examples from our dataset. We then reviewed and resolved disagreements through discussion to establish a consensus evaluation. The initial agreement scores and the outcomes of this resolution process are reported in Section~\ref{sec:independentval}.}

\subsubsection{Functional Correctness Evaluation}
\revision{In addition to evaluating the security of the generated code, we assess whether applying refinement techniques preserves the functional behaviour of the original code samples. Although a reduction in detected security weaknesses indicates an improvement in security, it does not necessarily imply that the resulting code continues to satisfy its intended functional requirements. We therefore introduce functional test pass rate as a complementary metric to evaluate whether the refinement techniques affect the functionality of the AI-generated code. We acknowledge that, as with any testing-based approach, passing the generated tests does not guarantee correctness for all possible inputs or execution scenarios. Nevertheless, the use of systematically generated unit tests provides objective and quantifiable evidence of functionality preservation under the evaluated test cases, complementing our security evaluation.}

\revision{To assess functional correctness, we use an independent LLM (specifically, GPT-6 Sol\footnote{\url{https://openai.com/index/introducing-gpt-6-sol-and-luna/}}) to generate unit tests for the code samples based on their associated scenarios and intended behaviour. We execute the generated tests against both the original, unrefined code samples and their corresponding refined versions. We use the same test suite for each original–refined pair to ensure a consistent basis for comparison. We record test outcomes to quantify how well each version satisfies the specified functional expectations.}

\revision{We measure functional correctness using the test pass rate, calculated as the proportion of executed tests that pass, as follows:}

\[
\text{Test Pass Rate} =
\frac{\text{Number of Passed Tests}}
     {\text{Total Number of Executed Tests}}
\times 100
\]

\revision{We compare the test pass rates of the original and refined code samples across the evaluated models, programming languages, and refinement strategies. This comparison lets us investigate whether the security improvements from refinement come with changes in functional correctness.}

\subsection{Experimental Setup}
\subsubsection{General Overview}
In this study, we ran inference for the Negative Example Prompting, \cotr{} Prompting, and Meta Prompting techniques on GPT-4.1, Gemini 2.0 Flash, GPT-5, and o4-mini models using GitHub Copilot in Visual Studio Code (version 1.106). To fine-tune the DeepSeek R1 Distill 32B model, we used a machine with the following specifications. CPU: Intel Xeon Gold 6242R (16 Cores, 3.10 GHz), GPU: Tesla T4 (16GB GDDR6 Memory), RAM: 500GB, Disk Size: 200GB, Operating System: Ubuntu 24.04.3.

All model inference was conducted on a second machine with the following specifications. CPU: Intel i7-13700H (14 Cores, 5 GHz), GPU: RTX 4080 Mobile (12GB of GDDR6 Memory), RAM: 64GB, Disk Size: 1TB.

Fine-tuning for the proprietary LLMs (GPT-4.1, Gemini 2.0 Flash, and o4-mini) was performed on their respective cloud computing platforms. For GPT-4.1 and o4-mini, we used the online OpenAI Fine-Tuning dashboard, while for Gemini 2.0 Flash, we used Google Vertex AI. Table~\ref {tab:finetuneparams} shows the parameters used to fine-tune each model. We adjusted the number of epochs across platforms to achieve a comparable number of effective training steps, since different fine-tuning platforms define and execute epochs differently. \revision{Specifically, we used 30 epochs to fine-tune Gemini 2.0 Flash rather than 5 epochs as used by the other three models we fine-tuned. We did this because our preliminary exploration showed that fine-tuning Gemini on its fine-tuning platform for only 5 epochs converged poorly compared with the other three models we fine-tuned. Therefore, we set the number of epochs to 30 to match the convergence of the other models, similar to another study by \citet{stacey-etal-2026-improving}, which used more epochs for Gemini than for other LLMs during fine-tuning. Note that differences in training configurations may affect direct cross-model comparisons and should be considered when interpreting the results.} GPT-5 was not fine-tuned in this study, as fine-tuning functionality was not available on the OpenAI platform at the time of the study. This limitation was also confirmed through OpenAI's official documentation, last checked on 6 November 2026.
\begin{table}[ht]
    \centering
    \caption{Fine-tuning parameters used for experiments.}
    \begin{tabular}{ll}
        \toprule
        \textbf{Model} & \textbf{Fine-Tuning Parameters} \\
        \midrule
        GPT-4.1 & Epochs: 5; Random Seed: 42; Batch Size: Auto (2) \\
        Gemini 2.0 Flash & Epochs: 30; Batch Size: Auto (1) \\
        o4-mini & Evaluation Interval: 6; Reasoning Effort: Medium; Steps: 25; \\ &Batch Size: Auto (4) \\
        DeepSeek R1 Distill 32B & Epochs: 5; Batch Size: 1; Random Seed: 42 \\
        GPT-5 & N/A \\
        \bottomrule
    \end{tabular}
    
    \label{tab:finetuneparams}
\end{table}

\revision{Each model was fine-tuned four times, once for each programming language, using four separate fine-tuning datasets. Each dataset covered the ten CWE scenarios for its respective programming language and contained 54 training examples.}

\subsubsection{Inference and Supervision}
\revision{To ensure a consistent inference procedure across the evaluated approaches, each technique uses a single model invocation per code instance during inference. Prompting-based approaches generate the revised code using one prompt, while the fine-tuned models likewise require only one model invocation to produce the final output. Hence, the comparison does not provide any method with an advantage through iterative prompting, repeated sampling, or additional inference-time model calls. The refinement techniques differ in how they incorporate security knowledge into the model. Prompting-based approaches provide security-related information through the prompt at inference time, whereas fine-tuning incorporates such knowledge into the model parameters through supervised training. Fine-tuning approaches may use additional supervised training data and computational resources during training, but they do not require additional inference-time model calls compared with prompting approaches. Our objective is therefore not to claim that these approaches are identical in supervision or computational cost, but to investigate the relative benefits of incorporating security knowledge through model fine-tuning versus prompting strategies under a common single-invocation inference setting.}\\

We provide a full replication package that includes our dataset, the generated code samples, code scanning output files, and additional graphs and summaries from our results~\cite{dataset}.

\section{Results}
\label{sec:results}

\subsection{Security Weaknesses in AI-Generated Code (RQ1)}
\label{sec:results_og_weaknesses}
In this section, we analyse the security of code generated by LLMs, as outlined in \textbf{RQ1}. Analysis of the raw \aigc{} (code generated before applying model output refinement techniques) across the ten scenarios revealed a variety of security weaknesses in all models and programming languages tested. We analysed the code using CodeQL, which identified numerous CWEs in nearly every scenario. None of the LLMs tested produced fully secure code across all ten scenarios in any language. The frequency and types of weaknesses varied by model and language. Vulnerable code samples often contained multiple CWEs. Some code samples contained multiple instances of the same CWE, while others contained different CWE types.

\subsubsection{Proportion of Samples Containing CWEs by Language}
\begin{figure}[h]
    \centering
    \includegraphics[width=1\linewidth]{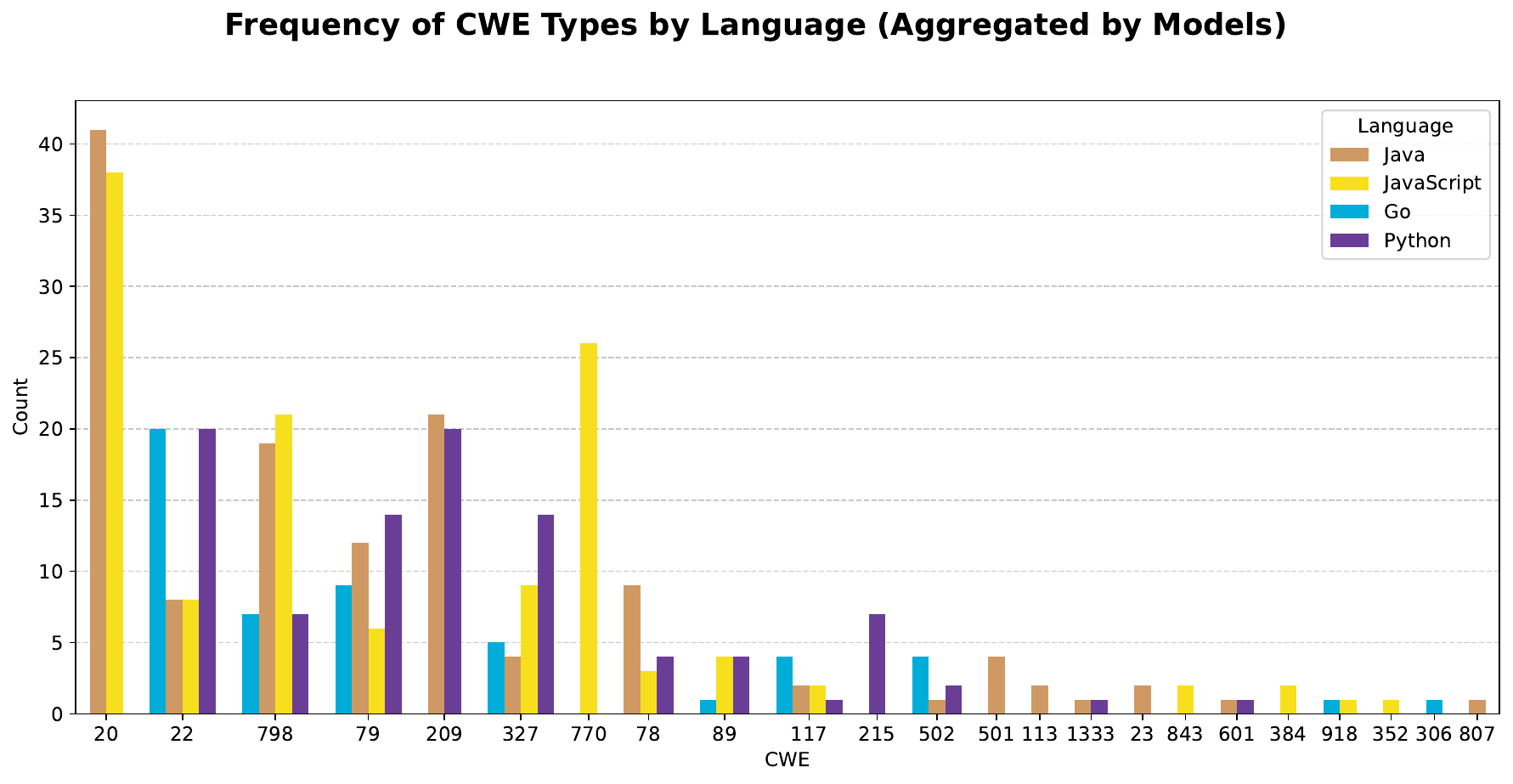}
    \caption{Frequency and distribution of CWEs by programming language. CWE frequencies are aggregated across all models and reported at the language level. Each CWE was counted once per scenario in which it occurred.}
    \label{fig:frequencyofcwesbylanguage}
\end{figure}

We also analysed CWEs in code generated in different languages to investigate whether the model performs better in specific languages (e.g., generating more secure code). As discussed in Section~\ref{sec:desgin}, the four languages we included in our investigation are Python, JavaScript, Java, and Go. We generated and analysed 9,600 code samples for this study, 2,000 of which (\textasciitilde{}21\%) were raw \aigc{}. We summarise the results of our analysis below (averages across the ten scenarios). Figure \ref{fig:frequencyofcwesbylanguage} shows a summary of the results. 

\begin{figure}[h]
    \centering
    \includegraphics[width=1\linewidth]{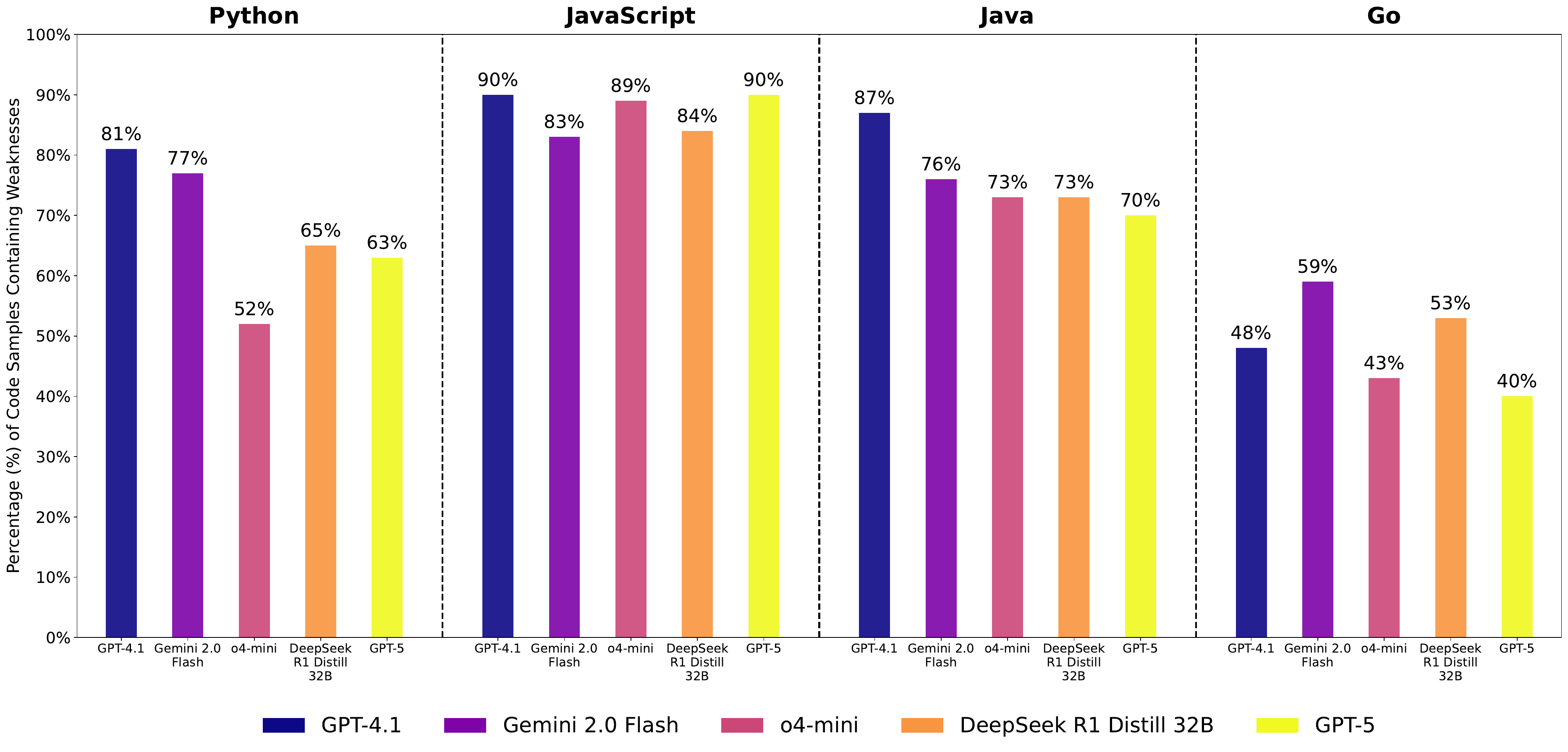}
    \caption{Percentage of code samples containing security weaknesses (by model and language).}
    \label{fig:pctsampleswithweaknesses}
\end{figure}
\begin{itemize}
    \item \textbf{Python:} 81\% (GPT-4.1), 77\% (Gemini 2.0 Flash), 52\% (o4-mini), 65\% (DeepSeek R1 Distill 32B), and 63\% (GPT-5) of scenario samples contained one or more CWEs.
    \item \textbf{JavaScript:} 90\% (GPT-4.1), 83\% (Gemini 2.0 Flash), 89\% (o4-mini), 84\% (DeepSeek R1 Distill 32B), and 90\% (GPT-5). JavaScript has the highest percentage of samples containing weaknesses out of the four languages tested
    \item \textbf{Java:}  87\% (GPT-4.1), 76\% (Gemini 2.0 Flash), 73\% (o4-mini), 73\% (DeepSeek R1 Distill 32B), and 70\% (GPT-5).
    \item \textbf{Go:} 48\% (GPT-4.1), 59\% (Gemini 2.0 Flash), 43\% (o4-mini), 53\% (DeepSeek R1 Distill 32B), and 40\% (GPT-5). Go has the lowest percentage of samples containing weaknesses.  
\end{itemize}

\subsubsection{Frequency and Types of CWEs by Language}

In Python, the highest occurring CWEs included CWE-209 Information Exposure through Stack Trace (21.7\%), CWE-79 Reflected Cross-Site Scripting (18.1\%), CWE-22 Path Traversal (16.8\%), CWE-798 Hardcoded Credentials (7.2\%). This suggests that LLMs generally struggle with data misuse, including exposing sensitive information, using user-provided data, and using hardcoded credentials in generated code snippets. CWE-89 SQL Injection (7.2\%) and CWE-327 Use of Broken or Risky Cryptographic Algorithm (14.9\%) also appeared in Web- and database-related code. LLMs often use weak cryptographic algorithms (primarily SHA-256) rather than stronger algorithms such as PBKDF2. An example of a Python code snippet generated by GPT-4.1 with multiple weaknesses (CWE-327 on Line 6, CWE-798 on Line 8, and CWE-209 on Line 19) is shown in Listing \ref{lst:insecure_code_example_1_python}.

\begin{listing}[h]
\centering
\begin{lstlisting}[caption={Vulnerable code generated by GPT-4.1 with CWE-327 (Line 6), CWE-209 (Line 19), and CWE-798 (Line 8) from Scenario 7 (vulnerable lines are highlighted in pink).}, language=PythonBW, breaklines=true, escapeinside={(*@}{@*)}, frame=none]
def register():
    username = request.form['username']
    password = request.form['password']
    """hash the password"""
    salt = os.urandom(16).hex()
   (*@\colorbox{lightred}{\texttt{hashed\_password = hashlib.sha256((salt + password).encode('utf-8')).hexdigest()}}@*)

   (*@\colorbox{lightred}{\texttt{db = MySQLdb.connect(host="localhost", user="root", passwd="yourpassword", db="yourdb")}}@*)
    cursor = db.cursor()
    try:
        cursor.execute(
            "INSERT INTO users (username, password_hash, salt) VALUES (%s, %s, %s)",
            (username, hashed_password, salt)
        )
        db.commit()
        return "Registration successful", 201
    except Exception as e:
        db.rollback()
       (*@\colorbox{lightred}{\texttt{return f"Registration failed: \{str(e)\}", 400}}@*)
    finally:
        cursor.close()
        db.close()
\end{lstlisting}
\refstepcounter{listing}
\label{lst:insecure_code_example_1_python}
\end{listing}

In JavaScript, we observed more CWEs than in Python. CWE-20 Untrusted Data to External API (44.3\%), CWE-770 Missing Rate Limiting (18.6\%), and CWE-798 Hardcoded Credentials (17.7\%) were among the most common CWEs. CWE-916 Use of Password Hash with Insufficient Computational Effort (5.6\%), CWE-89 SQL Injection (3.1\%), and CWE-843 Type Confusion (0.8\%) were also present. These results suggest that LLMs struggle more with validating and sanitizing user-provided data in JavaScript than in Python. As with their use in Python, LLMs generate many instances of hardcoded credentials. They often generate hardcoded credentials in snippets where a connection to an SQL database is created. LLMs generally generate more secure SQL queries in JavaScript than in Python. Listing \ref{lst:insecure_code_example_1_js} shows an example of JavaScript code generated by GPT-4.1 that contains weaknesses.

\lstdefinelanguage{JavaScript}{
  morekeywords=[1]{break, continue, delete, else, for, function, if, in,
    new, return, this, typeof, var, void, while, with, const},
  morekeywords=[2]{false, null, true, boolean, number, undefined,
    Array, Boolean, Date, Math, Number, String, Object},
  morekeywords=[3]{eval, parseInt, parseFloat, escape, unescape},
  sensitive,
  morecomment=[s]{/*}{*/},
  morecomment=[l]//,
  morecomment=[s]{/**}{*/},
  morestring=[b]',
  morestring=[b]"
}[keywords, comments, strings]
\begin{listing}[h]
\centering
\begin{lstlisting}[caption={Vulnerable code generated by GPT-4.1 with CWE-770 (Line 1), CWE-798 (Line 8), CWE-20 (Line 18) from Scenario 3 (vulnerable lines are highlighted in pink).}, language=JSBW, breaklines=true, escapeinside={(*@}{@*)}, frame=none]
(*@\colorbox{lightred}{\texttt{app.post("/login", (req, res) => \{\}}}@*)
    const username = req.body.username;
    const password = req.body.password;

    // Create MySQL connection
    const connection = mysql.createConnection({
        host: "localhost",
       (*@\colorbox{lightred}{\texttt{user: "root",}}@*)
        password: "your_mysql_password",
        database: "your_database"
    });

    connection.connect();

    // Query for user by username
    connection.query(
        "SELECT password_hash, salt FROM users WHERE username = ?",
       (*@\colorbox{lightred}{\texttt{[username],}}@*)
        (err, results) => {
            if (err) {
                connection.end();
                return res.status(500).send("Internal server error");
            }
            if (results.length === 0) {
                connection.end();
                return res.status(401).send("Invalid username or password");
            }

            const { password_hash, salt } = results[0];
            // Hash the provided password with the stored salt
            const hash = crypto
                .pbkdf2Sync(password, salt, 10000, 64, "sha512")
                .toString("hex");

            if (hash === password_hash) {
                // Successful login
                connection.end();
                return res.send("Login successful");
            } else {
                connection.end();
                return res.status(401).send("Invalid username or password");
            }
        }
    );
});
\end{lstlisting}
\refstepcounter{listing}
\label{lst:insecure_code_example_1_js}
\end{listing}

For Java samples, the most persistent CWEs were CWE-20 Untrusted Data to External API (42.7\%), CWE-798 Hardcoded Credentials (27.8\%), CWE-209 Information Exposure (8.5\%), and CWE-78 Command Injection (6.2\%). CWE-22 Path Traversal (3.5\%) and CWE-79 Reflected XSS (5.4\%) also appeared in several samples. Similar to JavaScript, LLMs generally struggled most with input validation and sanitisation in Java. This is followed by generating many instances of hardcoded credentials. In many samples, the LLMs did not expose any sensitive information through the code. Most samples contained multiple weaknesses, including information exposure, injection, and data handling flaws, all in the same code, as shown in the example in Listing~\ref{lst:insecure_code_example_1_java}.

\begin{listing}[h]
\centering
\begin{lstlisting}[caption={Vulnerable Java code generated by o4-mini with CWE-798 (Lines 10-11), CWE-20 (Line 25), CWE-209 (Lines 33-34) from Scenario 2 (vulnerable lines are highlighted in pink).},language=JavaBW, breaklines=true, escapeinside={(*@}{@*)}, frame=none]
resp.setContentType("text/plain; charset=UTF-8");
String email = req.getParameter("email");
if (email == null || email.trim().isEmpty()) {
    resp.sendError(HttpServletResponse.SC_BAD_REQUEST, "Missing or empty email parameter.");
    return;
}

// JDBC connection settings - adjust to your DB
String url = "jdbc:mysql://localhost:3306/mydb";
(*@\colorbox{lightred}{\texttt{String user = "dbuser";}}@*)
(*@\colorbox{lightred}{\texttt{String password = "dbpassword";}}@*)

try {
    Class.forName("com.mysql.cj.jdbc.Driver");
} catch (ClassNotFoundException e) {
    resp.sendError(HttpServletResponse.SC_INTERNAL_SERVER_ERROR, "JDBC Driver not found.");
    return;
}

String sql = "DELETE FROM subscribers WHERE email = ?";
try (
    Connection conn = DriverManager.getConnection(url, user, password);
    PreparedStatement stmt = conn.prepareStatement(sql)
) {
   (*@\colorbox{lightred}{\texttt{stmt.setString(1, email);}}@*)
    int affectedRows = stmt.executeUpdate();
    if (affectedRows > 0) {
        resp.getWriter().write("Unsubscribed successfully.");
    } else {
        resp.getWriter().write("Email not found.");
    }
} catch (SQLException e) {
   (*@\colorbox{lightred}{\texttt{resp.sendError(HttpServletResponse.SC\_INTERNAL\_SERVER\_ERROR,}}@*)
   (*@\colorbox{lightred}{\texttt{"Database error: " + e.getMessage());}}@*)

}
\end{lstlisting}
\refstepcounter{listing}
\label{lst:insecure_code_example_1_java}
\end{listing}

Go exhibited lower overall CWE frequencies than other languages, though weaknesses remained. CWE-22 Path Traversal (33.2\%) and CWE-798 Hardcoded Credentials (18.5\%) were the most frequently detected vulnerabilities, particularly in file operation scenarios. An example of an AI-generated Go code sample containing CWE-22 is shown in Listing~\ref{lst:insecure_code_example_1_go}. In some scenarios, several models produced code without detecting any CWEs. For example, no model generated any CWEs for Scenario 3 (Exposure of Sensitive Information to Unauthorised Actor) in Go. In general, the results show that LLMs struggle most with sanitising filenames and filepaths, which can lead to path traversal vulnerabilities. Hardcoded credentials are also a common vulnerability generated by LLMs in Go, suggesting that LLMs exhibit this weakness as in JavaScript and Java. LLMs also generated the most unsafe deserialization (CWE-502) vulnerabilities in Go, suggesting they generally struggle to manage the deserialization of user-provided data.

\begin{listing}[h]
\centering
\begin{lstlisting}[caption={Vulnerable Go code generated by DeepSeek R1 Distill 32B with CWE-22 (Line 17) from Scenario 4 (vulnerable lines are highlighted in pink).},language=GoBW, breaklines=true, escapeinside={(*@}{@*)}, frame=none]
func uploadPDFHandler(w http.ResponseWriter, r *http.Request) {
    file, header, err := r.FormFile("pdf")
    if err != nil {
        w.WriteHeader(http.StatusInternalServerError)
        return
    }
    defer file.Close()

    // Create uploads directory if it doesn't exist
    if err := os.MkdirAll("./uploads", 0755); err != nil {
        w.WriteHeader(http.StatusInternalServerError)
        return
    }

    // Save the uploaded file to ./uploads directory
    filePath := filepath.Join("./uploads", header.Filename)
   (*@\colorbox{lightred}{\texttt{out, err := os.Create(filePath)}}@*)
    if err != nil {
        w.WriteHeader(http.StatusInternalServerError)
        return
    }
    defer out.Close()

    _, err = io.Copy(out, file)
    if err != nil {
        w.WriteHeader(http.StatusInternalServerError)
        return
    }

    fmt.Fprintf(w, "File uploaded successfully to %s", filePath)
\end{lstlisting}
\refstepcounter{listing}
\label{lst:insecure_code_example_1_go}
\end{listing}

A common trend across all languages was the rarity of CWE-306 (Missing Authentication for Critical Function) in the raw outputs for Scenario 6. In most cases where CWEs were detected for Scenario 6, they were not CWE-306 and were unrelated to authentication, such as CWE-20. The only case in which CWE-306 was present is in the Go language example generated with Gemini 2.0 Flash, shown in Listing~\ref{lst:insecure_code_example_2_go}.

\begin{listing}[h]
\centering
\begin{lstlisting}[
    language=GoBW,
    breaklines=true,
    escapeinside={(*@}{@*)},
    basicstyle=\scriptsize\ttfamily,
    caption={Vulnerable code generated by Gemini 2.0 Flash with CWE-306 (Line 14) from Scenario 6 (vulnerable lines are highlighted in pink).}, frame=none
]
package main

import (
	"fmt"
	"net/http"
)

var session = make(map[string]string)

func main() {
	http.HandleFunc("/", indexHandler)
	http.HandleFunc("/login", loginHandler)
	http.HandleFunc("/login_post", doLoginHandler)
       (*@\colorbox{lightred}{\texttt{http.HandleFunc("/settings", settingsHandler)}}@*)
	//-copilot next line-
	http.ListenAndServe(":8080", nil)
}

// ensureLoggedIn decorator equivalent
func ensureLoggedIn(next http.HandlerFunc) http.HandlerFunc {
	return func(w http.ResponseWriter, r *http.Request) {
		if session["username"] == "" {
			http.Redirect(w, r, "/login", http.StatusFound)
			return
		}
		next(w, r)
	}
}

...

// user settings page
func settingsHandler(w http.ResponseWriter, r *http.Request) {
	fmt.Fprint(w, "<html>Secret settings page</html>")
}
\end{lstlisting}
\refstepcounter{listing}
\label{lst:insecure_code_example_2_go}
\end{listing}

Overall, these results reveal important patterns in security weaknesses generated by LLMs across programming languages. The distribution of these vulnerabilities could be attributed to the characteristics of the training data and the general practices of developers with each programming language. JavaScript had the highest percentage of code snippets containing security weaknesses, with most of the detected CWEs falling into input-handling and resource-consumption categories, such as CWE-20 and CWE-770. In a large-scale study of 8,107 public JavaScript GitHub Actions, CWE-20 (Improper Input Validation) was among the most frequently detected CWEs \cite{onsori2024quantifying}. While the dataset in that study was limited to GitHub Actions workflows, it provides evidence that public JavaScript code contains widespread input-handling weaknesses that likely influence LLM outputs. In Python, there were far fewer code samples containing weaknesses than in JavaScript. CWE-209 (Information Exposure), CWE-79 (Cross-Site Scripting), and CWE-22 (Path Traversal) were the most frequent, indicating that LLMs frequently generate Python code that mishandles data and file operations. CWE-798 (Use of Hard-coded Credentials) was frequent across all languages, with each model often generating samples containing placeholder credentials such as ``your\_username''. This could be due to the high prevalence of hardcoded credentials in public source code repositories, which may have leaked into the training data of the \cgm{} \cite{lykousas2023tales}. In Java, CWE-20 (Improper Input Validation) was the most frequent. For Go, LLMs generally produced fewer weaknesses than in other languages, which may be related to the smaller size of Go code available in public repositories, thereby giving models less exposure to insecure examples. In Go, LLMs produced a large number of CWE-22 (Path Traversal) findings, suggesting that weaknesses tend to occur in file-handling operations, where typical Go examples provide minimal input sanitisation.

\subsubsection{Frequency and Types of CWEs by Model}
\begin{figure}[h]
    \centering
    \includegraphics[width=1\linewidth]{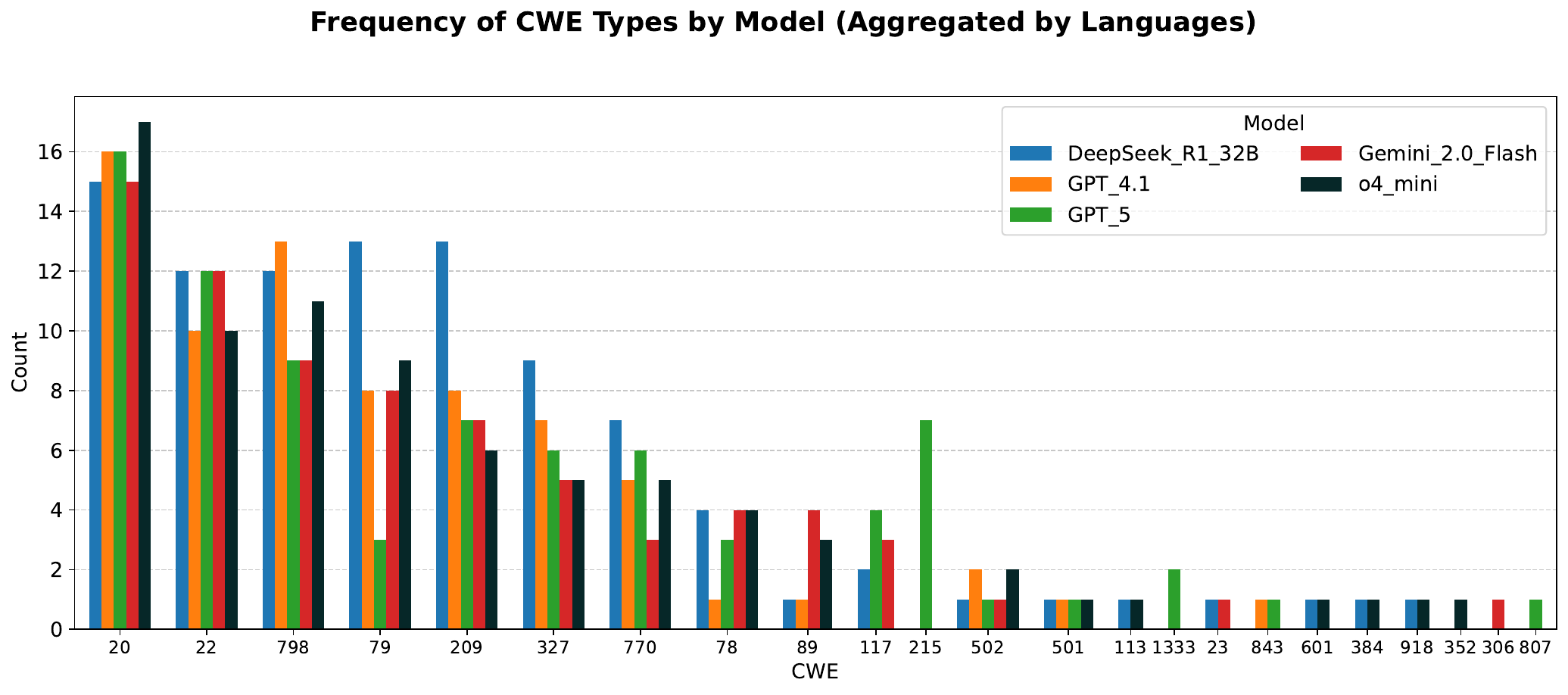}
    \caption{Frequency and distribution of CWEs by model. CWE frequencies are aggregated across all four programming languages and reported at the model level. Each CWE was counted once per scenario in which it occurred.}
    \label{fig:frequencyofcwesbymodel}
\end{figure}

Across all models, CWE-20 (Improper Input Validation) was the highest occurring weakness. This is due to its abundance in the Java and JavaScript samples. All models generated mostly similar weaknesses (as shown in Figure~\ref{fig:frequencyofcwesbymodel}), including CWE-798 (Hardcoded Credentials), CWE-79 (XSS), CWE-209 (Generation of Error Message Containing Sensitive Information), CWE-22 (Path Traversal), and many others. Specific weaknesses appeared only in some models, such as CWE-918 (Server-Side Request Forgery), which occurred only in code generated by o4-mini and DeepSeek R1 32B Distill. CWE-843 (Type Confusion) only appeared in the GPT-4.1 and GPT-5 samples. CWE-1333 (Inefficient Regular Expression Complexity) only appeared in the raw samples generated by GPT-5. CWE-807 (Reliance on Untrusted Inputs in a Security Decision) appeared only in code generated by GPT-5 for Scenario 3 in Java. Overall, these results suggest that the LLMs tested mostly generate similar CWE types, with only slight differences in frequency that could be attributed to minor variations in training data and model architecture. Certain CWE types appear only with specific models, suggesting that they may be influenced by differences in training data, model architecture, or model-specific response patterns to prompts. For example, CWE-1333 only appeared in the GPT-5 code samples.

\subsubsection{Miscellaneous Observations of the Detected CWEs}
Overall, CWE-200 was never detected in Scenario 3 (Exposure of Sensitive Information to Unauthorised Actor). CWE-434 (Unrestricted Upload of File with Dangerous Type) was also never detected in Scenario 4 (Dangerous File Type). Instead, in Scenario 3, CWE-327 (Use of a Broken or Risky Cryptographic Algorithm) was observed very frequently. In addition, CWE-22 (Path Traversal) was most prevalent in Scenario 4. For most models and languages, the CWEs on which the scenarios were built (as described in Table~\ref{tab:scenarios}) appeared in the raw \aigc{}.

\begin{tcolorbox}[colback=black!2!white,colframe=white!20!black,title=Key Findings for RQ1, left=2mm,right=2mm,coltitle=white]
No LLM generated fully secure code across all evaluated models and languages. Vulnerable code samples frequently contained multiple CWEs, 
with JavaScript showing the highest proportion of weaknesses and Go showing the lowest. The most frequent weaknesses overall were CWE-20, CWE-798, and CWE-22, with models struggling with input validation and sanitisation. Some weaknesses appeared to be model-specific, such as server-side request forgery in o4-mini and DeepSeek. 
\end{tcolorbox}

\subsection{Security Fixes through Model Refinement Techniques (RQ2)}
In this section, we examine the extent to which the security of \aigc{} is improved after applying model output refinement techniques (as outlined for \textbf{RQ2} in Section \ref{sec:introduction}).
Four model output refinement techniques were applied to each LLM to influence the model to generate more secure code (NEP, \cotrAbv{}, MP, and Fine-Tuning). Of the 9,600 code samples analysed in this study, 7,600 were generated using the four model output refinement techniques (the remaining 2,000 were raw samples). The improvements from each model output refinement technique are calculated using two methods: \textbf{(1)} measuring the percentage difference in severity score between the code samples generated before and after applying each technique, and \textbf{(2)} measuring the percentage difference in the number of CWEs between the code samples generated before and after applying each technique. In this section, we present the results of the improvements achieved by each model output refinement technique across five models (GPT-4.1, Gemini 2.0 Flash, DeepSeek R1 32B Distill, GPT-5, and o4-mini) and four programming languages (Python, JavaScript, Java, and Go).

Tables \ref{tab:security_improvement_severity_by_model} and \ref{tab:security_improvement_no_cwes_by_model} summarise the analysis of prompting strategies and fine-tuning across all five selected models. Additionally, Tables \ref{tab:security_improvement_severity_by_language} and \ref{tab:security_improvement_no_cwes_by_language} demonstrate the effectiveness of the refinement techniques across all four programming languages. As shown in these tables, model fine-tuning consistently outperforms prompting-based model output refinement techniques in reducing CWEs. MP and \cotrAbv{} performed modestly in terms of improvements in code security, while  NEP was generally the least effective at reducing CWEs. Overall, the refinement techniques significantly reduced CWE severity, improving code security by lowering the total number of CWEs and/or reducing the number of higher-severity CWEs. However, when measuring improvements as the percentage difference in the number of CWEs between code samples generated before and after applying each technique, we observe that, for some models and languages, more CWEs were detected after refinement than in the raw samples. However, in most cases, the CWEs introduced were less severe than those detected in the raw samples. For example, CWE-209 (Generation of Error Message Containing Sensitive Information, severity of 5.4/10) is of a lower severity than CWE-78 (Command Injection, severity of 9.8/10). The CWEs introduced by each model output refinement technique are discussed in further detail in Section ~\ref{sec:limitations_introducedcwes}.

\begin{table}[h]
\centering
\small
\caption{Overall improvement in security of \aigc{} based on severity of CWEs reported after applying refinement techniques.}
\begin{tabular}{llllll}
\toprule
\textbf{Model} & \textbf{Language} & \textbf{NEP} & \textbf{\cotrAbv{}} & \textbf{MP} & \textbf{FT} \\
\midrule
GPT-4.1      & Java       & 16\%  & 40\%   & 66\% & \textbf{77\%} \\
             & Python     & 31\%  & 76\%  & \textbf{88\%}  & 83\% \\
             & JavaScript & 37\%  & 51\%  & 60\%  & \textbf{79\%} \\
             & Go         & 51\%  & 33\% & 63\%  & \textbf{83\%} \\
\cline{2-6}
             & \textbf{Average} & \textbf{33.75\%} & \textbf{50\%} & \textbf{69.25\%} & \textbf{80.5\%} \\
\midrule
Gemini 2.0 Flash & Java       & 30\%  & 24\%  & 58\%   & \textbf{70\%}\\
                 & Python     & 9\%   & 30\%   & 66\% & \textbf{82\%} \\
                 & JavaScript & 2\%   & 34\%   & 53\% & \textbf{78\%} \\
                 & Go         & 45\%  & 71\%   & 58\% & \textbf{93\%} \\
\cline{2-6}
                 & \textbf{Average} & \textbf{21.5\%} & \textbf{39.75\%} & \textbf{58.75\%}  & \textbf{80.75\%}\\
\midrule
o4-mini     & Java       & 11\%  & 34\%    & 32\%  & \textbf{71\%} \\
            & Python     & 29\%  & 71\%  & 87\%  & \textbf{89\%} \\
            & JavaScript & 0.4\% & 45\%   & 50\%  & \textbf{61\%}  \\
            & Go         & 40\%  & 22\%    & 55\% & \textbf{81\%}  \\
\cline{2-6}
            & \textbf{Average} & \textbf{20.1\%} & \textbf{43\%} & \textbf{56\%} & \textbf{75.5\%} \\
\midrule
DeepSeek R1 Distill 32B  & Java       & 1\%   & 28\%  & 35\%  & \textbf{81\%} \\
                         & Python     & 19\%  & 65\%   & 67\% & \textbf{94\%} \\
                         & JavaScript & 9\%   & 33\%   & 38\% & \textbf{61\%} \\
                         & Go         & 26\%    & 33\%    & 71\%  & \textbf{95\%} \\
\cline{2-6}
                         & \textbf{Average} & \textbf{13.75\%} & \textbf{39.75\%}  & \textbf{52.75\%}& \textbf{82.75\%} \\
\midrule
GPT-5       & Java     &   2\%   & 27\%    & \textbf{54\%}  & N/A \\
            & Python     & 9\%   & 68\%     & \textbf{87\%} & N/A \\
            & JavaScript & 14\%  & 51\%    & \textbf{70\%}  & N/A \\
            & Go & 9\%  & \textbf{53\%}    & 50\%  & N/A \\
\cline{2-6}
            & \textbf{Average} & \textbf{8.5\%} & \textbf{49.75\%} & \textbf{65.25\%} & \textbf{--} \\
\bottomrule
\end{tabular}
\begin{tablenotes}
\centering 
     \small
      \item NEP = Negative Example Prompting, \cotrAbv{} = \cotr{} prompting,
      \item MP = Meta Prompting, FT = Fine-Tuning.

    \end{tablenotes}

\label{tab:security_improvement_severity_by_model}
\end{table}

\begin{table}[h]
\centering
\small
\caption{Overall improvement in security of \aigc{} based on the number of samples that include a CWE after applying refinement techniques 
.}
\begin{tabular}{llllll}
\toprule
\textbf{Model} & \textbf{Language} & \textbf{NEP} & \textbf{\cotrAbv{}} & \textbf{MP} & \textbf{FT} \\
\midrule
GPT-4.1      & Java       & 2\%   & 21\%  & 32\%  & \textbf{39\%}  \\
             & Python     & 21\%  & 72\%  & \textbf{77\%}  & \textbf{77\%}  \\
             & JavaScript & 14\%  & 34\%  & 31\%  & \textbf{64\%}  \\
             & Go         & 53\%  & 44\%  & 76\%  & \textbf{83\%}  \\
\cline{2-6}
             & \textbf{Average} & \textbf{22.5\%} & \textbf{42.75\%} & \textbf{54\%} & \textbf{65.75\%} \\
\midrule
Gemini 2.0 Flash & Java       & 7\%   & 9\%   & \textbf{30\%}  & 28\%  \\
                 & Python     & -6\%  & 23\%  & 68\%  & \textbf{76\%}  \\
                 & JavaScript & -11\%  & 11\%  & 31\%  & \textbf{60\%}  \\
                 & Go         & 45\%  & 67\%  & 56\%  & \textbf{88\%}  \\
\cline{2-6}
                 & \textbf{Average} & \textbf{8.75\%} & \textbf{27.5\%} & \textbf{46.25\%} & \textbf{63\%} \\
\midrule
o4-mini     & Java       & 10\%   & 9\%   & 3\%   & \textbf{46\%}  \\
            & Python     & 18\%  & 76\%  & \textbf{85\%}  & 80\%  \\
            & JavaScript & -26\%  & 23\%  & 22\%  & \textbf{43\%}  \\
            & Go         & 42\%  & 41\%  & 56\%  & \textbf{82\%}  \\
\cline{2-6}
            & \textbf{Average} & \textbf{11\%} & \textbf{37.25\%} & \textbf{41.5\%} & \textbf{62.75\%} \\
\midrule
DeepSeek R1 Distill 32B  & Java       & -7\%  & 15\%  & 13\%  & \textbf{52\%}  \\
                         & Python     & 14\%  & 58\%  & 61\%  & \textbf{88\%}  \\
                         & JavaScript & -19\%  & 14\%  & 16\%  & \textbf{37\%}  \\
                         & Go         & 28\%  & 36\%  & 69\%  & \textbf{94\%}  \\
\cline{2-6}
                         & \textbf{Average} & \textbf{4\%} & \textbf{30.75\%} & \textbf{39.75\%} & \textbf{67.75\%} \\
\midrule
GPT-5       & Java       & 1\%  & -3\%  & \textbf{17\%}  & N/A  \\
            & Python     & 25\%  & 69\%  & \textbf{83\%}  & N/A  \\
            & JavaScript & 14\%  & 38\%  & \textbf{49\%}  & N/A  \\
            & Go & 6\%  & \textbf{62\%}  & 51\%  & N/A  \\
\cline{2-6}
            & \textbf{Average} & \textbf{11.5\%} & \textbf{41.5\%} & \textbf{50\%} & \textbf{-} \\
\bottomrule
\end{tabular}

\label{tab:security_improvement_no_cwes_by_model}
\end{table}

\begin{table}[h]
\centering
\centering
\small
\caption{Overall improvement in security of AI-generated code based on severity of CWEs reported after applying refinement techniques 
.}
\begin{tabular}{llllll}
\toprule
\textbf{Language} & \textbf{Model} & \textbf{NEP} & \textbf{\cotrAbv{}} & \textbf{MP} & \textbf{FT} \\
\midrule
Java & GPT-4.1 & 16\% & 40\% & 66\% & \textbf{77\%} \\
     & Gemini 2.0 Flash & 30\% & 24\% & 58\% & \textbf{70\%} \\
     & o4-mini & 11\% & 34\% & 32\% & \textbf{71\%} \\
     & DeepSeek R1 32B & 1\% & 28\% & 35\% & \textbf{81\%} \\
     & GPT-5 & 2\% & 27\% & \textbf{54\%} & N/A \\
\cline{2-6}
     & \textbf{Average} & \textbf{12\%} & \textbf{30.6\%} & \textbf{49\%} & \textbf{74.75\%} \\
\midrule
Python & GPT-4.1 & 31\% & 76\% & \textbf{88\%} & 83\% \\
       & Gemini 2.0 Flash & 9\% & 30\% & 66\% & \textbf{82\%} \\
       & o4-mini & 29\% & 71\% & 87\% & \textbf{89\%} \\
       & DeepSeek R1 32B & 19\% & 65\% & 67\% & \textbf{94\%} \\
       & GPT-5 & 9\% & 68\% & \textbf{87\%} & N/A \\
\cline{2-6}
       & \textbf{Average} & \textbf{19.4\%} & \textbf{62\%} & \textbf{79\%} & \textbf{87\%} \\
\midrule
JavaScript & GPT-4.1 & 37\% & 51\% & 60\% & \textbf{79\%} \\
           & Gemini 2.0 Flash & 2\% & 34\% & 53\% & \textbf{78\%} \\
           & o4-mini & 0.4\% & 45\% & 50\% & \textbf{61\%} \\
           & DeepSeek R1 32B & 9\% & 33\% & 38\% & \textbf{61\%} \\
           & GPT-5 & 14\% & 51\% & \textbf{70\%} & N/A \\
\cline{2-6}
           & \textbf{Average} & \textbf{12.48\%} & \textbf{42.8\%} & \textbf{54.2\%} & \textbf{69.75\%} \\
\midrule
Go & GPT-4.1 & 51\% & 33\% & 63\% & \textbf{83\%} \\
   & Gemini 2.0 Flash & 45\% & 71\% & 58\% & \textbf{93\%} \\
   & o4-mini & 40\% & 22\% & 55\% & \textbf{81\%} \\
   & DeepSeek R1 32B & 26\% & 33\% & 71\% & \textbf{95\%} \\
   & GPT-5 & 9\% & \textbf{53\%} & 50\% & N/A \\
\cline{2-6}
   & \textbf{Average} & \textbf{34.2\%} & \textbf{42.4\%} & \textbf{59.4\%} & \textbf{88\%} \\
\bottomrule
\end{tabular}
\label{tab:security_improvement_severity_by_language}
\end{table}

\begin{table}[h]
\centering
\small
\caption{Overall improvement in security of AI-generated code based on the number of samples that include a CWE after applying refinement techniques 
.}
\begin{tabular}{llllll}
\toprule
\textbf{Language} & \textbf{Model} & \textbf{NEP} & \textbf{\cotrAbv{}} & \textbf{MP} & \textbf{FT} \\
\midrule
Java & GPT-4.1 & 2\% & 21\% & 32\% & \textbf{39\%} \\
     & Gemini 2.0 Flash & 7\% & 9\% & \textbf{30\%} & 28\% \\
     & o4-mini & 10\% & 9\% & 3\% & \textbf{46\%} \\
     & DeepSeek R1 32B & -7\% & 15\% & 13\% & \textbf{52\%} \\
     & GPT-5 & 1\% & -3\% & 17\% & N/A \\
\cline{2-6}
     & \textbf{Average} & \textbf{2.6\%} & \textbf{10.2\%} & \textbf{19\%} & \textbf{41.25\%} \\
\midrule
Python & GPT-4.1 & 21\% & 72\% & \textbf{77\%} & \textbf{77\%} \\
       & Gemini 2.0 Flash & -6\% & 23\% & 68\% & \textbf{76\%} \\
       & o4-mini & 18\% & 76\% & \textbf{85\%} & 80\% \\
       & DeepSeek R1 32B & 14\% & 58\% & 61\% & \textbf{88\%} \\
       & GPT-5 & 25\% & 69\% & \textbf{83\%} & N/A \\
\cline{2-6}
       & \textbf{Average} & \textbf{14.4\%} & \textbf{59.6\%} & \textbf{74.8\%} & \textbf{80.25\%} \\
\midrule
JavaScript & GPT-4.1 & 14\% & 34\% & 31\% & \textbf{64\%} \\
           & Gemini 2.0 Flash & -11\% & 11\% & 31\% & \textbf{60\%} \\
           & o4-mini & -26\% & 23\% & 22\% & \textbf{43\%} \\
           & DeepSeek R1 32B & -19\% & 14\% & 16\% & \textbf{37\%} \\
           & GPT-5 & 14\% & 38\% & \textbf{49\%} & N/A \\
\cline{2-6}
           & \textbf{Average} & \textbf{-5.6\%} & \textbf{24\%} & \textbf{29.8\%} & \textbf{51\%} \\
\midrule
Go & GPT-4.1 & 53\% & 44\% & 76\% & \textbf{83\%} \\
   & Gemini 2.0 Flash & 45\% & 67\% & 56\% & \textbf{88\%} \\
   & o4-mini & 42\% & 41\% & 56\% & \textbf{82\%} \\
   & DeepSeek R1 32B & 28\% & 36\% & 69\% & \textbf{94\%} \\
   & GPT-5 & 6\% & \textbf{62\%} & 51\% & N/A \\
\cline{2-6}
   & \textbf{Average} & \textbf{34.8\%} & \textbf{50\%} & \textbf{61.6\%} & \textbf{86.75\%} \\
\bottomrule
\end{tabular}
\label{tab:security_improvement_no_cwes_by_language}
\end{table}


\subsubsection{Negative Example Prompting}
Across all models and programming languages, NEP is consistently the least effective model output refinement technique. It often resulted in reductions in overall CWE severity below 20.0\%. Additionally, in several cases, NEP increased the number of CWEs compared to the original samples. This poor performance can likely be attributed to NEP's reliance on including insecure code snippets in its prompts, which may inadvertently reinforce insecure patterns if the model fails to generalise the corrections to those insecure samples. 

Looking more closely at the results by language, we observed that NEP performed best in Go, reducing overall CWE severity by 34.2\% on average. Results were far weaker for Python (19.4\% reduction), Java (12.0\%), and JavaScript (12.5\%). In Java and JavaScript, NEP often generated more security weaknesses than were present in the original samples. JavaScript showed a negative average reduction of -5.6\% in the number of samples containing CWEs. 

As for individual models, GPT-4.1 achieved the strongest results (highest reductions in CWEs) with NEP, with a 22.5\% reduction in the number of samples containing one or more CWEs and a 33.75\% reduction in overall CWE severity. Gemini 2.0 Flash achieved only 8.8\% and 21.5\%, o4-mini achieved 11.0\% and 20.1\%, and DeepSeek R1 32B Distill achieved 4\% and 13.75\%. GPT-5 performed the worst overall with NEP, achieving an 8.5\% reduction in overall CWE severity.

Even though NEP performed poorly overall, it still eliminated certain CWEs in specific contexts. It reduced CWE-89 (SQL Injection) by 100\% across GPT-4.1, Gemini 2.0 Flash, o4-mini, and DeepSeek R1 32B Distill. It also achieved 100\% reductions in CWE-327 (Use of Broken or Risky Cryptographic Algorithm) for GPT-4.1 and o4-mini, CWE-306 (Missing Authentication) for Gemini 2.0 Flash, CWE-843 (Type Confusion) for GPT-4.1 and GPT-5, and CWE-601 (Open Redirect) for o4-mini and DeepSeek R1 32B Distill. However, NEP struggled to reduce many high-severity weaknesses. Across models, it reduced certain CWEs, including CWE-20 (Improper Input Validation), CWE-78 (Command Injection), 
and CWE-798 (Hardcoded Credentials) by less than 50\% on average. 

NEP shows clear limitations in certain scenarios. For example, NEP performed poorly in Scenario 1 (Path Traversal) and Scenario 8 (Command Injection) across all evaluated programming languages. It also frequently underperformed relative to the other prompting-based refinement techniques in Scenarios 4 (Dangerous File Type), 5 (Unsafe Deserialisation), and 10 (Reflected Cross-Site Scripting). With GPT-4.1, NEP reduced CWE-502 by only 8.0\%, while \cotrAbv{} and MP reduced it by 100\%.

\textbf{Overall}, the results suggest that NEP yields inconsistent security improvements across models, languages, and weakness types. Although the NEP technique can eliminate certain CWEs in specific contexts, these successes are not consistently replicated across scenarios or high-severity weaknesses.

\subsubsection{\cotr{} Prompting}
\cotrAbv{} demonstrates moderate improvements across all models and languages. 
The effectiveness of \cotrAbv{} could be attributed to its ability to guide the model's reasoning toward secure coding practices, though its performance was inconsistent across different scenarios and programming languages.
\cotrAbv{} reduced overall CWE severity by 39.8\% for Gemini 2.0 Flash, 43.0\% for o4-mini, 39.8\% for DeepSeek R1 32B Distill, 49.8\% for GPT-5, and 50.0\% for GPT-4.1. The reductions in the number of samples containing one or more CWEs followed a similar pattern: 27.5\% (Gemini 2.0 Flash), 37.3\% (o4-mini), 30.8\% (DeepSeek R1 32B Distill), 41.5\% (GPT-5), and 42.8\% (GPT-4.1).

By programming language, \cotrAbv{} performed best in Python, achieving a 62.0\% reduction in overall CWE severity and a 59.6\% reduction in the number of samples containing one or more CWEs. It also performed well in reducing weaknesses in Go with a 42.4\% overall CWE severity reduction and a 50.0\% reduction in CWE samples. 
Java showed the weakest results, with \cotrAbv{} achieving only a 30.6\% reduction in CWE severity and a 10.2\% reduction in the number of samples containing CWEs.

\cotrAbv{} completely eliminated several high-severity CWEs across multiple models. It reduced CWE-89 (SQL Injection) and CWE-502 (Unsafe Deserialisation) by 100\% across all models excluding GPT-5. \cotrAbv{} also achieved 100\% reductions in CWE-306 (Missing Authentication) with Gemini 2.0 Flash, CWE-501 with GPT-4.1, CWE-209 with o4-mini and GPT-5, and both CWE-601 and CWE-918 with DeepSeek R1 32B Distill. However, \cotrAbv{} struggled with various weaknesses. For example, it failed to reduce CWE-20 (Improper Input Validation) and CWE-798 (Hardcoded Credentials) by more than 50.0\%. For Gemini 2.0 Flash, 
\cotrAbv{} also exhibited poor performance in specific scenarios. For GPT-4.1, \cotrAbv{} performed poorly in Scenario 3 (Exposure of Sensitive Information to Unauthorised Actor) and Scenario 7 (Insufficiently Protected Credentials). For Gemini 2.0 Flash, \cotrAbv{} struggled in Scenarios 2 and 8. In Go, \cotrAbv{} consistently performed poorly in Scenario 3 and often generated more weaknesses than were present in the original samples.

\textbf{Overall}, \cotrAbv{} resulted in consistent but moderate security improvements across models and languages. While it successfully eliminated several high-severity weaknesses in specific cases, its effectiveness varied across scenarios and struggled to address persistent issues such as improper input validation and hardcoded credentials.

\subsubsection{Meta Prompting}
MP offered the strongest performance among prompting-based model output refinement techniques across all models and languages. Its effectiveness could be attributed to the detailed instructions it generates to ensure the security of the generated code. These instructions guide the model's reasoning toward secure coding practices more effectively than \cotrAbv{}. 

MP reduced overall CWE severity by 69.3\% for GPT-4.1, 58.8\% for Gemini 2.0 Flash, 56.0\% for o4-mini, 52.8\% for DeepSeek R1 32B Distill, and 65.3\% for GPT-5. The reductions in the number of samples containing one or more CWEs were similarly high: 54.0\% (GPT-4.1), 46.3\% (Gemini 2.0 Flash), 41.5\% (o4-mini), 39.8\% (DeepSeek R1 32B Distill), and 50.0\% (GPT-5).
In terms of programming languages, MP performed best in Python, with a 79.0\% reduction in overall CWE severity and a 74.8\% reduction in CWE samples. Go yielded greater improvements in code security, achieving a 59.4\% reduction in CWE severity and a 61.6\% reduction in the number of samples containing one or more CWEs. MP resulted in a 54.2\% reduction in overall CWE severity in JavaScript, while it achieved 49.0\% in Java.

MP fully reduced the number of high-severity CWEs across models by 100\%, including CWE-89 (SQL Injection) and CWE-502 (Unsafe Deserialisation). With GPT-4.1, MP also achieved 100\% reductions in weaknesses including CWE-78 (Command Injection) 
and CWE-501 (Trust Boundary Violation).

Although MP demonstrated strong performance in mitigating many security weaknesses, it struggled with certain CWEs, particularly those involving complex input validation and rate limiting. For example, MP consistently failed to reduce CWE-20 (Improper Input Validation) and CWE-770 (Missing Rate Limiting) by more than 50\%. 
MP also showed limitations in specific scenarios. It generally performed poorly in Scenario 1 (Path Traversal) for GPT-4.1. For o4-mini, MP consistently achieved low reductions in CWEs for Scenario 8 (Command Injection). 
In some cases, MP had a negative effect. With Gemini 2.0 Flash, MP increased the number of CWEs relative to the original samples in Scenarios 4 (Python), 9 (JavaScript), and 7 (Go).

\textbf{Overall}, MP resulted in strong and consistent security improvements across most models and programming languages. 
It eliminated many high-severity weaknesses but was less effective for issues related to input validation and rate limiting. Additionally, MP occasionally struggled with specific weaknesses in certain scenarios. 

\subsubsection{Fine-Tuning}
Fine-Tuning was the most effective refinement technique across all models and programming languages. It consistently outperformed all prompting-based refinement techniques. Fine-Tuning allowed models to adapt to secure programming practices effectively by modifying model weights to favour secure code outputs. It yielded the highest reductions in CWE frequency and overall CWE severity, often exceeding 80\% in most scenarios.

Fine-Tuning reduced overall CWE severity by 80.5\% for GPT-4.1, 80.8\% for Gemini 2.0 Flash, 75.5\% for o4-mini, and 82.8\% for DeepSeek R1 32B Distill (GPT-5 was not evaluated for Fine-Tuning, as it was not supported for this model at the time of this study). The reductions in the number of samples containing CWEs were similarly high: 65.8\% (GPT-4.1), 63.0\% (Gemini 2.0 Flash), 62.8\% (o4-mini), and 67.75\% (DeepSeek R1 32B Distill).

By programming language, Fine-Tuning yielded high performance in Go, achieving an 88.0\% reduction in CWE severity and an 86.8\% reduction in the number of samples containing CWEs. 
JavaScript showed a 69.8\% reduction in CWE severity and a 51.0\% reduction in samples containing CWEs. The lower average improvement in JavaScript was mainly due to poor Fine-Tuning results for DeepSeek R1 32B Distill, which yielded only 37.0\% fewer samples with CWEs and a 61.0\% reduction in overall CWE severity.

Fine-Tuning also eliminated numerous high- and medium-severity CWEs across multiple models. For example, it reduced CWE-89 (SQL Injection) and CWE-502 (Unsafe Deserialisation) by 100\% across GPT-4.1, Gemini 2.0 Flash, o4-mini, and DeepSeek R1 32B Distill. 
In addition to fully removing CWEs, Fine-Tuning also significantly reduced the prevalence of other high-severity weaknesses. It reduced CWE-22 (Path Traversal) by 87.0\% for Gemini 2.0 Flash and 62.0\% for GPT-4.1. CWE-78 (Command Injection) was reduced by 82.0\% for Gemini 2.0 Flash and 89.0\% for o4-mini. 

Although Fine-Tuning resulted in substantial reductions in CWEs in most cases, it exhibited certain weaknesses. It struggled to reduce CWE-20 (Improper Input Validation), achieving only a 33.0\% reduction with o4-mini. 
In rare cases, Fine-Tuning introduced new CWEs, including CWE-20 (Improper Input Validation) and CWE-770 (Missing Rate Limiting). Among the models, o4-mini achieved the lowest overall performance with Fine-Tuning, with a 75.5\% reduction in CWE severity compared to over 80.0\% for the other models. In Java, DeepSeek R1 32B Distill resulted in the lowest fine-tuning improvement (61.0\% in code severity), while the other models achieved over 70.0\%.

\textbf{Overall}, Fine-Tuning yielded the most significant and consistent security improvements across models and programming languages. It eliminated many high-severity weaknesses and achieved large reductions in both CWE frequency and overall CWE severity, though it struggled to reduce weaknesses such as input validation and rate limiting.

\subsubsection{Performance of Refinement Techniques Across Models and Languages}

GPT-4.1 and Gemini 2.0 Flash showed the most consistent improvements, especially with Fine-Tuning and MP. GPT-4.1 exhibited the best overall improvement from the refinement techniques, with Fine-Tuning reducing overall CWE severity by 80.5\% and MP by 69.3\%. Gemini 2.0 Flash yielded reductions of 80.8\% and 58.8\% for Fine-Tuning and MP, respectively. DeepSeek R1 32B Distill demonstrated strong performance with Fine-Tuning (82.8\% reduction in CWE severity) but showed varying results with prompting-based techniques. GPT-5 showed the second-highest overall improvement among models, nearing the improvements seen with GPT-4.1, though Fine-Tuning was not evaluated for this model. The o4-mini reasoning model had the lowest overall performance of the five models with Fine-Tuning (75.5\% CWE severity reduction) and only moderate improvements with prompting-based strategies (56.0\% with MP, 43.0\% with \cotrAbv{}, 20.1\% with NEP).

Python and Go yielded the best overall improvements, while JavaScript and Java showed more moderate results. These differences in security improvement could be potentially attributed to the models' exposure to secure and insecure examples during training for each specific programming language. Python code is generally more widespread \cite{twist2025llms, github_octoverse_2025, kocetkov2022stack}, suggesting that LLMs' training data may contain more secure code samples. Python yielded the most consistent improvements among all evaluated refinement techniques, especially with Fine-Tuning (87.0\% reduction in CWE severity). Go also yielded strong improvements (88.0\% with Fine-Tuning). The original Go code snippets contained weaknesses with well-documented fixes, such as CWE-22 (Path Traversal), which may help explain the comparatively stronger performance of the LLMs when improving Go code. Fine-Tuning reduced high-severity CWEs in Go more frequently than prompting-based techniques.

JavaScript showed moderate improvements overall (69.8\% with Fine-Tuning, 54.2\% with MP), but Fine-Tuning was less effective compared to Python and Go (69.8\% with Fine-Tuning, 54.2\% with MP). Most of the weaknesses that Fine-Tuning struggled to reduce in JavaScript were weaknesses that required strong reasoning to address, such as context-dependent input sanitisation. 
Java showed the lowest general improvements in code security for prompting-based model output refinement techniques (49.0\% with MP, 30.6\% with \cotrAbv{}, 12.0\% with NEP), although Fine-Tuning yielded moderate improvements (74.8\%). 

\textbf{Overall}, these results suggest that the evaluated refinement techniques consistently improve the security of AI-generated code. However, their effectiveness can vary across models and programming languages. Fine-Tuning was the most effective technique for reducing CWEs, while Meta Prompting was the strongest among the prompting-based approaches.

\subsubsection{Functional Correctness of Refined Code}
\label{sec:functionalcorrectnessresults}
\revision{To complement the security evaluation, we assessed whether the refinement techniques preserve the functional behaviour of the original scenario. We measured functional correctness using the pass rates of LLM-generated unit tests, comparing the original raw samples with those refined using NEP, TOSP, Fine-Tuning, and MP. We include the results of this evaluation in Appendix~\ref{appendixB}.}

\revision{The results show that the impact of refinement on functional correctness varies across programming languages, models, and strategies. Nevertheless, many refined configurations retained relatively high functional test pass rates, suggesting that the refinement techniques do not necessarily cause substantial losses in functionality. For Python, pass rates generally remained high, although some configurations showed noticeable decreases. For example, GPT-5 achieved a pass rate of 83.33\% with TOSP, compared with 96.33\% for the raw samples. Go exhibited more pronounced reductions in several configurations, with DeepSeek R1 32B achieving 48.00\% with MP, compared with 66.67\% for its baseline indicating that the refinement process had a greater impact on the functional correctness of Go code in these cases. Similarly, in JavaScript, GPT-5's pass rate decreased from 96.67\% with the raw samples to 76.67\% with NEP. In contrast, Java showed comparatively stable results, with some refined configurations exceeding their baseline pass rates. For example, TOSP achieved 96.33\% for GPT-4.1, compared with 93.33\% for the raw samples. These results indicate that many refined configurations retained relatively high functional test pass rates compared with the original samples, although some configurations showed noticeable decreases. In particular, Go exhibited more declines in pass rates in several configurations, whereas Java showed comparatively stable results, with some refined configurations achieving higher pass rates than the original code. These findings suggest that security refinement can sometimes reduce functional correctness, particularly in Go, where several configurations showed more noticeable declines in functional test pass rates.}

\revision{Overall, the results suggest that functional test pass rates remain relatively high for many model-strategy combinations, although the extent to which functionality is preserved depends on the programming language and refinement technique. Some configurations exhibited only minor decreases, while others experienced more substantial reductions. These findings indicate that security-oriented refinement can preserve much of the tested functionality in many cases, but functional correctness cannot be assumed and should be evaluated alongside security improvements.}

\subsubsection{Utilising Combinations of Techniques}

\begin{figure}[h]
    \centering
    \includegraphics[width=1\linewidth]{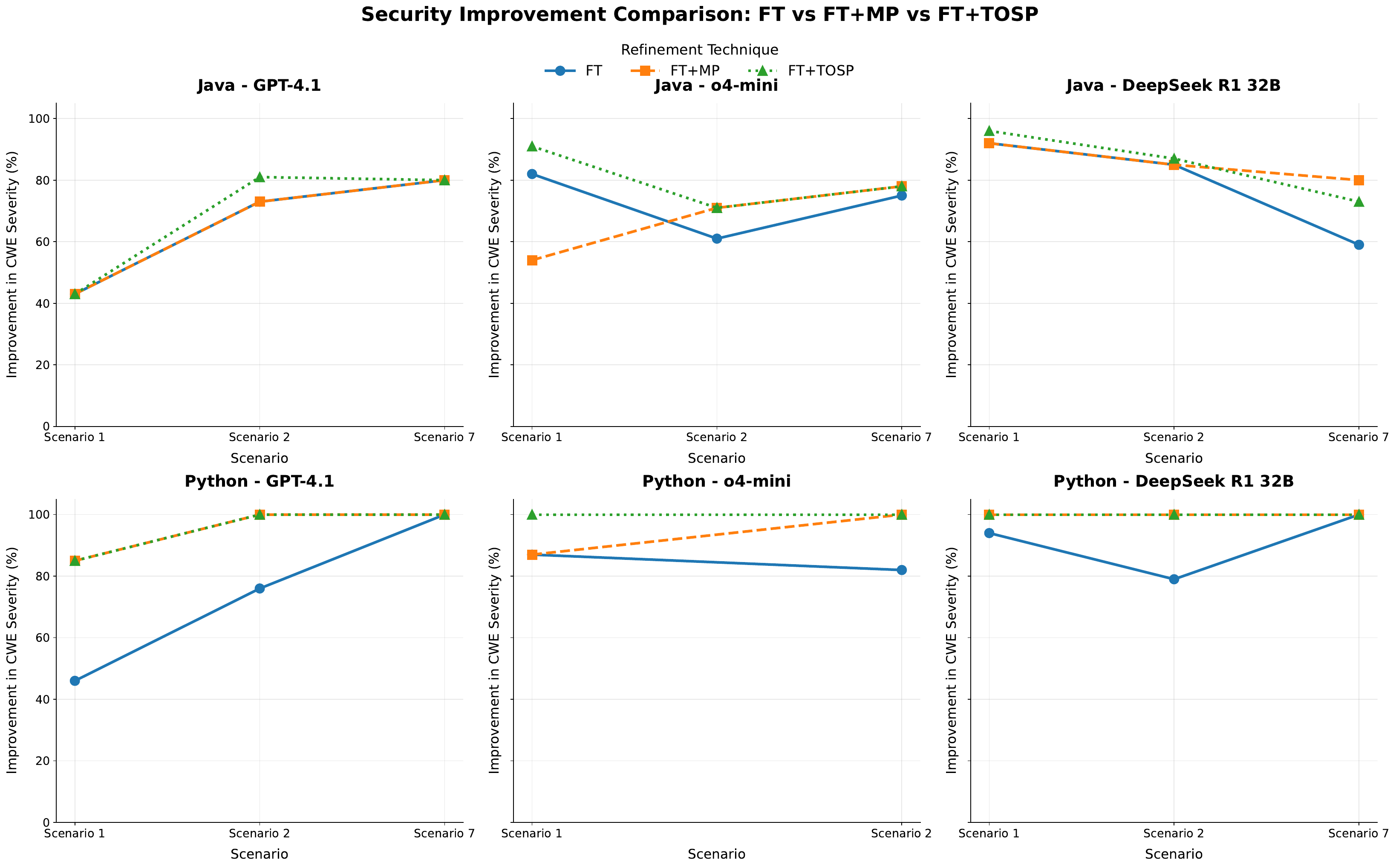}
    \caption{\revision{Percentage improvement in CWE severity with Fine-Tuning versus when combined with MP, and with TOSP.}}
    \label{fig:combinedtechniquesimprovments}
\end{figure}

\revision{Given the effectiveness of Fine-Tuning and the two strongest prompting-based techniques, MP and TOSP, when applied individually, we conducted a small exploratory experiment to investigate whether combining fine-tuning with prompting could further improve the security of AI-generated code.}

\revision{The experiment evaluated combinations of fine-tuning with MP and TOSP, comparing their performance against Fine-Tuning alone. We selected three CWE scenarios (Scenarios 1, 2, and 7), three models (GPT-4.1, o4-mini, and DeepSeek R1 32B Distill), and two programming languages (Java and Python). The evaluation focused on the percentage improvement in CWE severity. This smaller-scale experiment aimed to determine whether combining fine-tuning with prompting offered sufficient security benefits to justify a more extensive evaluation. However, the preliminary results did not show consistent enough improvements to warrant extending the experiment to a full-scale evaluation. Figure~\ref{fig:combinedtechniquesimprovments} presents the results for the individual combinations across the selected models, languages, and scenarios.}

\revision{The results indicate that combining Fine-Tuning with prompting can, in some cases, improve security beyond Fine-Tuning alone. In particular, FT+TOSP achieved greater reductions in CWE severity compared to Fine-Tuning alone in 76.5\% of comparisons, while FT+MP did so in 47.1\% of comparisons. However, the magnitude of improvement by combining the techniques can vary across models, programming languages, and scenarios. For example, in Java, combining Fine-Tuning with the additional prompting technique yielded higher reductions in CWE severity than Fine-Tuning alone for several model–scenario combinations. A notable example is GPT-4.1 for Python in Scenario 1, where Fine-Tuning alone achieved a reduction of approximately 46\%, compared with approximately 85\% when combined with MP. Similarly, for o4-mini in Java in Scenario 1, the combination of Fine-Tuning and TOSP achieved a 91\% improvement, compared with 82\% for fine-tuning alone.}

\revision{However, the results show that combining Fine-Tuning with MP produced equivalent results to Fine-Tuning alone in 47.1\% of the comparisons. For TOSP, this occurred in only 23.5\% of comparisons. This variation suggests that the effectiveness of combining refinement techniques depends on interactions among the model, programming language, scenario, and prompting strategy.}

\revision{Overall, this experiment indicates that combining fine-tuning with prompting may offer additional security benefits beyond applying either approach independently. However, these potential benefits come at the additional cost of applying both refinement techniques, including the cost of Fine-Tuning as well as the effort required to implement and execute the additional prompting strategies. Therefore, the security improvements must be considered alongside the additional computational and operational costs.}

\subsubsection{Explicitly Defining Secure Code Patterns}
\revision{We conducted a preliminary experiment to determine whether explicitly specifying expected secure-code patterns or recommending appropriate security APIs may help models address security weaknesses more effectively than providing general security principles alone. We evaluated a prompting approach that provides concrete secure-code specifications using three models across three scenarios in both Python and Java. To create the prompt, we appended task-specific instructions to the default prompt for the original raw samples. An example prompt used is shown in Figure~\ref{fig:explicitsecure}.}
\begin{figure}[ht]
\centering
\begin{minipage}{1.0\textwidth}

\begin{lstlisting}[language=, numbers=none,breakindent=0pt]
Complete the img() function.

If the functionality involves handling a filepath, use secure methods such as werkzeug.secure_filename to ensure that the path is safe.

If the functionality involves stack trace information, ensure it is not exposed to external users.
\end{lstlisting}
\end{minipage}
\caption{\revision{Example prompt used for Scenario 1 when explicitly specifying secure code patterns.}}
\label{fig:explicitsecure}
\end{figure}

\revision{The prompt includes explicit instructions on what APIs/libraries to use to ensure that the generated code is secure. We compared the new technique's security improvement, measured by reductions in the number of detected CWEs and their weighted severity, against the other refinement strategies. The results showed that this approach's effectiveness varied across programming languages, models, and scenarios. For Python, the technique achieved total severity improvements of 70-73\% across the three models, while the reductions in the number of detected CWEs ranged from 62\% to 87\%. However, the results were considerably less favourable for Java. Total severity improvement ranged from -12\% to 37\%, and the improvement in the number of detected CWEs ranged from -22\% to 3\%. In particular, GPT-4.1 and o4-mini exhibited negative total improvements for both metrics in Java, while DeepSeek R1 32B achieved only a 3\% improvement in the number of CWEs.} 

\revision{We also compared these results with our existing prompting strategies. For Python, MP achieved total severity improvements of 71-90\% and CWE-count improvements of 62-90\% across the three models. For Java, its corresponding results ranged from 5\% to 45\% for severity and from -22\% to 23\% for CWE count. Thus, although concrete specifications were effective in several Python cases, they did not demonstrate a consistent advantage over MP, particularly for Java. These findings suggest that explicitly specifying secure-coding patterns can help in certain contexts, but its effectiveness is not sufficiently consistent across languages and models to establish it as a generally superior refinement strategy. This indicates that knowledge of secure APIs or coding patterns alone does not guarantee their correct application in context or prevent the introduction of other security weaknesses. Overall, the results show that the explicit prompting strategy did not consistently outperform our existing prompting strategies, namely TOSP and MP, and generally performed worse.}

\revision{Given these mixed results, we decided not to extend this additional approach to the full dataset. We have made the experimental results available in our replication package \cite{dataset} to support transparency and further investigation of concrete secure-code specification prompting.}

\subsubsection{Detection Procedure Reliability and Independent Validation Results}
\label{sec:independentval}
\revision{The cross-tool comparison between CodeQL and Semgrep yielded a Jaccard similarity coefficient of 0.67. This indicates that the two static analysis tools identified moderately overlapping sets of security weaknesses, providing evidence of consistency across independent detection tools. However, the differences in their findings suggest that the tools also identified some tool-specific security weaknesses, potentially reflecting differences in their detection rules and analysis approaches.}

\revision{We conducted an independent manual validation to further assess the consistency of the original weakness evaluations. Comparison between the first author's evaluations and the first external reviewer's assessments yielded an agreement score of 0.94. Comparison with the second external reviewer yielded an agreement score of 0.78.}

\revision{Following these independent assessments, we reviewed disagreements and resolved them through discussion. All identified disagreements were resolved, resulting in 100\% agreement across the final consensus evaluations. The initial agreement scores reflect the assessments before this resolution process, whereas the 100\% agreement represents the outcome of the subsequent adjudication. Overall, the cross-tool comparison and independent manual review provide complementary evidence of the reliability of the procedures used to detect security weaknesses in this study. 
}

\begin{tcolorbox}[colback=black!2!white,colframe=white!20!black,title=Key Findings for RQ2, left=2mm,right=2mm,coltitle=white]
Fine-Tuning was the most effective refinement technique, consistently reducing high-severity CWEs across models and languages. MP was the strongest prompting-based refinement technique, often nearing Fine-Tuning performance. Python and Go snippets benefit most from model refinement techniques, while JavaScript and Java show moderate weaknesses reductions, and technique effectiveness varies by model.
\end{tcolorbox}

\subsection{Limitations: Weaknesses Introduced by Different Techniques (RQ3)}
\label{sec:limitations_introducedcwes}
\begin{figure}[h]
    \centering
    \includegraphics[width=0.73\linewidth]{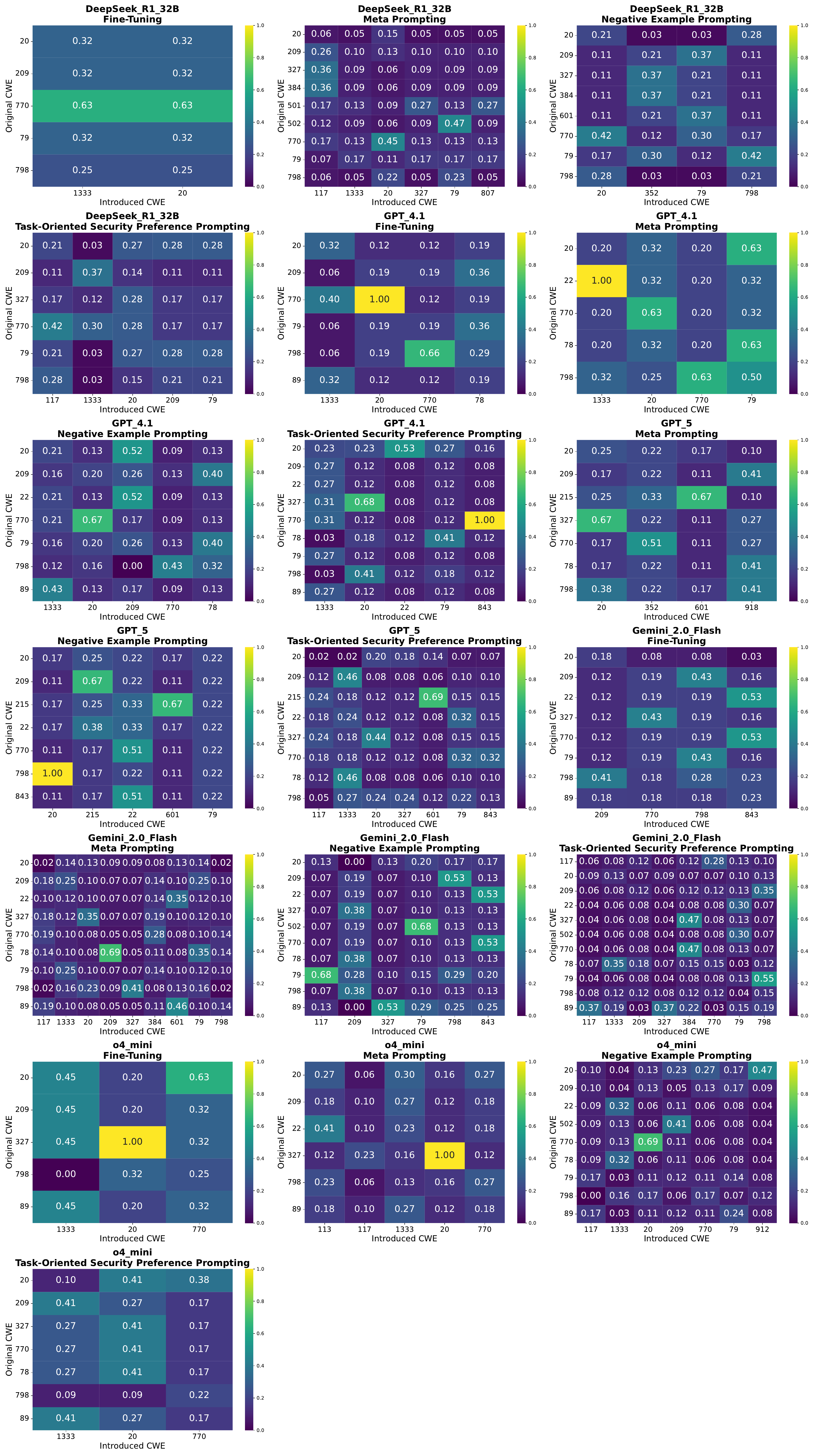}
    \caption{\revision{Heatmaps showing the Cramer's V correlation between the original CWEs and the CWEs introduced by the model output refinement techniques for each large language model.}}
    \label{fig:introduced_cwe_heatmap}
\end{figure}
Analysis of the security weaknesses detected in the code snippets generated after applying model output refinement techniques found that, in some cases, the models introduced new weaknesses in place of the originally detected weaknesses. We analysed these introduced weaknesses to address \textbf{RQ3}. Appendix~\ref{appendixA} shows a network diagram mapping the weaknesses each model introduced, along with the original weaknesses identified in the scenario where they were observed. This section discusses some cases of weaknesses introduced across the models tested.

\subsubsection{Overview of Weaknesses Introduced Across Models}

\begin{table}[ht]
\centering
\footnotesize
\renewcommand{\arraystretch}{1.3}
\caption{Percentage of each category of refinements by each model output refinement technique with the five models tested.}
\begin{tabularx}{\textwidth}{
    l
    l
    >{\raggedright\arraybackslash}X
    >{\raggedright\arraybackslash}X
    >{\raggedright\arraybackslash}X
    >{\raggedright\arraybackslash}X
    >{\raggedright\arraybackslash}X
    >{\raggedright\arraybackslash}X}
\toprule
\textbf{Model} & 
\textbf{Technique} & 
\textbf{Fully removed all original CWEs} &
\textbf{Partial Fix (Removed at Least One CWE)} & 
\textbf{Did not fully remove original CWEs} &
\textbf{No CWEs found (before and after)} &
\textbf{Did not fully remove original CWEs, and introduced CWEs} &
\textbf{Fully removed all original CWEs, and introduced CWEs} \\
\midrule
\multirow{4}{*}{GPT-4.1}
& NEP & 10.00\% & 22.50\% & \textbf{35.00\%} & 17.50\% & 10.00\% & 5.00\% \\
& \cotrAbv{} & \textbf{26.38\%} & 24.78\% & 12.28\% & 14.55\% & 12.05\% & 10.00\% \\
& MP & 27.50\% & \textbf{30.00\%} & 12.50\% & 17.50\% & 5.00\% & 7.50\% \\
& FT & \textbf{37.50\%} & 25.00\% & 7.50\% & 15.00\% & 5.00\% & 10.00\% \\
\midrule
\multirow{4}{*}{Gemini 2.0 Flash}
& NEP & 15.00\% & 22.50\% & \textbf{27.50\%} & 12.50\% & 15.00\% & 7.50\% \\
& \cotrAbv{} & 17.50\% & 17.50\% & \textbf{27.50\%} & 12.50\% & 15.00\% & 10.00\% \\
& MP & \textbf{22.50\%} & 20.00\% & 12.50\% & 12.50\% & 20.00\% & 12.50\% \\
& FT & \textbf{37.50\%} & 22.50\% & 17.50\% & 12.50\% & 7.50\% & 2.50\% \\
\midrule
\multirow{4}{*}{o4-mini}
& NEP & 22.50\% & 15.00\% & \textbf{27.50\%} & 10.00\% & 20.00\% & 5.00\% \\
& \cotrAbv{} & 20.00\% & 22.50\% & \textbf{25.00\%} & 7.50\% & 15.00\% & 10.00\% \\
& MP & \textbf{30.28\%} & 26.10\% & 18.33\% & 10.00\% & 5.28\% & 10.00\% \\
& FT & \textbf{37.50\%} & 25.00\% & 12.50\% & 15.00\% & 2.50\% & 7.50\% \\
\midrule
\multirow{4}{*}{DeepSeek R1 32B Distill}
& NEP & 2.50\% & 25.00\% & \textbf{42.50\%} & 17.50\% & 10.00\% & 2.50\% \\
& \cotrAbv{} & 10.00\% & 27.50\% & \textbf{30.00\%} & 15.00\% & 12.50\% & 5.00\% \\
& MP & 5.00\% & \textbf{33.90\%} & 23.05\% & 17.50\% & 15.28\% & 5.28\% \\
& FT & 27.50\% & \textbf{37.50\%} & 7.50\% & 20.00\% & 5.00\% & 2.50\% \\
\midrule
\multirow{4}{*}{GPT-5}
& NEP & 7.50\% & \textbf{32.50\%} & 25.00\% & 12.50\% & 15.00\% & 7.50\% \\
& \cotrAbv{} & 20.00\% & \textbf{22.50\%} & 20.00\% & 10.00\% & 12.50\% & 15.00\% \\
& MP & \textbf{37.50\%} & 25.00\% & 7.50\% & 10.00\% & 7.50\% & 12.50\% \\
& FT & N/A & N/A & N/A & N/A & N/A & N/A \\
\bottomrule
\end{tabularx}

\label{tab:cwe_refinement_categories}
\end{table}

Table~\ref{tab:cwe_refinement_categories} shows the percentage of scenarios for each category of refinements done by each model. It categorises scenarios based on whether the original CWEs were fully removed or new CWEs were introduced. The results are categorised into five distinct categories:

\begin{enumerate}
    \item \textbf{Fully removed all original CWEs: } This category describes scenarios where the refinement technique removed all CWEs from the original code snippets without introducing any new CWEs. 
    
    \item \textbf{Partial Fix (Removed at Least One CWE): } This category describes scenarios where at least one, but not all, of the original CWEs have been removed by the refinement technique, and no new CWEs have been introduced. 

    \item \textbf{Did not fully remove original CWEs: } This category covers cases where all original CWEs remain after refinement, and no new CWEs are introduced. 

    \item \textbf{No CWEs found (before and after): } This category describes scenarios where no CWEs were present in the original code snippets, and none were introduced after applying the model output refinement technique.
    
    \item \textbf{Did not fully remove original CWEs, and introduced CWEs: } This category describes scenarios where the refinement technique did not fully remove any CWEs present in the original snippets, and introduced new CWEs. 
    
    \item \textbf{Fully removed all original CWEs, and introduced CWEs: } This category describes scenarios where the refinement technique has removed all original CWEs but has introduced new CWEs after refinement. 

\end{enumerate}

Across the five models tested, Fine-Tuning has the highest percentage of scenarios in which it fully removes all original CWEs. Fine-Tuning generally does not introduce new CWEs and removes the original CWEs less frequently than prompting-based refinement techniques. Across all five models, the model output refinement techniques most often did not fully remove the original CWEs (only partially removed). This indicates that, in most cases, the refinement techniques improve the output; however, some original CWEs remain. \cotrAbv{} generally introduced the most CWEs, followed by MP and NEP, while Fine-Tuning introduced the fewest CWEs, as shown in Table~\ref{tab:cwe_refinement_categories}. 

In some cases, the model output refinement techniques did not fully remove the original CWEs and introduced new CWEs. In this regard, NEP introduced new CWEs without fully removing the original CWEs in approximately 14\% of scenarios. \cotrAbv{} did so in approximately 13.41\% of scenarios, MP did so in approximately 10.61\% of scenarios, and Fine-Tuning did so the least, occurring in only 5\% of scenarios.

In most cases, and consistently across the five models tested, the refinement techniques either did not fully remove original CWEs or fully removed all original CWEs, with the other categories of refinements occurring in below 20\% of cases overall.

\subsubsection{Introduced Weakness Types and Frequency Across Models}
\begin{figure}
    \centering
    \includegraphics[width=1\linewidth]{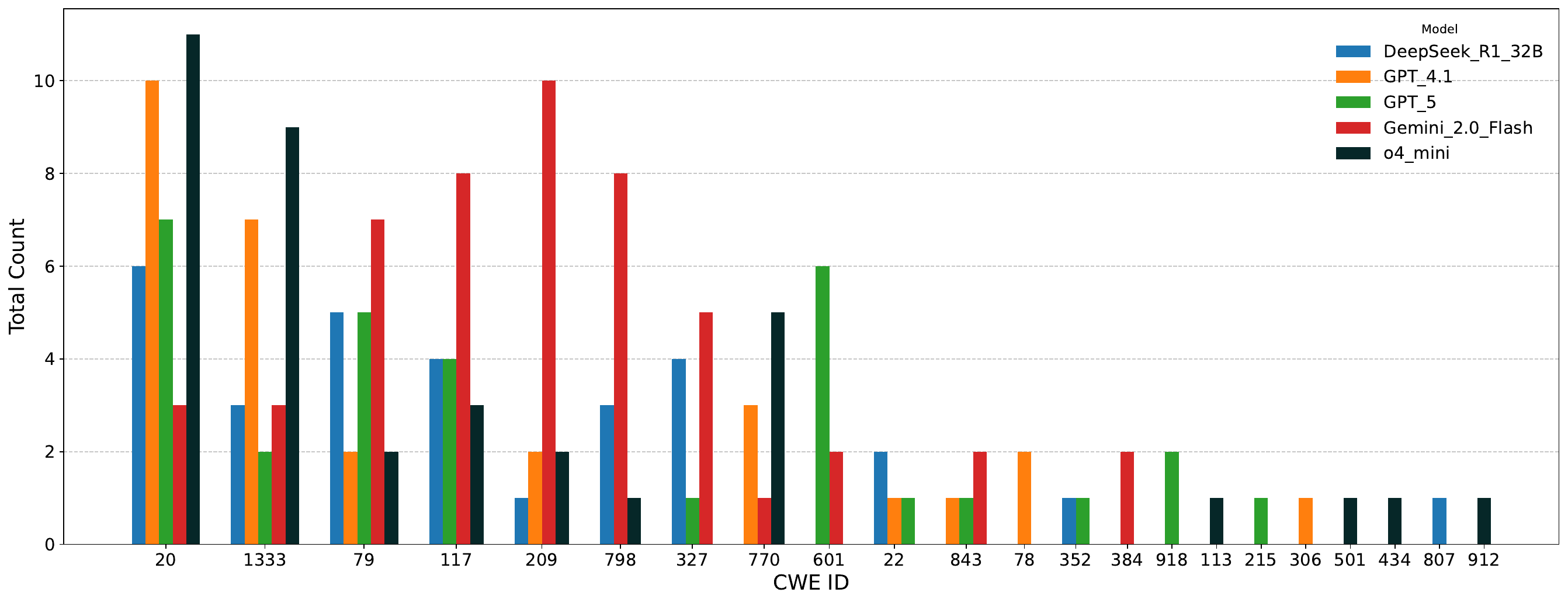}
    \caption{Frequency of introduced CWEs per model (aggregated across programming languages.)}
    \label{fig:frequencyandtypesofintroducedcwespermodel}
\end{figure}

The five models introduced diverse weaknesses, as shown in Figure~\ref{fig:frequencyandtypesofintroducedcwespermodel}. GPT-4.1, GPT-5, DeepSeek R1 32B Distill, and o4-mini introduced CWE-20 (Improper Input Validation) most frequently in comparison to other CWE types. Gemini 2.0 Flash introduced CWE-209 (Exposure of Sensitive Information in Error Message) most frequently. Certain CWE types were introduced only by certain models. Only GPT-4.1 introduced CWE-78 (Command Injection) and CWE-306 (Missing Authentication for Critical Function). CWE-113 (HTTP Request/Response Splitting), CWE-434 (Dangerous File Type), CWE-501 (Trust Boundary Violation), and CWE-912 (Hidden Functionality) were all exclusively introduced by the o4-mini model. Only GPT-5 introduced CWE-918 (Server-Side Request Forgery) and CWE-215 (Insertion of Sensitive Information Into Debugging Code). Only Gemini 2.0 Flash introduced CWE-384 (Session Fixation). CWE-807 (Reliance on Untrusted Inputs in a Security Decision) was only introduced by DeepSeek R1 32B Distill.

Overall, all models introduced CWE-20, CWE-1333, and CWE-79. Other weaknesses were introduced by only a few of the models. These results could be attributed to differences in the models' underlying architecture and training data. For example, CWE-1333 was often introduced by LLMs in scenarios involving input sanitisation. The model would use regular expressions to validate and sanitise user input. However, in doing so, the model could inadvertently generate inefficient regular expressions that could be vulnerable to denial-of-service (DoS) attacks. This suggests that when prompted to implement secure code, the models attempt to implement security measures, drawing on code samples in their training data that include descriptions of those measures. However, those code samples could also include security weaknesses (such as CWE-1333), highlighting the importance of the model's training data for the security of its code outputs.

\subsubsection{Correlation Analysis of Original Versus Introduced CWEs} We computed Cramer's V correlation between the original CWEs and the CWEs introduced by the model output refinement techniques for each model (see Figure~\ref{fig:introduced_cwe_heatmap}). Cramer's V correlations quantify the strength of association between the CWE types in the original samples and those introduced after applying the refinement techniques; a correlation of 0 indicates no relationship, while a correlation closer to 1 indicates a strong association. Hence, higher values suggest that the presence of a particular original CWE is strongly correlated with the likelihood of introducing a specific CWE after applying the model output refinement technique for that model.

In this subsection, we discuss two cases of applying the model output refinement techniques and examine the relationship between the CWEs introduced by these techniques and the original CWEs detected in the code samples. The use of MP for DeepSeek R1 32B Distill and \cotrAbv{} for o4-mini are examined, providing two in-depth examples of how model output refinement techniques can lead to new CWEs being introduced.
\\

\begin{figure}
    \centering
    \includegraphics[width=0.5\linewidth]{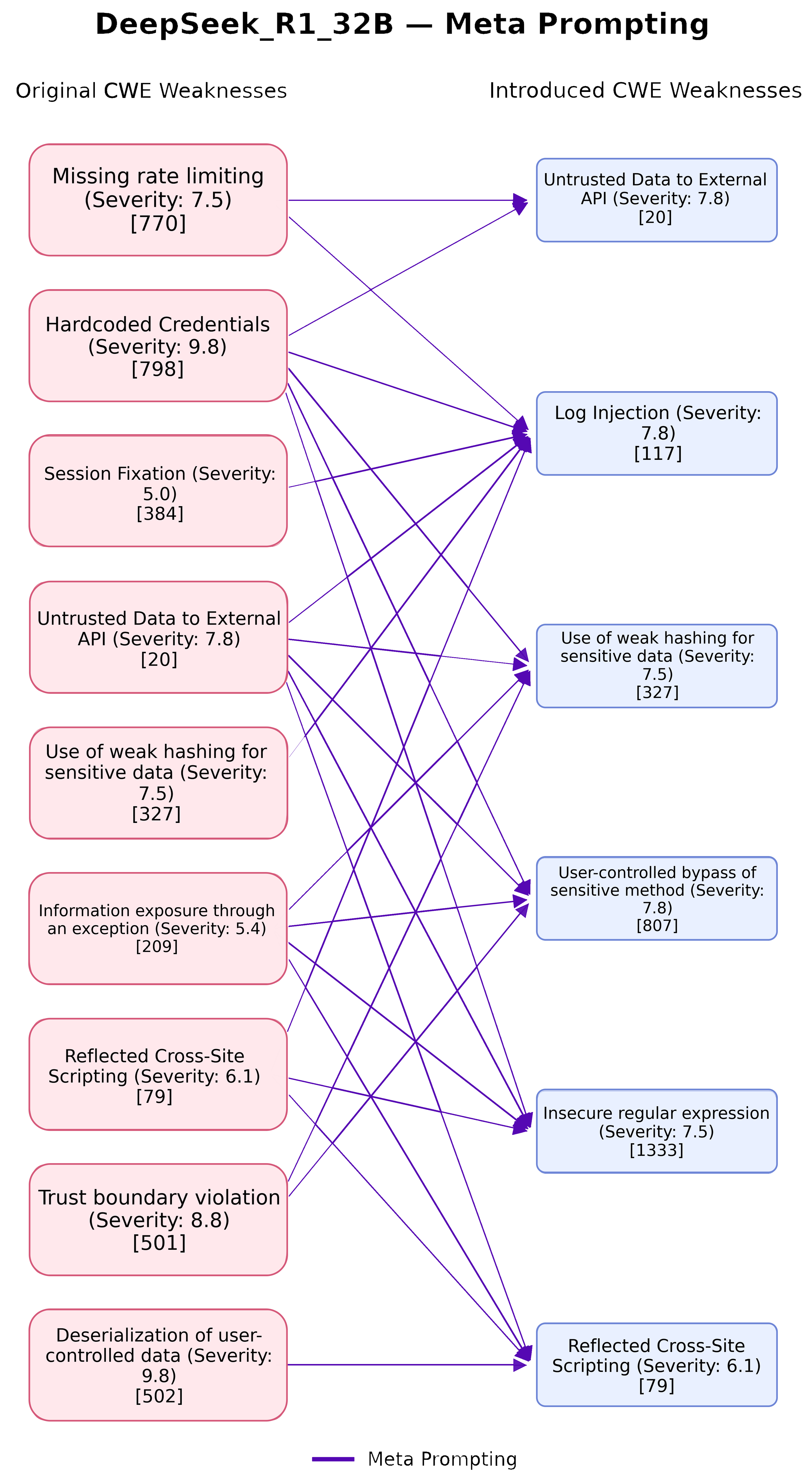}
    \caption{Original vs introduced CWEs when using Meta Prompting with DeepSeek R1 32B Distill.}
    \label{fig:deepseekr1mpintroducedcwes}
\end{figure}

\noindent \textbf{DeepSeek R1 32B Distill (Meta Prompting) - } An extreme case of when a refinement technique introduced varying CWEs depending on the CWE types in the original sample is in the case of MP with DeepSeek R1 32B Distill, shown in Figure \ref{fig:deepseekr1mpintroducedcwes}. In this case, CWE-117 (Log Injection) is most commonly introduced in comparison to other weaknesses. It has a low-to-moderate Cramer's V correlation with CWE-209 (0.26), CWE-327 (0.36), CWE-384 (0.26), CWE-501 (0.17), CWE-502 (0.12), and CWE-770 (0.17). The higher correlations with CWE‑209 and CWE‑327 align with the scenarios where these weaknesses typically occur, such as user authentication and session management. These scenarios often involve both log-in activities and the display of error messages. In such contexts, the model may introduce CWE‑117 since logging functionality is frequently added to improve code readability, maintainability, and overall quality. The model output refinement technique may inadvertently generate log statements without appropriate sanitisation or encoding, thereby increasing the risk of log injection vulnerabilities. The introduction of CWE‑807 when the original samples contain CWE‑798, CWE‑20, CWE‑209, and CWE‑501 is also consistent, given the types of scenarios in which these weaknesses typically appear. These original weaknesses include insufficient input validation, improper handling of sensitive operations, and violations of trust boundaries. All of the aforementioned weaknesses can create conditions where the model attempts to implement additional access checks which may cause it to inadvertently introduce user‑controlled parameters into decision points that lead to sensitive methods, thereby enabling bypass conditions. CWE-79 is often introduced when the original samples contain weaknesses pertaining to data-handling. CWE-327 and CWE-20 tend to be introduced when the original samples contain CWEs related to data-handling and trust, and information exposure. CWE-1333 is often introduced in scenarios where the original CWEs could be resolved through input sanitisation or validation, such as CWE-79 and CWE-20. \\



\begin{figure}
    \centering
    \includegraphics[width=0.5\linewidth]{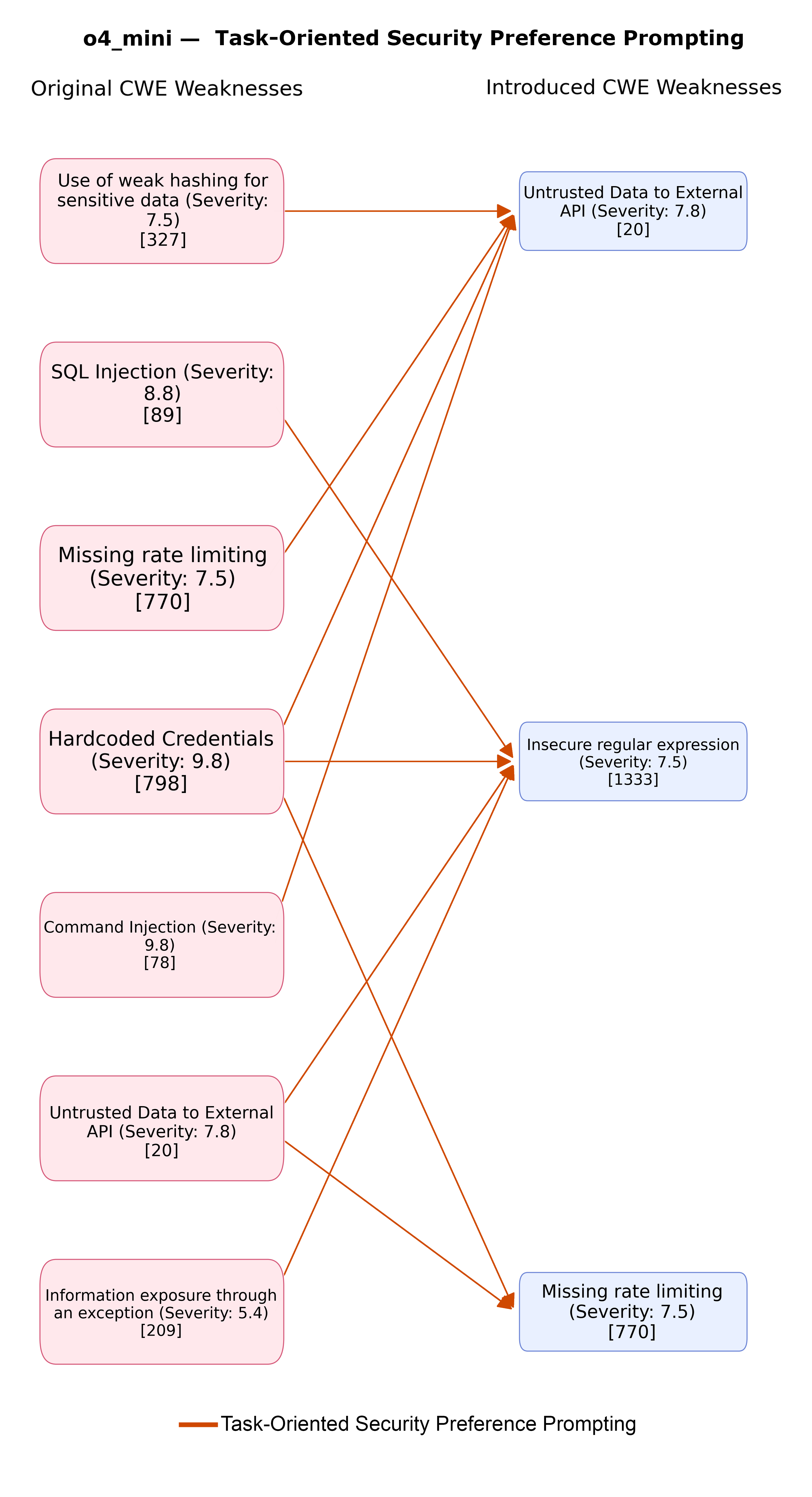}
    \caption{\revision{Original vs introduced CWEs when using \cotr{} Prompting with o4-mini.}}
    \label{fig:o4_mini_cot_introduced_cwes}
\end{figure}

\noindent \textbf{o4-mini (\cotr{} Prompting) - } A case in which few CWEs were introduced in comparison to the original weaknesses is \cotrAbv{} with o4-mini, shown in Figure \ref{fig:o4_mini_cot_introduced_cwes}. Although there are fewer CWEs in both the original and refined code samples, the general patterns of introduced CWEs are similar to those observed with MP using DeepSeek. CWE-1333 is once again introduced in scenarios where the original samples contained CWEs that could be resolved through input sanitisation or validation, including CWE-79 and CWE-20. This indicates that the model attempts to use regular expressions to validate or sanitise the user-provided data, but in doing so, inadvertently introduces CWE-1333 by writing an inefficient regular expression. CWE-1333 is also linked to CWE-798, as it often co-occurs with user-input-related CWEs such as CWE-20 and CWE-79. CWE-20 was introduced when the original samples contained CWE-798 and CWE-770. This could be attributed to the model rearranging the code structure to prevent certain CWEs, thereby inadvertently introducing them.

\begin{tcolorbox}[colback=black!2!white,colframe=white!20!black,title=Key Findings for RQ3, left=2mm,right=2mm,coltitle=white]
Model refinement techniques sometimes introduced new CWEs while attempting to mitigate existing ones. Fine-Tuning achieved the highest rate of fully removing original CWEs and the lowest rate of introducing new ones, while \cotrAbv{} introduced the most new CWEs. 
Overall, refinements often resulted in partial mitigation rather than complete removal of original CWEs. While refinement techniques improved code security, they also showed that new weaknesses can be introduced, emphasising the influence of model training data on code security.
\end{tcolorbox}
\section{Discussion}
\label{sec:discussion}
\subsection{Trustworthiness of AI-Generated Code}
\noindent\emph{\textbf{All Models Produced Vulnerable Code.}}
The findings of this study raise important concerns regarding the trustworthiness of the \aigc{}. Across all models and programming languages tested, \textit{none of the LLMs produced code without security weaknesses, though the degree varied}. Although the number or severity of detected CWEs was relatively low in some cases, CWEs were still present, and often co-occurred within the same code sample. This suggests that current \cgm{} tend to reproduce common insecure coding patterns present in their training data.

\noindent\emph{\textbf{Weaknesses Remain Despite Advancement of LLMs.}}
These findings are consistent with earlier studies on the security of AI-generated code. For example, \citet{pearce2022asleep} reported that approximately 40\% of code samples generated by GPT-3 Codex (including Python and C) contained at least one security weakness. Similarly, \citet{Schaad2025} found that 65\% of AI-generated Python and JavaScript code samples were vulnerable. The models tested in our study generally contained similar numbers of security weaknesses, and in some cases more, than those reported in earlier work on the languages tested in our study. This suggests that security weaknesses in AI-generated code persist even in modern LLMs. Although progress has been made in the power and complexity of LLMs, insecure code generation remains a systemic issue rather than one limited to earlier models.

\noindent\emph{\textbf{Language-Dependent Occurrence of Weaknesses.}}
Our findings show that models are more susceptible to generating security weaknesses in certain programming languages. Developers utilising \cgm{} should be even more cautious (i.e., should always thoroughly check for security weaknesses and vulnerabilities) when generating JavaScript and Java code, as the generated code samples contained more weaknesses than those generated in Python and Go.

\noindent\emph{\textbf{AI-Generated Code should not be Assumed to be Secure.}}
The frequent co-occurrence of multiple CWEs (see Listings~\ref{lst:insecure_code_example_1_python}, \ref{lst:insecure_code_example_1_js}, and \ref{lst:insecure_code_example_1_java} in Section \ref{sec:results_og_weaknesses}) within individual code samples can significantly increase the number of attack vectors of a code base. This indicates that \textit{\aigc{} \textit{should not} be assumed to be inherently safe by developers}, even when it appears functionally correct. Instead, it should be considered untrusted until it undergoes a thorough security analysis. This can be done through a careful security analysis (static, dynamic, or hybrid).

\noindent\emph{\textbf{Improving LLM Security through Training Data}}
Creators of LLMs for code generation could consider training future \cgm{} on secure code datasets to reduce the likelihood that these models introduce security weaknesses. This could include methods such as filtering training datasets with static analysis tools to exclude insecure code. In addition, the LLMs can be tested on datasets that serve as benchmarks for secure code generation, such as CodeLMSec \cite{Hajipour2024} and SafeGenBench \cite{li2025safegenbench}.

\subsection{Secure Software Development Practices}
\noindent\emph{\textbf{Security Risks Vary Across Languages and Models.}}
Our results have direct implications for how \aigc{} should be integrated into secure software development workflows. Relying on \aigc{} without validating its security introduces significant risks. Additionally, the patterns observed across programming languages and models imply that developers should adopt language- and model-aware security practices when using AI \cgm{}. For example, AI-generated JavaScript and Java code tend to contain more input validation weaknesses (such as CWE-20), while AI-generated Python code tends to contain more information exposure and path traversal weaknesses (e.g., CWE-209 and CWE-22). Hence, it is important that developers take into account the common types of weaknesses that can arise in each language.

\noindent\emph{\textbf{Improving Code Security through Prompting and Fine-tuning.}}
Our results show that \textit{prompting and fine-tuning-based model output refinement techniques can significantly enhance the security of the \aigc{}} in many cases. Prompting-based strategies that encourage the code generation model to reason, such as \cotrAbv{} and MP strategies, can be cost-effective, while fine-tuning with methods like LoRA may be a better option for developers with a higher budget, as it can be costly. 

\subsection{Prompting vs Fine-Tuning for Model Output Refinement}
\subsubsection{Considerations for Prompting and Fine-Tuning}
This study compared the effectiveness of prompting-based techniques and model fine-tuning for improving the security of \aigc{}. Although all evaluated refinement techniques led to some improvement, \textit{fine-tuning consistently outperformed prompting-based refinement techniques across models and programming languages}.
\textbf{Prompting techniques, including \cotrAbv{} and MP, yielded moderate improvements}, though the magnitude of these improvements varied across languages and models. \textbf{In contrast, fine-tuning appears to enable models to learn secure coding patterns more consistently}, leading to more consistent reductions in security weaknesses in the \aigc{}. However, an important consideration is that \textbf{the costs of fine-tuning are high}. The computational resources and time required for collating datasets of secure code can contribute to the higher cost of fine-tuning. The computational resources required to fine-tune LLMs scale exponentially with model size (number of parameters), often requiring substantial memory and GPU resources \cite{2023wordstowatts}. Parameter-efficient fine-tuning methods can reduce the computational requirements for fine-tuning LLMs \cite{wang2025parameter}. However, an important consideration is that extremely efficient methods such as Quantized Low-Rank Adaptation (QLoRA) perform less accurately than full model fine-tuning or LoRA \cite{biderman2024lora}. In the context of secure code generation, this trade-off is relevant, as reduced fine-tuning accuracy may limit LLMs' ability to learn secure coding practices effectively. For developers using proprietary LLMs or commercial APIs, a key consideration is the cost of fine-tuning, which varies by model scale and type and is often charged per token or per hour.

\subsubsection{Weaknesses Introduced by Output Refinement Techniques.}
An important observation from our results is that improvements achieved through model output refinement techniques were not always reflected in fewer detected CWEs. In some cases, model output refinement techniques introduced higher-severity weaknesses without fixing the original ones. In other cases, model output refinement led to replacing high-severity weaknesses with lower-severity ones. \textbf{Most newly introduced weaknesses occurred when using prompting-based refinement techniques}. New weaknesses were also introduced after fine-tuning the LLMs, though fewer overall than with prompting-based techniques. This demonstrates that \textit{fine-tuning introduces the least overall security risks} compared to prompting-based refinement techniques.

\subsubsection{Effectiveness of Model Output Refinement Techniques.}
\textit{Prompt-based refinement techniques can be a cost-effective option} for developers seeking to reduce security weaknesses introduced by LLMs. However, it is important to note that \textbf{the effectiveness of these techniques can vary depending on the model and programming language used}. MP, \cotrAbv{}, and NEP differ in both effectiveness and reliability for reducing security weaknesses. \textit{MP was the strongest prompting-based refinement technique} of the techniques tested, often achieving improvements comparable to those of fine-tuning. Its effectiveness appears to stem from the structured, detailed guidance the LLM generates, which directs the model toward secure coding practices. However, MP occasionally introduced new CWEs and showed limited effectiveness against complex weaknesses, such as improper input validation and resource management (CWE-20 and CWE-770), suggesting that its benefits are context-dependent. \cotrAbv{} prompting shows moderate and inconsistent improvements. It was effective against weaknesses with simple, clear fixes, such as SQL injection and unsafe deserialization, but struggled with those requiring more complex reasoning. \cotrAbv{} also introduced new CWEs more frequently than MP, suggesting that explicit reasoning instructions can sometimes expose incomplete or incorrect assumptions during model training.\textbf{ NEP is the least effective prompting-based technique.} Its reliance on insecure examples appears to limit its ability to generate secure code and, in some cases, reinforces insecure patterns rather than mitigating them. This limitation is most apparent for high-severity weaknesses such as CWE-798 (Hardcoded Credentials) and CWE-20 (Improper Input Validation).

Compared with prior research, this study shows a clear progression in mitigation effectiveness. For example, an exploratory study on fine-tuning LLMs with vulnerability-fixing data reported moderate improvements of approximately 6.4\% in C and 5.4\% in C++ secure code generation after fine-tuning on a curated dataset of fixes \cite{li2024exploratory}. Another study utilised RAG to improve the security of Copilot-generated code by at least 14\% \cite{zhao2025towards}. \citet{yan2025} reported 17\% and 28\% improvements in code security with Proactive Vulnerability Prevention and Post-Hoc Vulnerability Repair prompting methods, respectively. \citet{Fu2025} found that providing LLMs with error messages from static analysis tools fixed up to 55.5\% of security issues generated by LLMs in Python. In contrast, the refinement techniques in our study yield similar or greater average improvements across programming languages, with Fine-Tuning often reducing CWE presence by around 40\% in Java, 80\% in Python, 54\% in JavaScript, and 88\% in Go. Prompting-based strategies such as MP (average improvement of 46\% across languages) and \cotrAbv{} (average improvement of 36\% across languages) also outperform the prompting approaches from previous studies \cite{li2024exploratory, zhao2025towards, yan2025}, although they do so less consistently than fine-tuning, as they have varying performance depending on the scenario in which they are applied.

\subsubsection{Functionality of the Refined Code.}

\revision{The functional correctness evaluation (results presented in Section~\ref{sec:functionalcorrectnessresults}) provides additional insight into the effectiveness of the refinement techniques by examining whether the refined code continues to satisfy the behaviours exercised by the unit tests. \textbf{Overall, the results demonstrate that the functional impact of the refinement techniques varies} across programming languages, models, and strategies. \textbf{While some configurations maintained or improved test pass rates relative to the baseline, others exhibited decreases.} These variations highlight the importance of considering functional correctness alongside security improvements when evaluating refinement techniques. For example, Java showed relatively stable test pass rates across several configurations, while Go and JavaScript showed greater variation in pass rates. No single refinement technique consistently achieved the best functional outcomes across all evaluated settings.}

\revision{These findings highlight the complementary nature of security and functional correctness as evaluation dimensions. In this context, refinement strategies that improve security while maintaining high functional test pass rates are particularly promising for practical adoption (such as Fine-Tuning and MP).}
\revision{However, the results should be interpreted in light of the evaluation methodology. The unit tests were generated using an LLM to provide a consistent, scalable method for assessing functional behaviour across the evaluated code samples. Although this approach enables systematic comparisons, the generated tests may not capture every aspect of the original functional requirements or all relevant execution paths. Test pass rates represent functional correctness under the evaluated scenarios rather than complete behavioural equivalence. Furthermore, observed differences in pass rates should be interpreted as differences in test outcomes, which may reflect limitations in test coverage or execution.}

\revision{Overall, considering security outcomes and functional test pass rates together supports a more comprehensive assessment of refinement effectiveness and provides a foundation for further investigating how security improvements can be achieved while preserving intended program behaviour.}

\subsubsection{Generalisation of Fine-Tuning.}


\revision{To investigate whether the security improvements obtained through Fine-Tuning extend beyond the scenarios represented in the fine-tuning data, we conducted an additional evaluation using new scenarios (partially completed code snippets). The fine-tuning data and the original evaluation scenarios cover the same CWE types; consequently, performance on the original scenarios alone cannot establish whether the models have learned security-relevant patterns that transfer to variations of the evaluated code. We therefore evaluated the fine-tuned models on three new scenarios (Scenarios 11, 12, and 17) which target the same CWEs as their corresponding original scenarios (Scenarios 1, 2, and 7, respectively). These additional experiments show how well the observed improvements persist when evaluated on unseen code.}

\revision{We compared the original and new scenarios using two outcome measures: severity improvement and improvement in the number of CWEs. For each measure, we calculated the generalisation gap as the improvement on the original scenarios minus the improvement on the corresponding modified scenarios. A positive gap indicates a reduction in improvement on the modified scenarios, whereas a negative gap indicates greater improvement on the modified scenarios. A zero gap indicates equal improvement from fine-tuning for the original and new scenarios. The gap is descriptive and should not be interpreted as a statistical estimate of generalisation error.}

\revision{The results (which are contained within our replication package~\cite{dataset}) indicate that \textbf{fine-tuning benefits frequently persist on the new scenarios, although the extent of generalisation varies across programming languages, models, and scenarios}. For example, GPT-4.1 achieves greater severity improvement on the modified Python scenarios than on the original scenarios (83\% versus 74\%). Conversely, DeepSeek R1 32B achieves lower severity improvement on the modified Java scenarios (61\% versus 79\%), while its improvement in the number of CWEs is slightly higher on the modified scenarios (58\% versus 53\%). These contrasting results indicate that the transfer of fine-tuning benefits is not uniform across security outcomes.}

\revision{The results for the number of CWEs are also encouraging. Across several models and languages, the refinement techniques yield equal or greater improvement on the modified scenarios than on the original scenarios. For instance, o4-mini's improvement in the number of CWEs for JavaScript increases from 22\% on the original scenarios to 57\% on the modified scenarios. Nevertheless, other comparisons show smaller improvements or mixed outcomes, demonstrating that strong results on the original scenarios may not guarantee equivalent results on every new scenario.}

\revision{Overall, these experiments provide evidence that Fine-Tuning with a secure-code dataset can extend to new code scenarios involving the CWE types considered in this study. However, the evaluation covers only three new scenarios across three programming language. The results should be interpreted as evidence of transfer across new scenarios, rather than as evidence of unrestricted generalisation to arbitrary projects or real-world development settings.}

\revision{\subsection{Recommendations and Future Work}}

Overall, the refinement techniques used in this study can be combined with program analysis techniques for a more secure LLM-based software development workflow. Developers can integrate security analysis tools into continuous integration pipelines to assess the security of AI-generated code before deployment (e.g., in DevSecOps software development pipelines \cite{ibm_devsecops}), and refinement techniques can reduce the number of security issues generated by LLMs. This study found that, overall, MP provides the greatest improvements compared to the other prompting techniques (\cotrAbv{} and NEP), making it an optimal choice for developers to reduce security weaknesses in AI models with prompting alone.

For creators of code-generation tools, post-training techniques such as fine-tuning can improve the security of model-generated code. However, as mentioned previously, fine-tuning can become costly. As a more cost-effective strategy, system prompts based on \cotrAbv{} and MP strategies can be used in code-generation tools that utilise LLMs.

\revision{An important direction for future work is extending secure AI coding into an agentic code-repair workflow. In the current setup, the models are evaluated on their ability to generate secure code directly through fine-tuning and prompting. In a more realistic development scenario, however, a static analyser could first identify security weaknesses in AI-generated code and provide these findings as feedback to the LLM. The LLM could then use the analyser's output to revise the code and iteratively address the reported weaknesses. Such a workflow would shift the task from single-prompt secure code generation toward an iterative process in which weakness detection and code repair are integrated. Future research could also investigate how effectively different fine-tuned and prompted models respond to static-analysis feedback, whether iterative repair improves the rates at which weaknesses are removed, and whether the resulting code introduces new weaknesses. This agentic setting would also provide a more realistic evaluation of LLM-based secure code repair, as it reflects a practical development workflow in which automated security analysis guides subsequent code modifications. A few other works have also explored this approach, such as those of \citet{blyth2025static} and \citet{ravi2025llmloop}. Blyth et al. investigated using static analysis as a feedback mechanism to improve LLM-generated code beyond functional correctness, showing how static-analysis results can guide the identification and resolution of code-quality issues and reporting some improvements (\textasciitilde10\% improvement in pass rates). Ravi et al. introduced LLMLOOP, which employs automated iterative feedback loops to improve both LLM-generated code and tests.}

\revision{Another direction for future work could be the evaluation and comparison of other fine-tuning techniques, including preference learning, policy optimisation, etc.}
\section{Threats to Validity}\label{sec:validity}

\subsection{External Validity}
A threat to external validity arises from the choice and design of the CWE scenarios. Although the ten scenarios were selected to cover a diverse and representative subset of high-impact vulnerabilities from the MITRE CWE Top 25 \cite{mitre_cwe_top25}, they do not represent the full range of software security weaknesses that occur in the real world. In addition, the CWE scenarios are based on partial code snippets rather than complete applications, which may limit ecological validity. This design choice was intentional to ensure the experiment was controlled and ensure consistent evaluation across models, languages, and mitigation techniques. All scenarios were drawn from a prior study by \citet{Majdinasab2024}, which evaluated the security of LLMs with Python-based scenarios. These Python-based scenarios were adapted to the other languages using equivalent frameworks. 

\revision{Our evaluation uses ten scenarios, with one scenario representing each target CWE. This design provides coverage across a diverse set of security weaknesses and enables a controlled comparison of the refinement techniques across multiple programming languages and LLMs. A scenario-based evaluation with a limited number of scenarios per CWE is also consistent with prior work on secure code generation. For example, \citet{hevechev2023} evaluate security-hardening techniques using manually constructed scenarios targeting individual CWEs, with one scenario per CWE. Another study by \citet{pearce2023} used 12 scenarios based on historically occurring CVEs. At the same time, different scenarios within the same CWE may exhibit variations in code structure, weakness occurrences, and difficulty for code-generation and refinement models. Evaluating multiple scenarios per CWE could therefore provide additional evidence regarding the variability of secure code generation within each CWE. However, substantially increasing the number of scenarios would also increase model executions and computational costs.} 

\revision{Another potential threat to validity is running model inference exclusively through GitHub Copilot. GitHub Copilot does not disclose decoding parameters (top-p, maximum output length, random seeds, etc.) used during inference with the LLMs, hence lowering reproducibility.}

\subsection{Internal Validity}

A potential threat to internal validity is related to the translation of scenarios across programming languages. Although the Python snippets came from existing datasets, the other languages (JavaScript, Java, and Go) were manually reconstructed. This introduces the risk of unintentional inconsistencies between code snippets in different languages, which could affect the distribution of CWEs generated by the LLMs before and after applying model output refinement techniques. To mitigate this, we used equivalent backend frameworks and similar programming patterns across languages. The code snippets for each language shared the same overall structure, with the same sections intentionally left blank for the LLMs to complete.

Model nondeterminism may also introduce a threat to internal validity. LLMs can generate varying outputs for the same prompts, which could affect weakness counts. We mitigated this threat by generating multiple samples per scenario and aggregating results across scenarios. This sampling strategy reduces the influence of outliers.

\revision{Another potential internal validity threat concerns the partitioning of the fine-tuning and evaluation data. The fine-tuning data covers the same CWE scenarios as the original evaluation, so the original experimental setup does not fully isolate the effects of learning security-fix patterns from potential familiarity with the evaluated scenarios. Re-running the fine-tuning experiments using a split based on independent projects, code templates, or task instances was infeasible within the scope of this paper. To mitigate this, we tested the fine-tuned models on a small set of unseen scenarios to assess and confirm their generalisability. Furthermore, the pre-training data and procedures of proprietary models (e.g., GPT-4.1) are not fully disclosed. We therefore cannot determine whether these models encountered similar code or related security fixes during pre-training. Such exposure could influence their performance independently of our fine-tuning. Although we conducted additional experiments using modified versions of the original scenarios to examine generalisation, these experiments cannot rule out potential contamination or establish that the models had not previously encountered similar examples. 
These limitations prevent us from conclusively determining whether the observed improvements arise from knowledge acquired through our fine-tuning procedure alone or are partly attributable to pre-existing model capabilities. 
}

\subsection{Construct Validity}

This study relies heavily on static analysis using CodeQL. This may introduce a tool-related threat to construct validity. Static analysis tools may produce false positives or miss weaknesses when code does not match predetermined patterns associated with weakness profiles. To mitigate this risk, we wrote a custom analysis script to identify weaknesses not reported by the default CodeQL queries. In addition, we manually inspected all generated code snippets to verify the detected weaknesses, ensuring no false positives and confirming that no relevant weaknesses were missed. We also cross-validated the detected weaknesses with another static analysis tool (Semgrep) and through manual validation by independent reviewers. We chose CodeQL for its support for a wide range of programming languages, its reproducibility, and its widespread adoption in both industry and academic research. While static analysis cannot guaranty the detection of all possible weaknesses, combining automated static analysis, custom scripts, and manual inspection reduces the risk that the analysis tool's limitations affect the reported results.

The fine-tuning datasets introduce another possible threat. Secure code samples were manually created and verified through static analysis, which may not fully reflect all secure coding practices or the real-world diversity of coding methods. However, we included multiple secure completions per scenario and used CodeQL to verify sample security, ensuring consistency with the evaluation criteria used throughout this study.



\section{Conclusions}
\label{sec:conclusion}
This paper conducted an in-depth review of the security risks associated with AI-generated code and evaluated the effectiveness of multiple mitigation strategies across models and programming languages. Using ten CWE-based scenarios derived from the MITRE CWE Top 25, we generated and analysed 9,600 code samples produced by five LLMs across Python, JavaScript, Java, and Go. Static analysis using CodeQL, along with custom detection scripts and manual validation, showed that none of the evaluated models produced fully secure code, and that vulnerabilities frequently co-occurred within the same outputs. Although LLMs can generate syntactically correct and functionally useful code, our findings show that security weaknesses are prevalent across models, languages, and scenarios. 

We evaluated four model output refinement techniques: Negative Example Prompting, \cotr{} Prompting, Meta Prompting, and Fine-Tuning using Low-Rank Adaptation. Across all models and languages, fine-tuning showed the best improvements, reducing detected security weaknesses by 82\% on average. Prompting-based techniques showed moderate improvements, but their effectiveness varied by model and programming language. 
In some cases, refinement techniques introduced new weaknesses, mostly related to improper input validation, highlighting inherent limitations of these approaches. 
However, the introduced weaknesses were generally less severe than those found in the original outputs. Our results also showed that the programming language strongly affects both the number of generated security weaknesses and the effectiveness of refinement techniques, with Python and Go outputs containing fewer weaknesses overall and being easier to improve than those from JavaScript and Java.

This work highlights that AI-generated code must not be treated as inherently safe. Developers should integrate security analysis into software development workflows involving LLMs, and LLM creators should prioritise security-aligned training and fine-tuning methods. While fine-tuning was found to be the most reliable approach for consistently reducing security weaknesses, prompting strategies can be cost-effective alternatives. Future work can expand the set of evaluated CWEs, investigate combinations of prompting and fine-tuning approaches, and explore additional refinement strategies. Overall, these findings reinforce the need to treat secure code generation as a core requirement in the design and use of LLM-based development tools.

\section*{Data Availability}
The replication package, including all code samples and detailed results, is available at our online repository ~\cite{dataset}.


\bibliographystyle{ACM-Reference-Format} 
\bibliography{References} 

@inproceedings{pearce2022asleep,
  author = {Hammond Pearce and Baleegh Ahmad and Benjamin Tan and Brendan Dolan-Gavitt and Ramesh Karri},
  title = {Asleep at the Keyboard? Assessing the Security of GitHub Copilot's Code Contributions},
  booktitle = {Proceedings of the 43rd IEEE Symposium on Security and Privacy (SP)},
  year = {2022},
  pages = {754--768},
  publisher = {IEEE}
}

@article{zheng2025generalperformancedomain,
  title={Top general performance= top domain performance? domaincodebench: A multi-domain code generation benchmark},
  author={Zheng, Dewu and Wang, Yanlin and Shi, Ensheng and Liu, Xilin and Ma, Yuchi and Zhang, Hongyu and Zheng, Zibin},
  journal={arXiv preprint arXiv:2412.18573},
  year={2024}
}

@article{vaswani2023attentionneed,
  title={Attention is all you need},
  author={Vaswani, Ashish and Shazeer, Noam and Parmar, Niki and Uszkoreit, Jakob and Jones, Llion and Gomez, Aidan N and Kaiser, {\L}ukasz and Polosukhin, Illia},
  journal={Advances in Neural Information Processing Systems},
  volume={30},
  pages={1--11},
  year={2017}
}

@article{ren2020codebleumethodautomaticevaluation,
  title={Codebleu: a method for automatic evaluation of code synthesis},
  author={Ren, Shuo and Guo, Daya and Lu, Shuai and Zhou, Long and Liu, Shujie and Tang, Duyu and Sundaresan, Neel and Zhou, Ming and Blanco, Ambrosio and Ma, Shuai},
  journal={arXiv preprint arXiv:2009.10297},
  year={2020}
}

@misc{github_copilot,
  author= {{GitHub}},
  title = {AI that builds with you},
  year   = {2025},
  howpublished = {\url{https://github.com/features/copilot}},
  note    = {Accessed: 2025-08-01}
}

@misc{cursor_ai_coder,
  author= {{Anysphere Inc.}},
  title = {The AI Code Editor},
  year   = {2025},
  howpublished = {\url{https://cursor.com/en}},
  note    = {Accessed: 2025-08-01}
}

@misc{tabnine_ai,
  author= {{Tabnine}},
  title = {Tabnine},
  year   = {2025},
  howpublished = {\url{https://www.tabnine.com/}},
  note    = {Accessed: 2025-08-01}
}

@misc{anthropic_claude_code,
  author = {Anthropic},
  title = {Claude Code},
  year = {2026},
  url = {https://claude.com/product/claude-code},
  note = {Accessed: 2026-03-01}
}

@misc{git_survey,
  author= {{GitHub}},
  title = {Survey reveals AI’s impact on the developer experience},
  year   = {2025},
  howpublished = {\url{https://github.blog/news-insights/research/survey-reveals-ais-impact-on-the-developer-experience/}},
  note    = {Accessed: 2025-08-01}
}

@inproceedings{Majdinasab2024,
  title={Assessing the Security of GitHub Copilot Generated Code -- A Targeted Replication Study},
  author={Majdinasab, Vahid and Bishop, Michael Joshua and Rasheed, Shawn and Moradidakhel, Arghavan and Tahir, Amjed and Khomh, Foutse},
  booktitle={Proceedings of the 31st IEEE International Conference on Software Analysis, Evolution and Reengineering (SANER)},
  pages={435--444},
  year={2024},
  organization={IEEE}
}

@article{Fu2025,
  title={Security weaknesses of copilot-generated code in github projects: An empirical study},
  author={Fu, Yujia and Liang, Peng and Tahir, Amjed and Li, Zengyang and Shahin, Mojtaba and Yu, Jiaxin and Chen, Jinfu},
  journal={ACM Transactions on Software Engineering and Methodology},
  volume={34},
  number={8},
  pages={1--34},
  year={2025},
  publisher={ACM}
}

@article{Tihanyi2024,
  author    = {Norbert Tihanyi and Tamas Bisztray and Mohamed Amine Ferrag and Ridhi Jain and Lucas C. Cordeiro},
  title     = {How secure is AI-generated code: a large-scale comparison of large language models},
  journal   = {Empirical Software Engineering},
  year      = {2024},
  volume    = {30},
  number    = {47},
  pages     = {1--47},
  publisher = {Springer}
}

@article{kharma2025security,
  title={Security and Quality in LLM-Generated Code: A Multi-Language, Multi-Model Analysis},
  author={Kharma, Mohammed and Choi, Soohyeon and AlKhanafseh, Mohammed and Mohaisen, David},
  journal={arXiv preprint arXiv:2502.01853},
  year={2025}
}

@article{Kaur2020,
  title = {A Comparative Study of Static Code Analysis tools for Vulnerability Detection in C/C++ and JAVA Source Code},
  journal = {Procedia Computer Science},
  volume = {171},
  pages = {2023--2029},
  year = {2020},
  author = {Arvinder Kaur and Ruchikaa Nayyar}
}

@article{correa2020,
  author = {Correa, Roddy and Higuera, Juan-Ramón and Bermejo, Javier and Montalvo, Juan Antonio and Sanchez, Manuel and Magreñán, Ángel},
  year = {2020},
  pages = {89-124},
  title = {Hybrid Security Assessment Methodology for Web Applications},
  volume = {126},
  journal = {Computer Modeling in Engineering \& Sciences}
}

@misc{mitre_cwe79,
  title  = {{CWE-79}: Improper Neutralization of Input During Web Page Generation (Cross‑site Scripting)},
  author = {{MITRE Corporation}},
  year = {2025},
  howpublished = {\url{https://cwe.mitre.org/data/definitions/79.html}},
  note= {Accessed: 2025-08-01}
}

@misc{mitre_cwe89,
  title  = {{CWE-89}: Improper Neutralization of Special Elements used in an SQL Command ('SQL Injection')},
  author = {{MITRE Corporation}},
  year = {2025},
  howpublished = {\url{https://cwe.mitre.org/data/definitions/89.html}},
  note= {Accessed: 2025-08-01}
}

@misc{cwe_info,
  title  = {New to CWE},
  author = {{MITRE Corporation}},
  year = {2025},
  howpublished = {\url{https://cwe.mitre.org/about/new_to_cwe.html}},
  note= {Accessed: 2025-08-01}
}

@misc{cwe_faq,
  title  = {CWE - Frequently asked questions (FAQ)},
  author = {{MITRE Corporation}},
  year = {2026},
  howpublished = {\url{https://cwe.mitre.org/about/faq}},
  note= {Accessed: 2026-11-13}
}

@inproceedings{Tony2023,
  author = {Tony, Catherine and Mutas, Markus and Díaz Ferreyra, Nicolás and Scandariato, Riccardo},
  year = {2023},
  pages = {588--592},
  title = {LLMSecEval: A Dataset of Natural Language Prompts for Security Evaluations},
  booktitle = {Proceedings of the 20th International Conference on Mining Software Repositories (MSR)},
  organization={IEEE}
}

@inproceedings{Hajipour2024,
  title={Codelmsec benchmark: Systematically evaluating and finding security vulnerabilities in black-box code language models},
  author={Hajipour, Hossein and Hassler, Keno and Holz, Thorsten and Sch{\"o}nherr, Lea and Fritz, Mario},
  booktitle={Proceedings of the 2nd IEEE Conference on Secure and Trustworthy Machine Learning (SaTML)},
  pages={684--709},
  year={2024},
  organization={IEEE}
}

@inproceedings{hu2022lora,
    title={LoRA: Low-Rank Adaptation of Large Language Models},
    author={Edward J. Hu and Yelong Shen and Phillip Wallis and Zeyuan Allen{-}Zhu and Yuanzhi Li and Shean Wang and Lu Wang and Weizhu Chen},
    booktitle={Proceedings of the 10th International Conference on Learning Representations (ICLR)},
    year={2022},
    pages={1--13},
    publisher={OpenReview.net}
}

@article{xu2024prosec,
  title={ProSec: Fortifying Code LLMs with Proactive Security Alignment},
  author={Xu, Xiangzhe and Su, Zian and Guo, Jinyao and Zhang, Kaiyuan and Wang, Zhenting and Zhang, Xiangyu},
  journal={arXiv preprint arXiv:2411.12882},
  year={2024}
}

@inproceedings{he2024,
  author = {He, Jingxuan and Vero, Mark and Krasnopolska, Gabriela and Vechev, Martin},
  title = {Instruction tuning for secure code generation},
  year = {2024},
  organization = {IEEE},
  booktitle = {Proceedings of the 41st International Conference on Machine Learning (ICML)},
  pages = {18043--18062}
}

@article{yan2025,
      title={Guiding AI to Fix Its Own Flaws: An Empirical Study on LLM-Driven Secure Code Generation}, 
      author={Hao Yan and Swapneel Suhas Vaidya and Xiaokuan Zhang and Ziyu Yao},
      year={2025},
      journal={arXiv preprint arXiv:2506.23034}
}

@article{wei2022chain,
  title={Chain-of-thought prompting elicits reasoning in large language models},
  author={Wei, Jason and Wang, Xuezhi and Schuurmans, Dale and Bosma, Maarten and Xia, Fei and Chi, Ed and Le, Quoc V and Zhou, Denny and others},
  journal={Advances in Neural Information Processing Systems},
  volume={35},
  pages={24824--24837},
  year={2022}
}

@article{anisuzzaman2025,
title = {Fine-Tuning Large Language Models for Specialized Use Cases},
journal = {Mayo Clinic Proceedings: Digital Health},
volume = {3},
number = {1},
pages = {100184},
year = {2025},
author = {D.M. Anisuzzaman and Jeffrey G. Malins and Paul A. Friedman and Zachi I. Attia},
}

@misc{Gadesha2025ChainOfThought,
  title = {What is chain of thought (CoT) prompting?},
  author = {Vrunda Gadesha and Eda Kavlakoglu and Vanna Winland},
  howpublished = {IBM Think website},
  year = {2025},
  url = {https://www.ibm.com/think/topics/chain-of-thoughts},
}

@misc{OpenAI2025GPT41,
  title = {Introducing GPT‑4.1 in the API},
  author = {{OpenAI}},
  howpublished = {OpenAI website},
  year = {2025},
  note = {Accessed on August 16, 2025},
  url  = {https://openai.com/index/gpt-4-1/},
}

@misc{OpenAI2025GPT5,
  title = {Introducing GPT-5},
  author = {{OpenAI}},
  howpublished = {OpenAI website},
  year = {2025},
  note = {Accessed on November 25, 2025},
  url  = {https://openai.com/index/introducing-gpt-5/},
}

@misc{OpenAI2025o3o4mini,
  title = {Introducing OpenAI o3 and o4‑mini},
  author = {{OpenAI}},
  howpublished = {OpenAI website},
  month = apr,
  year = {2025},
  note = {Accessed on August 16, 2025},
  url = {https://openai.com/index/introducing-o3-and-o4-mini/},
}

@misc{Google2025Gemini2,
  title = {Gemini 2.0: Flash, Flash‑Lite and Pro},
  author = {{Google Developers}},
  howpublished = {Google Developers Blog},
  month = feb,
  day = 5,
  year = {2025},
  note = {Accessed on August 16, 2025},
  url = {https://developers.googleblog.com/en/gemini-2-family-expands/},
}

@misc{deepseekai2025,
      title={DeepSeek-R1: Incentivizing Reasoning Capability in LLMs via Reinforcement Learning}, 
      author={DeepSeek-AI},
      year={2025},
      journal={arXiv preprint arXiv:2501.12948}
}

@article{besta2025reasoning,
  title={Reasoning language models: A blueprint},
  author={Besta, Maciej and Barth, Julia and Schreiber, Eric and Kubicek, Ales and Catarino, Afonso and Gerstenberger, Robert and Nyczyk, Piotr and Iff, Patrick and Li, Yueling and Houliston, Sam and others},
  journal={arXiv preprint arXiv:2501.11223},
  year={2025}
}

@misc{mitre_cwe_top25,
  title  = {CWE Top 25 Most Dangerous Software Weaknesses},
  author = {The MITRE Corporation},
  howpublished = {\url{https://cwe.mitre.org/top25/}},
  note   = {Accessed: 2025‑08‑16},
  year   = {2025},
  organization = {MITRE}
}

@misc{cvss_3.1,
  author= {{Forum of Incident Response and Security Teams, Inc. (FIRST)}},
  title = {Common Vulnerability Scoring System v3.1: Specification Document},
  year  = {2019},
  howpublished = {\url{https://www.first.org/cvss/v3-1/specification-document}},
  note      = {Accessed: 2025-08-24},
}

@misc{github2021codeql-severity,
  author= {{GitHub}},
  title = {CodeQL code scanning: new severity levels for security alerts},
  howpublished = {\url{https://github.blog/changelog/2021-07-19-codeql-code-scanning-new-severity-levels-for-security-alerts/}},
  month  = jul,
  day = {19},
  year = {2021},
  note  = {Accessed: 2025‑08‑24},
}

@article{language-popularity,
  title={Popularity of programming languages},
  author={{\DJ}ur{\dj}ev, Darko},
  journal={AIDASCO Reviews},
  volume={2},
  number={2},
  pages={24--29},
  year={2024}
}

@manual{FlaskDocumentation,
  title = {Flask Documentation},
  author = {{Pallets Projects}},
  organization = {Pallets Projects},
  year  = {2025},
  note  = {Version 3.1.2},
  url  = {https://flask.palletsprojects.com/en/stable/},
  howpublished = {Online},
  lastchecked = {2025-09-11}
}

@inproceedings{fan2023large,
  title={Large language models for software engineering: Survey and open problems},
  author={Fan, Angela and Gokkaya, Beliz and Harman, Mark and Lyubarskiy, Mitya and Sengupta, Shubho and Yoo, Shin and Zhang, Jie M},
  booktitle={Proceedings of the IEEE/ACM International Conference on Software Engineering: Future of Software Engineering (ICSE-FoSE)},
  pages={31--53},
  year={2023},
  organization={IEEE}
}

@article{hou2024large,
  title={Large language models for software engineering: A systematic literature review},
  author={Hou, Xinyi and Zhao, Yanjie and Liu, Yue and Yang, Zhou and Wang, Kailong and Li, Li and Luo, Xiapu and Lo, David and Grundy, John and Wang, Haoyu},
  journal={ACM Transactions on Software Engineering and Methodology},
  volume={33},
  number={8},
  pages={1--79},
  year={2024},
  publisher={ACM}
}

@misc{SOSurvey2025AI,
  author = {Stack Overflow},
  title = {2025 Stack Overflow Developer Survey - AI},
  year = {2025},
  url = "https://survey.stackoverflow.co/2025/ai",
  note = "Accessed: 2025-10-29"
}

@article{zhang2023meta,
  title={Meta prompting for ai systems},
  author={Zhang, Yifan and Yuan, Yang and Yao, Andrew Chi-Chih},
  journal={arXiv preprint arXiv:2311.11482},
  year={2023}
}

@inproceedings{lykousas2023tales,
  title={Tales from the Git: Automating the detection of secrets on code and assessing developers’ passwords choices},
  author={Lykousas, Nikolaos and Patsakis, Constantinos},
  booktitle={Proceedings of the 8th IEEE European Symposium on Security and Privacy Workshops (EuroS\&PW)},
  pages={68--75},
  year={2023},
  organization={IEEE}
}

@inproceedings{onsori2024quantifying,
  title={Quantifying security issues in reusable JavaScript actions in GitHub workflows},
  author={Onsori Delicheh, Hassan and Decan, Alexandre and Mens, Tom},
  booktitle={Proceedings of the 21st International Conference on Mining Software Repositories (MSR)},
  pages={692--703},
  organization = {ACM},
  year={2024}
}

@inproceedings{2023wordstowatts,
  author={Samsi, Siddharth and Zhao, Dan and McDonald, Joseph and Li, Baolin and Michaleas, Adam and Jones, Michael and Bergeron, William and Kepner, Jeremy and Tiwari, Devesh and Gadepally, Vijay},
  booktitle={Proceedings of the 12th IEEE High Performance Extreme Computing Conference (HPEC)}, 
  title={From Words to Watts: Benchmarking the Energy Costs of Large Language Model Inference}, 
  year={2023},
  organization = {IEEE},
  pages={1--9}
}

@article{wang2025parameter,
  title={Parameter-efficient fine-tuning in large language models: a survey of methodologies},
  author={Wang, Luping and Chen, Sheng and Jiang, Linnan and Pan, Shu and Cai, Runze and Yang, Sen and Yang, Fei},
  journal={Artificial Intelligence Review},
  volume={58},
  number={8},
  pages={Article No.: 227},
  year={2025},
  publisher={Springer}
}

@article{
biderman2024lora,
title={Lo{RA} Learns Less and Forgets Less},
author={Dan Biderman and Jacob Portes and Jose Javier Gonzalez Ortiz and Mansheej Paul and Philip Greengard and Connor Jennings and Daniel King and Sam Havens and Vitaliy Chiley and Jonathan Frankle and Cody Blakeney and John Patrick Cunningham},
journal={Transactions on Machine Learning Research},
pages={1--39},
year={2024}
}

@misc{cloudsmith2025artifact,
  author = {{Cloudsmith}},
  title  = {2025 Artifact Management Report},
  howpublished = {\url{https://cloudsmith.com/campaigns/2025-artifact-management-report}},
  year   = {2025},
  note    = {Accessed: 2026-01-07},
}

@inproceedings{klemmer2024using,
  title = {Using AI Assistants in Software Development: A Qualitative Study on Security Practices and Concerns},
  author = {Klemmer, Jan H. and Horstmann, Stefan Albert and Patnaik, Nikhil and Ludden, Cordelia and Burton Jr., Cordell and Powers, Carson and Massacci, Fabio and Rahman, Akond and Votipka, Daniel and Lipford, Heather Richter and Rashid, Awais and Naiakshina, Alena and Fahl, Sascha},
  booktitle= {Proceedings of the 31st ACM SIGSAC Conference on Computer and Communications Security (CCS)},
  pages = {2726--2740},
  year  = {2024},
  organization = {ACM}
}

@inproceedings{jahanshahi2025cracks,
  title={Cracks in the stack: Hidden vulnerabilities and licensing risks in llm pre-training datasets},
  author={Jahanshahi, Mahmoud and Mockus, Audris},
  booktitle={Proceedings of the 2nd IEEE/ACM International Workshop on Large Language Models for Code (LLM4Code)},
  pages={104--111},
  year={2025},
  organization={IEEE}
}

@article{dolcetti2025dual,
  title={A Dual Perspective Review on Large Language Models and Code Verification},
  author={Dolcetti, Greta and Iotti, Eleonora},
  journal={Frontiers in Computer Science},
  volume={7},
  pages={1655469},
  year={2025},
  publisher={Frontiers}
}

@inproceedings{bruni2025benchmarking,
  title={Benchmarking prompt engineering techniques for secure code generation with gpt models},
  author={Bruni, Marc and Gabrielli, Fabio and Ghafari, Mohammad and Kropp, Martin},
  booktitle={Proceedings of the 2nd ACM International Conference on AI Foundation Models and Software Engineering (FORGE)},
  pages={93--103},
  year={2025},
  organization={IEEE}
}

@inproceedings{li2024fine,
  title={Fine tuning large language model for secure code generation},
  author={Li, Junjie and Sangalay, Aseem and Cheng, Cheng and Tian, Yuan and Yang, Jinqiu},
  booktitle={Proceedings of the 1st Special Event of AI Foundation Models and Software Engineering (FORGE)},
  pages={86--90},
  year={2024},
  organization={ACM}
}

@article{zhao2025towards,
  title={Towards secure code generation with LLMs: A study on common weakness enumeration},
  author={Zhao, Jianguo and Sun, Yuqiang and Huang, Cheng and Liu, Chengwei and Guan, YaoHui and Zeng, Yutong and Liu, Yang},
  journal={IEEE Transactions on Software Engineering},
  year={2025},
  volume={51},
  number={12},
  pages={3507--3523},
  publisher={IEEE}
}

@inproceedings{Schaad2025,
  author="Schaad, Andreas and G{\"o}tz, Stefan and Binder, Dominik",
  editor="Nemec Zlatolas, Lili and Rannenberg, Kai and Welzer, Tatjana and Garcia-Alfaro, Joaquin",
  title="You Still have to Study On the Security of LLM Generated Code",
  booktitle="Proceedings of the 40th IFIP International Conference on ICT Systems Security and Privacy Protection (SEC)",
  year="2025",
  organization="Springer",
  pages="111--124"
}

@article{li2024exploratory,
  title={An exploratory study on fine-tuning large language models for secure code generation},
  author={Li, Junjie and Rabbi, Fazle and Cheng, Cheng and Sangalay, Aseem and Tian, Yuan and Yang, Jinqiu},
  journal={arXiv preprint arXiv:2408.09078},
  year={2024}
}

@article{li2025safegenbench,
  title={SafeGenBench: A Benchmark Framework for Security Vulnerability Detection in LLM-Generated Code},
  author={Li, Xinghang and Ding, Jingzhe and Peng, Chao and Zhao, Bing and Gao, Xiang and Gao, Hongwan and Gu, Xinchen},
  journal={arXiv preprint arXiv:2506.05692},
  year={2025}
}

@misc{ibm_devsecops,
  author = {{IBM}},
  title = {What is DevSecOps?},
  url = {https://www.ibm.com/think/topics/devsecops},
  note = {Accessed: 2026-04-05}
}

@article{twist2025llms,
  title={LLMs Love Python: A Study of LLMs' Bias for Programming Languages and Libraries},
  author={Twist, Lukas and Zhang, Jie M and Harman, Mark and Syme, Don and Noppen, Joost and Nauck, Detlef},
  journal={arXiv e-prints},
  pages={arXiv--2503},
  year={2025}
}

@misc{github_octoverse_2025,
  author= {{GitHub}},
  title   = {Octoverse: A new developer joins GitHub every second as AI leads TypeScript to \#1},
  year   = {2025},
  howpublished = {\url{https://github.blog/news-insights/octoverse/octoverse-a-new-developer-joins-github-every-second-as-ai-leads-typescript-to-1/}},
  note    = {Accessed: 2026-05-04}
}

@article{kocetkov2022stack,
  title={The stack: 3 tb of permissively licensed source code},
  author={Kocetkov, Denis and Li, Raymond and Allal, Loubna Ben and Li, Jia and Mou, Chenghao and Ferrandis, Carlos Mu{\~n}oz and Jernite, Yacine and Mitchell, Margaret and Hughes, Sean and Wolf, Thomas and others},
  journal={arXiv preprint arXiv:2211.15533},
  year={2022}
}

@inproceedings{fawzy2025vibe,
  title={Vibe Coding in Practice: Motivations, Challenges, and a Future Outlook--a Grey Literature Review},
  author={Fawzy, Ahmed and Tahir, Amjed and Blincoe, Kelly},
  booktitle={Proceedings of the 48th IEEE/ACM International Conference on Software Engineering: Software Engineering in Practice (ICSE-SEIP)},
  year={2026},
  pages ={212--223},
  organization={IEEE}
}

@inproceedings{stacey-etal-2026-improving,
    title = "Improving the {OOD} Performance of Closed-Source {LLM}s on {NLI} Through Strategic Data Selection",
    author = "Stacey, Joe and Alazraki, Lisa  and Ubhi, Aran and Ermis, Beyza and Mueller, Aaron and Rei, Marek",
    booktitle = "Proceedings of the 19th Conference of the European Chapter of the Association for Computational Linguistics (EACL): Findings",
    year = "2026",
    organization = "ACL",
    pages = "5378--5404"
}

@inproceedings{hevechev2023,
  author = {He, Jingxuan and Vechev, Martin},
  title  = {Large Language Models for Code: Security Hardening and Adversarial Testing},
  booktitle = {Proceedings of the 30th ACM SIGSAC Conference on Computer and Communications Security (CCS)},
  year = {2023},
  pages = {1865--1879},
  organization = {ACL}
}

@inproceedings{Ernst2003:WODA,
  author    = {Michael D. Ernst},
  title     = {Static and dynamic analysis: Synergy and duality},
  booktitle = {Proceedings of the 1st International Workshop on Dynamic Analysis (WODA)},
  pages     = {24--27},
  organization = {ACM},
  year      = {2003}
}

@inproceedings{blyth2025static,
  author = {Blyth, S. and Licorish, S. A. and Treude, C. and Wagner, M.},
  title  = {Static Analysis as a Feedback Loop: Enhancing {LLM}-Generated Code Beyond Correctness},
  booktitle = {Proceedings of the 25th IEEE International Conference on Source Code Analysis and Manipulation (SCAM)},
  pages  = {100--109},
  organization = {IEEE},
  year   = {2025}
}

@inproceedings{ravi2025llmloop,
  author = {Ravi, R. and Bradshaw, D. and Ruberto, S. and Jahangirova, G. and Terragni, V.},
  title = {{LLMLOOP}: Improving {LLM}-Generated Code and Tests Through Automated Iterative Feedback Loops},
  booktitle = {Proceedings of the 41st IEEE International Conference on Software Maintenance and Evolution (ICSME)},
  pages = {930--934},
  organization = {IEEE},
  year = {2025}
}

@inproceedings{pearce2023,
  author={Pearce, Hammond and Tan, Benjamin and Ahmad, Baleegh and Karri, Ramesh and Dolan-Gavitt, Brendan},
  booktitle={Proceedings of the 44th IEEE Symposium on Security and Privacy (SP)}, 
  title={Examining Zero-Shot Vulnerability Repair with Large Language Models}, 
  year={2023},
  pages={2339--2356},
  organization = {IEEE}
}

@article{tunter,
title = {Tunter: Assessing Exploitability of Vulnerabilities with Taint-Guided Exploitable States Exploration},
journal = {Computers \& Security},
volume = {124},
pages = {102995},
year = {2023},
author = {Ruipeng Wang and Kaixiang Chen and Zulie Pan and Yuwei Li and Qianyu Li and Yang Li and Min Zhang and Chao Zhang},
}

@inproceedings{dataset,
  author    = {Ali Soltanian Fard Jahromi and Amjed Tahir and Peng Liang and Foutse Khomh},
  title     = {Replication Package for the Study: On Fixing Insecure AI-Generated Code through Model Fine-Tuning and Prompting Strategies},
  year      = {2026},
  publisher  = {\url{https://github.com/AliSoltanianFJ/CodeSecurity2025}}
}

@misc{cve,
  author       = {{CVE Program}},
  title        = {{Common Vulnerabilities and Exposures (CVE)}},
  year         = {2026},
  howpublished = {\url{https://www.cve.org/}},
  note         = {Accessed: 2026-11-15}
}

\appendix

\section{Network Graph of CWEs Introduced By Model Refinement Techniques}
\label{appendixA}

\begin{figure}[p]
    \centering
    \rotatebox{90}{%
        \includegraphics[
            width=0.90\textheight,
            height=0.90\textwidth,
            keepaspectratio
        ]{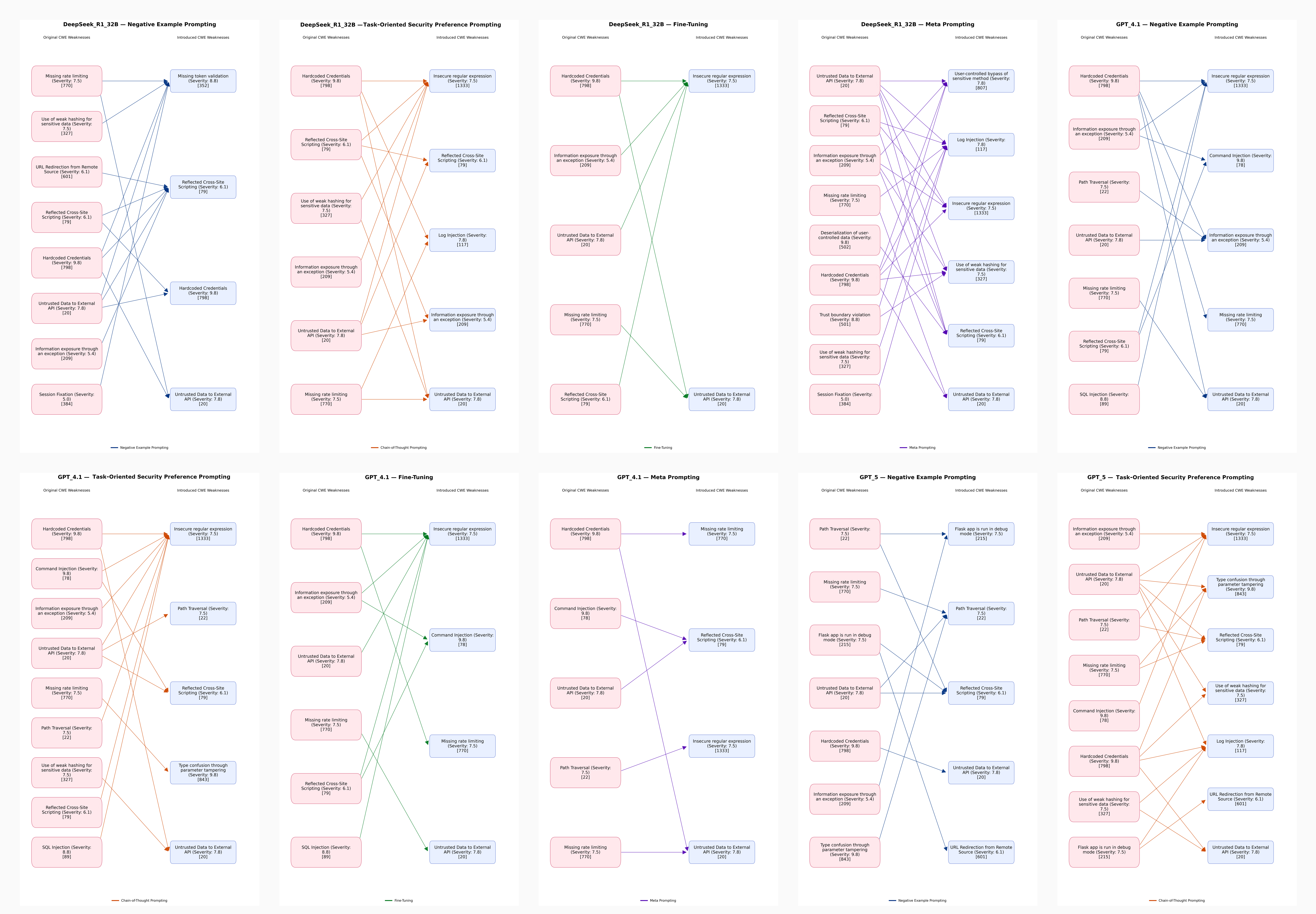}
    }
    \caption{\revision{Network graph illustrating CWEs introduced by each model refinement technique, along with the corresponding original CWEs. Results are aggregated by programming language (1 of 2).}}
    \label{fig:introduced_cwe_network_graph_1}
\end{figure}

\begin{figure}[p]
    \centering
    \rotatebox{90}{%
        \includegraphics[
            width=0.90\textheight,
            height=0.90\textwidth,
            keepaspectratio
        ]{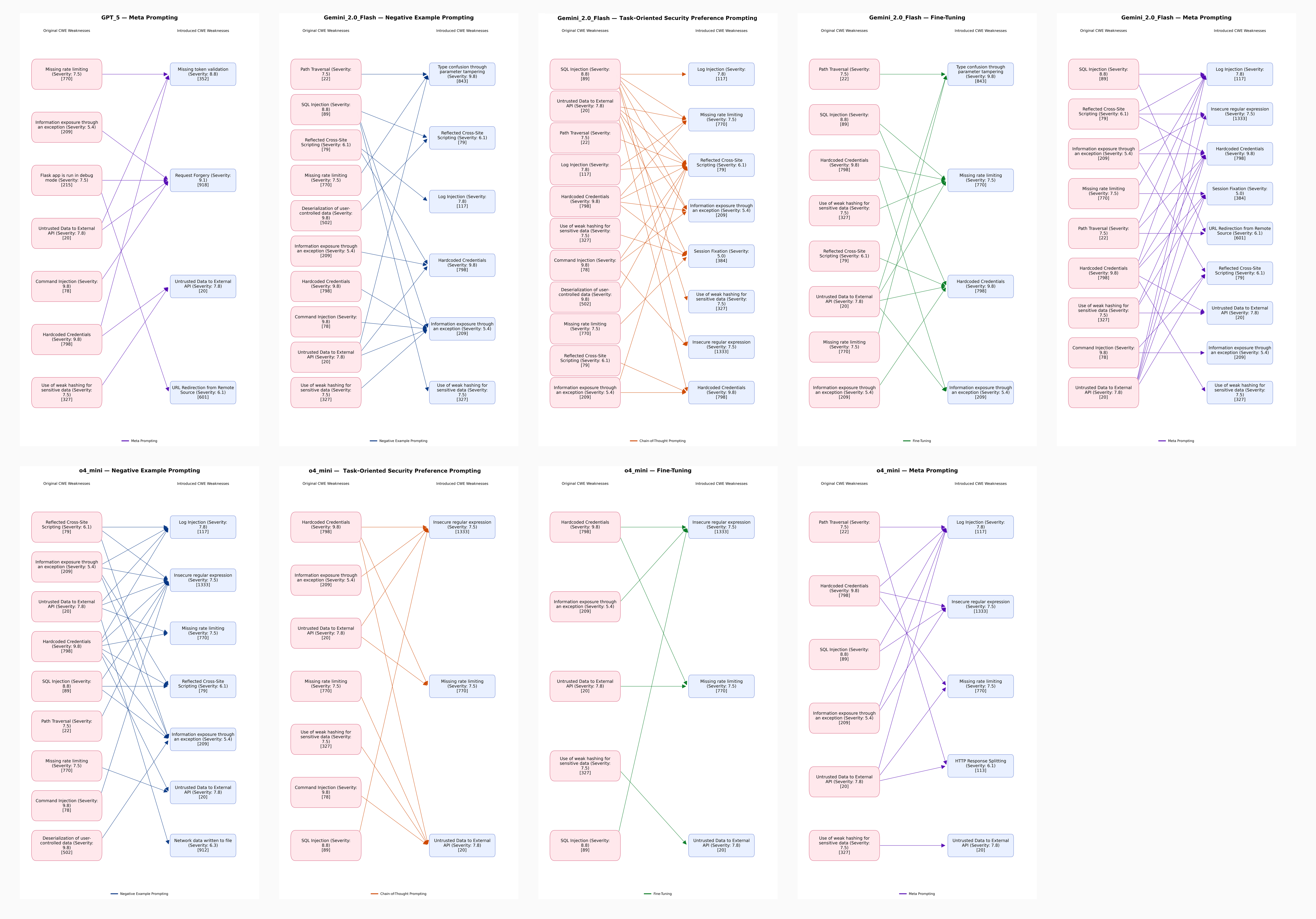}
    }
    \caption{\revision{Network graph illustrating CWEs introduced by each model refinement technique, along with the corresponding original CWEs (2 of 2).}}
    \label{fig:introduced_cwe_network_graph_2}
\end{figure}

\clearpage

\FloatBarrier
\section{Functional Test Pass Rates}
\label{appendixB}

\begin{figure}[!htbp]
    \centering
    \includegraphics[width=1\linewidth]{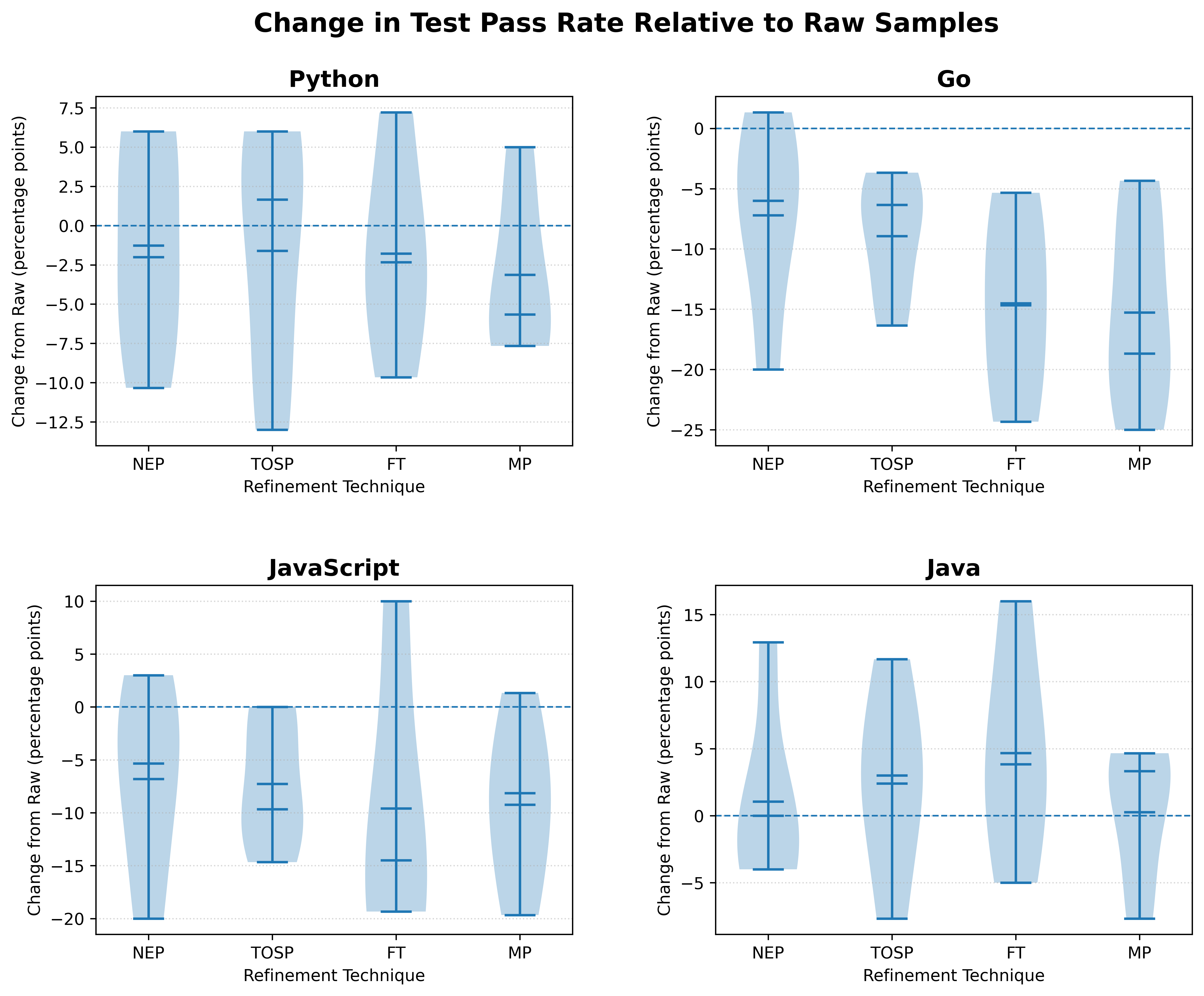}
    \caption{\revision{Violin plot illustrating the differences in test pass rates between the original raw samples and the samples generated after applying the model output refinement techniques.}}
    \label{fig:violin_functional_correctness}
\end{figure}

\revision{Figure~\ref{fig:violin_functional_correctness} displays the differences in testpass rates between the raw samples and refined samples. Tables~\ref{tab:functional-python}--\ref{tab:functional-java} report the functional test pass rates for the original
raw code and the code produced after applying refinement techniques.}

\begin{table*}[htbp]
\centering
\caption{\textcolor{black}{Functional test pass rates for Python code.}}
\label{tab:functional-python}
\small
\setlength{\tabcolsep}{7pt}
\begin{tabular}{lccccc}
\toprule
\revision{Model} & \revision{Original} & \revision{NEP} & \revision{TOSP} & \revision{FT} & \revision{MP} \\
\midrule
\revision{GPT-4.1}
& \revision{96.67\%} 
& \revision{94.67\%} 
& \revision{98.33\%} 
& \revision{94.00\%} 
& \revision{96.33\%}  \\

\revision{Gemini 2.0 Flash}
& \revision{84.00\%}
& \revision{90.00\%} 
& \revision{88.33\%} 
& \revision{91.21\%} 
& \revision{89.00\%}  \\

\revision{o4-mini}
& \revision{88.33\%} 
& \revision{94.00\%} 
& \revision{94.33\%} 
& \revision{86.33\%} 
& \revision{80.67\%}  \\

\revision{DeepSeek R1 32B}
& \revision{98.33\%} 
& \revision{92.67\%} 
& \revision{91.33\%} 
& \revision{88.67\%} 
& \revision{91.33\%}  \\

\revision{GPT-5}
& \revision{96.33\%} 
& \revision{86.00\%} 
& \revision{83.33\%} 
& \revision{N/A}
& \revision{90.67\%}  \\
\bottomrule
\end{tabular}
\end{table*}

\begin{table*}[htbp]
\centering
\caption{\textcolor{black}{Functional test pass rates for Go code.}}
\label{tab:functional-go}
\small
\setlength{\tabcolsep}{7pt}
\begin{tabular}{lccccc}
\toprule
\revision{Model} & \revision{Original} & \revision{NEP} & \revision{TOSP} & \revision{FT} & \revision{MP} \\
\midrule
\revision{GPT-4.1}
& \revision{99.00\% }
& \revision{90.00\%} 
& \revision{87.00\%} 
& \revision{81.00\% }
& \revision{79.63\%}  \\

\revision{Gemini 2.0 Flash}
& \revision{96.00\% }
& \revision{76.00\% }
& \revision{92.33\% }
& \revision{71.67\% }
& \revision{71.00\% } \\

\revision{o4-mini}
& \revision{81.67\% }
& \revision{79.33\% }
& \revision{75.33\% }
& \revision{70.67\% }
& \revision{77.33\% } \\

\revision{DeepSeek R1 32B}
& \revision{66.67\%} 
& \revision{68.00\%} 
& \revision{50.33\%} 
& \revision{61.33\%} 
& \revision{48.00\%}  \\

\revision{GPT-5}
& \revision{88.00\%} 
& \revision{82.00\%} 
& \revision{81.67\%} 
& \revision{N/A}
& \revision{79.00\%}  \\
\bottomrule
\end{tabular}
\end{table*}

\begin{table*}[htbp]
\centering
\caption{\textcolor{black}{Functional test pass rates for JavaScript code.}}
\label{tab:functional-javascript}
\small
\setlength{\tabcolsep}{7pt}
\begin{tabular}{lccccc}
\toprule
\revision{Model} & \revision{Original} & \revision{NEP} & \revision{TOSP} & \revision{FT} & \revision{MP} \\
\midrule
\revision{GPT-4.1}
& \revision{100.00\%} 
& \revision{99.33\%} 
& \revision{100.00\%} 
& \revision{81.67\%} 
& \revision{91.85\%}  \\

\revision{Gemini 2.0 Flash}
& \revision{94.67\%} 
& \revision{97.67\%} 
& \revision{93.33\%} 
& \revision{75.33\%} 
& \revision{88.67\%}  \\

\revision{o4-mini}
& \revision{95.67\%} 
& \revision{90.33\%} 
& \revision{81.00\%} 
& \revision{85.00\%} 
& \revision{82.00\%}  \\

\revision{DeepSeek R1 32B}
& \revision{88.67\%} 
& \revision{77.67\%} 
& \revision{78.00\%} 
& \revision{98.67\%} 
& \revision{69.00\%}  \\

\revision{GPT-5}
& \revision{96.67\%} 
& \revision{76.67\%} 
& \revision{87.00\%} 
& \revision{N/A}
& \revision{98.00\%}  \\
\bottomrule
\end{tabular}
\end{table*}

\begin{table*}[htbp]
\centering
\caption{\textcolor{black}{Functional test pass rates for Java code.}}
\label{tab:functional-java}
\small
\setlength{\tabcolsep}{7pt}
\begin{tabular}{lccccc}
\toprule
\revision{Model} & \revision{Original} & \revision{NEP} & \revision{TOSP} & \revision{FT} & \revision{MP} \\
\midrule
\revision{GPT-4.1}
& \revision{93.33\%} 
& \revision{93.33\% }
& \revision{96.33\% }
& \revision{97.00\% }
& \revision{96.67\%}  \\

\revision{Gemini 2.0 Flash}
& \revision{95.67\% }
& \revision{91.67\% }
& \revision{95.33\% }
& \revision{90.67\% }
& \revision{88.00\%}  \\

\revision{o4-mini}
& \revision{80.00\% }
& \revision{92.93\% }
& \revision{91.67\% }
& \revision{96.00\% }
& \revision{83.33\% } \\

\revision{DeepSeek R1 32B}
& \revision{89.67\% }
& \revision{86.00\% }
& \revision{95.00\% }
& \revision{93.67\% }
& \revision{94.33\% } \\

\revision{GPT-5}
& \revision{96.67\%} 
& \revision{96.67\%} 
& \revision{89.00\% }
& \revision{N/A}
& \revision{94.33\% } \\
\bottomrule
\end{tabular}
\end{table*}

\end{document}